\documentclass[trackchanges, twocolumn]{aastex701}

\usepackage{amsmath}
\usepackage{amssymb}
\usepackage{amsthm}
\usepackage{CJK}
\usepackage{natbib,url,bm}
\usepackage{array}
\usepackage{graphicx}
\usepackage{subfigure}
\usepackage{color}
\usepackage{longtable}
\usepackage{threeparttable} 
\usepackage{booktabs}
\usepackage{enumitem}
\usepackage{hyperref}

\begin{document}

\begin{CJK*}{UTF8}{gbsn}

\title{Unveiling Multimessenger Emission from Hidden Cores of Microquasars}

\author[orcid=0000-0002-9775-2692]{Yu-Jia Wei (魏煜佳)}
\affiliation{Department of Astronomy and Astrophysics, The Pennsylvania State University, 525 Davey Laboratory, University Park, PA 16802, USA}
\affiliation{Institute for Gravitation and the Cosmos, The Pennsylvania State University, University Park, PA 16802, USA}
\email[show]{yjw5518@psu.edu}

\author[orcid=0000-0002-5358-5642]{Kohta Murase}
\affiliation{Department of Physics, The Pennsylvania State University, University Park, PA 16802, USA}
\affiliation{Department of Astronomy and Astrophysics, The Pennsylvania State University, 525 Davey Laboratory, University Park, PA 16802, USA}
\affiliation{Institute for Gravitation and the Cosmos, The Pennsylvania State University, University Park, PA 16802, USA}
\affiliation{Center for Gravitational Physics and Quantum Information, Yukawa Institute for Theoretical Physics, Kyoto University, Kyoto, Kyoto 606-8502, Japan}
\email[show]{murase@psu.edu}

\author[orcid=0000-0003-2478-333X]{B. Theodore Zhang (张兵)}
\affiliation{Key Laboratory of Particle Astrophysics and Experimental Physics Division and Computing Center, Institute of High Energy Physics, Chinese Academy of Sciences, 100049 Beijing, China}
\affiliation{TIANFU Cosmic Ray Research Center, Chengdu, Sichuan, China}
\email[show]{zhangbing@ihep.ac.cn}

\begin{abstract}
Microquasars are radio-emitting X-ray binaries with relativistic jets and are established sources of $\sim 100$~TeV gamma rays, making them promising candidates for cosmic-ray acceleration.
Motivated by recent detections of $\sim 100~$TeV photons from Cygnus~X-1 and multi-PeV photons from Cygnus~X-3, 
we employ the Astrophysical Multimessenger Emission Simulator (AMES) to model their multimessenger emission from radio to ultrahigh-energy gamma rays.
Our modeling suggests that particle acceleration is extremely efficient in the jet region. The observed $\gtrsim 0.1$~PeV gamma rays can originate from either $p\gamma$ or $pp$ interactions, depending on the location and physical conditions of the emission region, while also reproducing the lower-energy spectra. These configurations yield observationally testable predictions. In the poorly constrained $0.1$-$10$~TeV band, the models predict either a deep valley, a mild suppression, or a power-law spectrum.
Additionally, models involving compact emission regions comparable to the orbital separation predict strong variability, while those invoking more extended and static external zones show more stable behavior.
We also provide a possible qualitative explanation for the energy-dependent modulation patterns, relying primarily on changes in the Doppler factor and external $\gamma\gamma$ absorption. In particular, explaining the PeV emission from Cygnus~X-3 favors a compact emission region, for which a magnetic field strength of order $10^2~\mathrm{G}$ is required.
Finally, after accounting for pion and muon cooling, the predicted neutrino flux is suppressed, implying that detection is more challenging than previously thought. 
\end{abstract}

\keywords{\uat{X-ray binary stars}{1811} -- \uat{Jets}{870} -- \uat{Non-thermal radiation sources}{1119} -- \uat{Cosmic rays}{329} -- \uat{Neutrino astronomy}{1100}}


\section{Introduction}
\label{sect:intro}
Microquasars are radio-emitting X-ray binaries (XRBs) characterized by the presence of relativistic jets, typically observed within the Milky Way or in nearby galaxies~\citep[e.g.,][for a review]{Mirabel-1999ARAA..37..409M, Fender-2001AIPC..558..221F, Done-2007AARv..15....1D, Zhang-2013FrPhy...8..630Z, Yuan-2014ARAA..52..529Y}. These systems consist of a compact object, such as a stellar-mass black hole (BH) or a neutron star (NS), accreting material from a companion star. As the accreted matter carries angular momentum, it forms an accretion disk around the compact object. The disk is the primary source of X-ray emission and drives the formation and collimation of relativistic jets. The jets, in turn, produce non-thermal radiation across a broad range of wavelengths, from radio to gamma-rays.

Previous studies have shown that microquasars exhibit two classical accretion states: the high/soft state and the low/hard state, in order of decreasing luminosity, originally identified in Cygnus~X-1~\citep{Tananbaum-1972ApJ...177L...5T}.
Persistent jets are generally expected during the low/hard state, while transient ejections occur when the source transitions from the low/hard to the high/soft state~\citep[e.g.,][]{Fender-2003MNRAS.343L..99F}. In addition to these compact (AU-scale) jets, extended large-scale jets are thought to arise from the long-term interaction of steady jets with the interstellar medium (ISM)~\citep[e.g.,][for a review]{Ribo-2005ASPC..340..269R, Fender-2006csxs.book..381F}.
Also, it is noted that weak jet activity in the soft state cannot be entirely ruled out~\citep{Rushton-2012MNRAS.419.3194R}.

XRBs have long been proposed as potential sources of high-energy neutrinos and gamma rays~\citep[e.g.,][]{Eichler-1984Natur.307..613E, Stecker-1985Natur.316..418S, Gaisser-1985PhRvL..54.2265G, Levinson-2001PhRvL..87q1101L, Distefano-2002ApJ...575..378D, Reynoso-2009AA...493....1R}.
Recently, microquasars have been established as 100 TeV photon emitters. Gamma-ray emission above $100$~TeV from the extended jet of V4641~Sgr has been detected by the High Altitude Water Cherenkov (HAWC) Observatory~\citep{Alfaro-2024Natur.634..557A} and H.E.S.S.~\citep{Acharyya-2025arXiv251110537A}. Furthermore, the Large High Altitude Air Shower Observatory (LHAASO) has reported gamma-ray emission up to $\sim 100~$TeV from the microquasars SS~433, V4641~Sgr, GRS~1915+105, MAXI~J1820+070, and Cygnus~X-1~\citep{LHAASO-2024arXiv241008988L} and up to ∼PeV from Cygnus~X-3~\citep{LHAASO-2025arXiv251216638L}. These observations indicate that microquasars are highly efficient particle accelerators capable of energizing particles beyond one PeV ($>10^{15}~$eV). 
Theoretical studies further suggest that microquasars may contribute to cosmic-ray spectra near the knee ($\sim 3~$PeV)~\citep[e.g.,][]{Cooper-2020MNRAS.493.3212C, Zhang-2025arXiv250620193Z,Kachelriess-2025AA...701A..22K}. 

However, the origin of the non-thermal gamma rays in microquasars remains uncertain. Two primary mechanisms have been proposed: leptonic processes~\citep[e.g.,][]{Dubus-2010MNRAS.404L..55D, Zdziarski-2012MNRAS.421.2956Z,Zdziarski-2018MNRAS.479.4399Z}, including synchrotron radiation and inverse Compton (IC) scattering from relativistic electrons, and hadronic processes~\citep[e.g.,][]{Bednarek-2005ApJ...631..466B, Reynoso-2008MNRAS.387.1745R, Baerwald-2013ApJ...773..159B}, involving proton synchrotron radiation, inelastic hadronuclear ($pp$) collisions, Bethe-Heitler (BH) pair production, and photomeson ($p\gamma$) interactions. 
Hadronic emission is of particular interest because photons above $0.1$~PeV, which are linked to cosmic rays (CRs) and neutrinos, are not easily explained by purely leptonic processes. 
While some lepto-hadronic models~\citep[e.g.,][]{Pepe-2015AA...584A..95P, Kimura-2020ApJ...904..188K, Kantzas-2021MNRAS.500.2112K, Kuze-2025ApJ...985..139K, Carpio-2025arXiv250622550C} have been developed to identify the dominant mechanism, most neglect crucial cooling processes such as muon and pion cooling, which can substantially affect the production of high-energy photons and neutrinos.

To address these limitations, we extend the Astrophysical Multimessenger Emission Simulator (AMES), which has been developed in a sequence of papers by a part of the authors~\citep[e.g.,][]{Murase-2012ApJ...745L..16M, Murase-2015ApJ...805...82M, Murase-2018PhRvD..97h1301M, Murase-2022ApJ...941L..17M, Zhang-2023MNRAS.524...76Z, Murase-2024PhRvD.109j3020M}, to self-consistently model both leptonic and hadronic processes responsible for multimessenger emission from microquasars. 
Motivated by recent detections of $\sim 100$~TeV photons from Cygnus~X-1~\citep{LHAASO-2024arXiv241008988L} and $\sim$~PeV photons from Cygnus~X-3~\citep{LHAASO-2025arXiv251216638L}, we investigate their potential multimessenger emission in this work using AMES.
To comprehensively model both the $0.1$-$1$~PeV and lower-energy band photons, we develop three distinct scenarios, each representing different physical conditions of the emission region: the jet-core model (Scenario~A), the stellar-wind interaction model (Scenario~B), and the extended-jet model (Scenario~C).
In each scenario, we assume two uniform emission zones, focusing on compact or effectively unresolved emission sites, which provide a framework for predicting the broadband photon and neutrino spectra from microquasars.

In Sect.~\ref{sect:model}, we present the theoretical framework, including the binary geometry, three physical scenarios, the numerical methods implemented in AMES, the origins of external seed photons, external absorption processes, and particle escape timescales.
In Sect.~\ref{sect:results}, we model the multimessenger emission from the microquasars Cygnus~X-1 and Cygnus~X-3 to investigate the production of high-energy photons and neutrinos, assess their potential for accelerating CRs, and constrain the magnetic field in the emission region.
In Sect.~\ref{sect:modulation}, we investigate the energy-dependent orbital modulation arising from variations in the Doppler factor and external $\gamma\gamma$ annihilation.
In Sect.~\ref{sect:neutrino}, we further examine the influence of muon and pion cooling in each scenario and predict the expected number of neutrino events for current and next-generation neutrino detectors.
Finally, we summarize our works in Sect.~\ref{sect:summary}.

We use notations as $Q_x=Q/10^x$ in the CGS unit. 
Quantities in the laboratory (lab) frame are unprimed, while those in the co-moving frame are primed.
Especially, we adopt the notation $E = \delta \varepsilon'$, where $E$ is the lab frame energy, $\varepsilon'$ is the co-moving frame energy, and $\delta$ is the Doppler factor. In this work, we do not distinguish between the lab and observer frames since microquasars reside in the Milky Way or nearby galaxies.

\section{Model}
\label{sect:model}

\begin{figure}[ht!]
    \centering
    \includegraphics[width=0.45\textwidth]{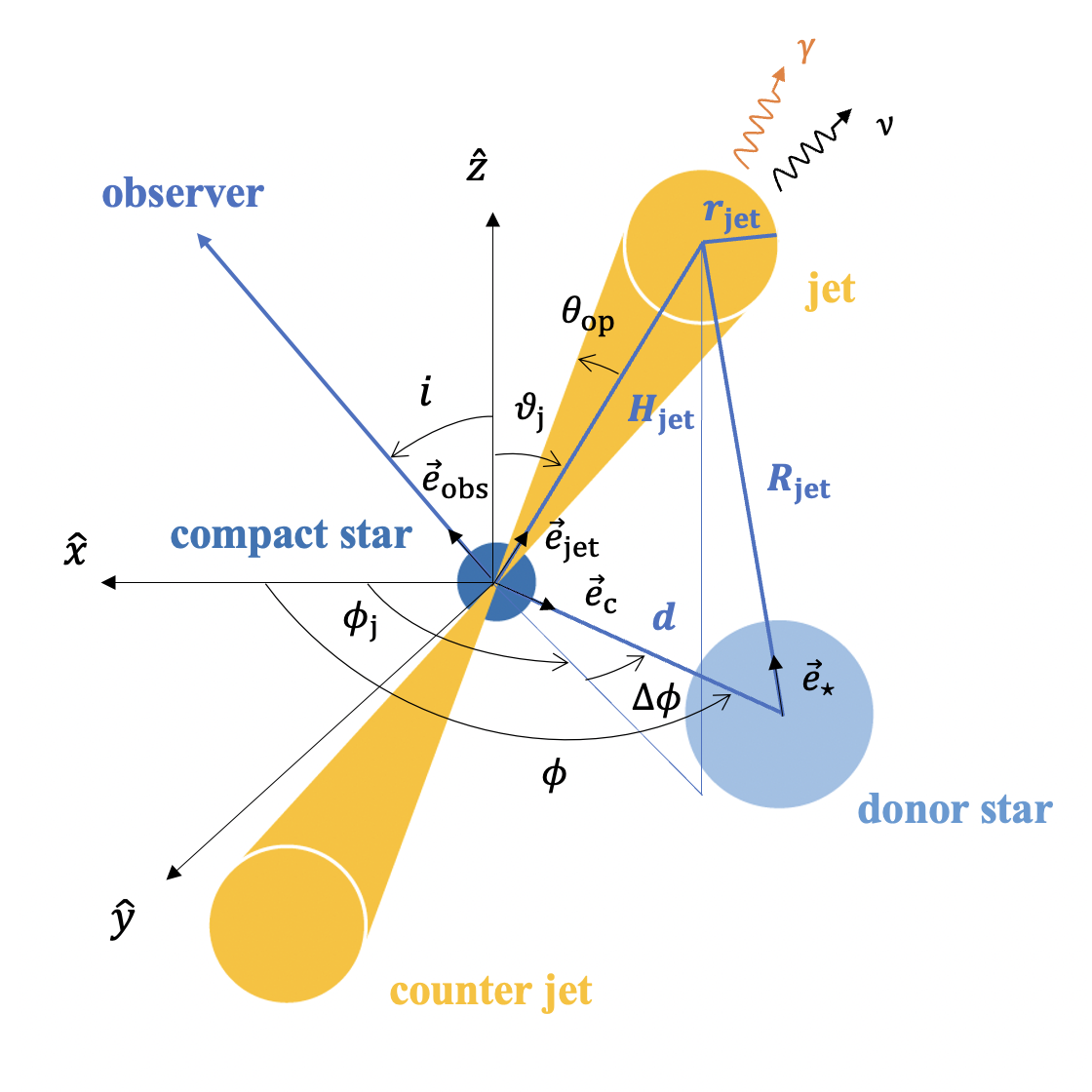}
    \caption{A schematic picture of the geometry, assuming that neutrinos and photons are produced within the jet blob.}
    \label{fig:model}
\end{figure}

\begin{figure*}[ht!]
	\centering
        \subfigure[]{
	    \begin{minipage}[b]{0.3\textwidth}
	    \begin{center}
	    \includegraphics[width=1.0\textwidth]{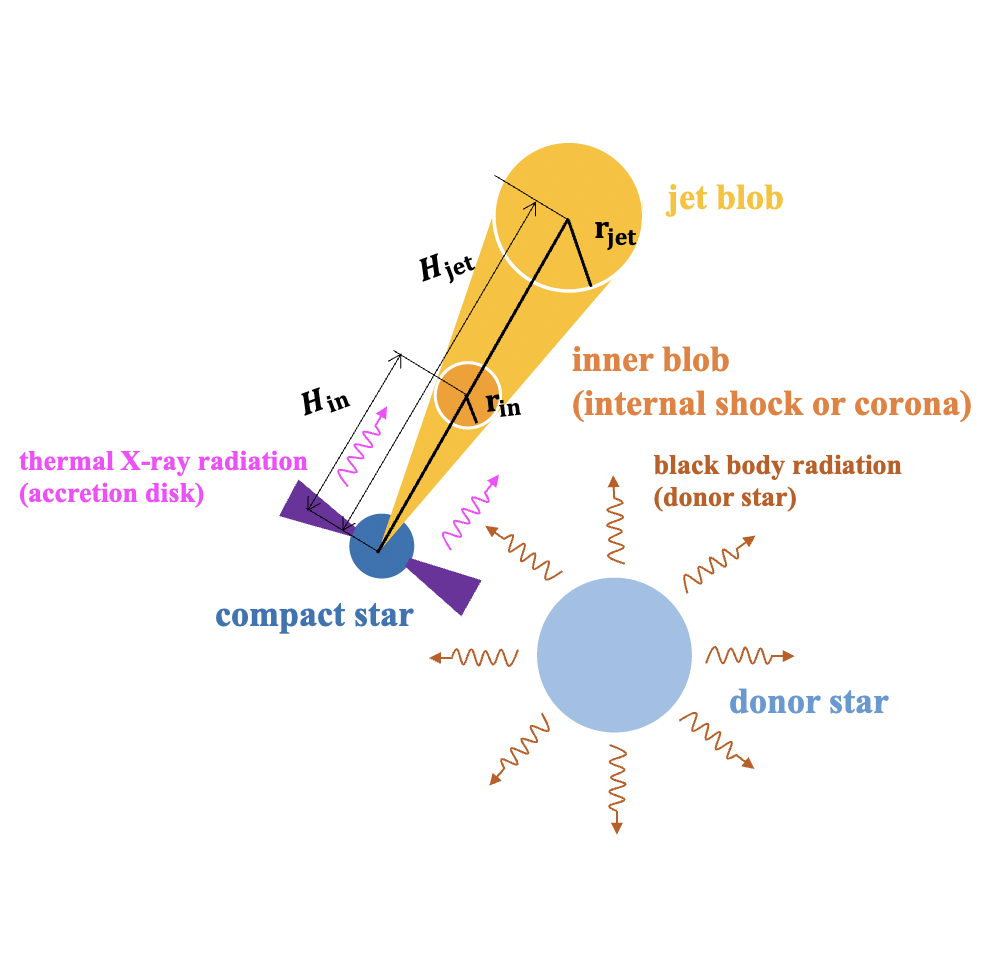}
	    \end{center}
	    \end{minipage}	 
        \label{fig:model-pgamma}
	}%
	\subfigure[]{
	    \begin{minipage}[b]{0.3\textwidth}
	    \begin{center}
	    \includegraphics[width=1.0\textwidth]{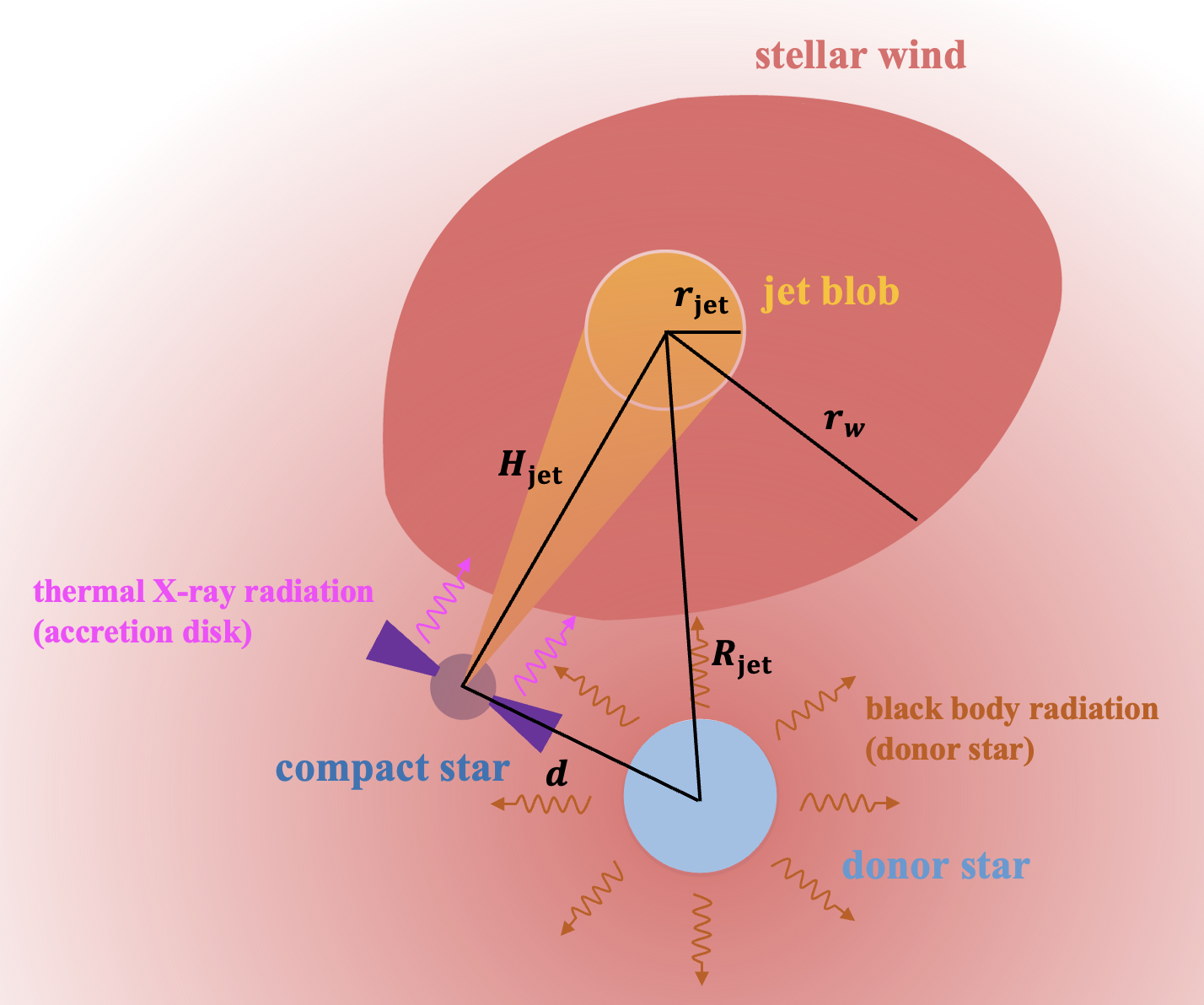}
	    \end{center}
	    \end{minipage}	 
        \label{fig:model-pp1}
	}%
 	\subfigure[]{
	    \begin{minipage}[b]{0.3\textwidth}
	    \begin{center}
	    \includegraphics[width=1.0\textwidth]{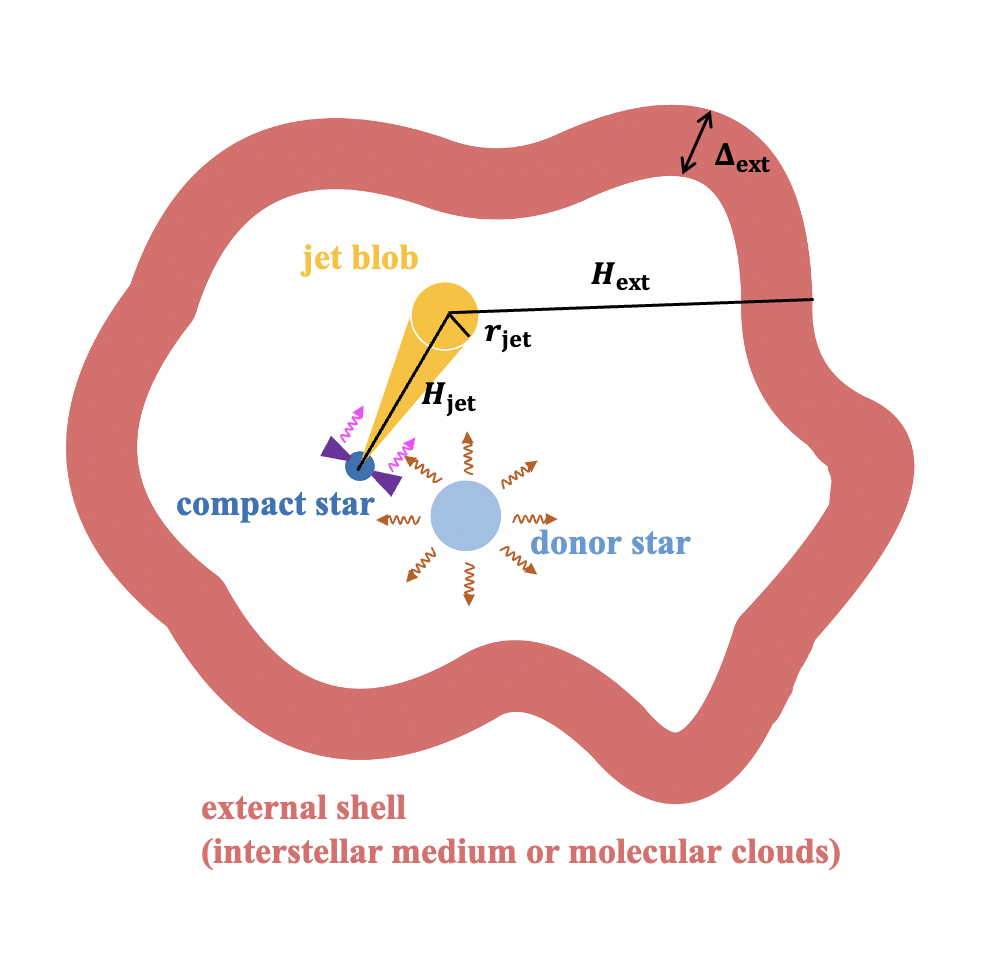}
	    \end{center}
	    \end{minipage}
        \label{fig:model-pp2}
	}%
	\caption{Schematic pictures of the three physical configurations modeled in this work. \textbf{Panel (a):} Scenario~A consists of an inner blob, originating from either the (outflowing) corona or an internal shock, and a jet blob, produced by either an internal or termination shock. Both blobs lie along the same direction but are located at different heights and move with distinct velocities. \textbf{Panel (b):} Scenario~B features a primary jet emission region surrounded by the dense stellar wind environment of the companion star. In the static external emission zone, charged particles escaping from the jet interact with the stellar wind. \textbf{Panel (c):} Scenario~C considers the contribution from the extended jet. 
    Within the large-scale, static external shell, particles that have escaped from the jet blob interact with the surrounding parsec-scale environment.}
\label{fig:model-2zone}
\end{figure*}

The binary system related to the microquasar consists of a donor star (e.g., a massive O-type star or Wolf-Rayet star) and a compact object (e.g., a black hole or neutron star). 
Fig.~\ref{fig:model} illustrates this system's geometry. For clarity, the schematic picture depicts radiation produced via leptonic and hadronic processes within a single spherical relativistic blob (a jet blob) moving with Lorentz factor $\Gamma_j$. In our calculations, we consider two distinct emission zones: the jet blob and a second emission region, as described below and illustrated in Fig.~\ref{fig:model-2zone}. We neglect the contribution from the counter jet, which is negligible compared with that from the primary jet in our models.

We define a Cartesian coordinate system centered on the compact object, with the binary orbital axis along the $z$-axis and the line of sight in the $xz$-plane. The observer direction is specified by the inclination angle $i$, while the donor-star position in the orbital plane is described by the azimuthal angle $\phi$. The jet direction is characterized by the polar and azimuthal angles $(\vartheta_j,\phi_j)$. Assuming a conical jet geometry, the co-moving radius of the jet blob is estimated as $r'_{\rm jet}\simeq \theta_{\rm op} H_{\rm jet}$, where $\theta_{\rm op}$ is the jet half-opening angle and $H_{\rm jet}$ is the blob height measured in the lab frame.

The relevant unit vectors are
\begin{equation}
    \begin{aligned}
        & \vec{e}_{\rm obs} = \left( \sin i, 0, \cos i \right), \\
        & \vec{e}_c = \left( \cos \phi, \sin \phi, 0 \right), \\
        & \vec{e}_{\rm jet} = \left( \cos \phi_j \sin \vartheta_j,  \sin \phi_j \sin \vartheta_j, \cos \vartheta_j \right), \\ 
        & \vec{e}_\star = \left( H_{\rm jet} \vec{e}_{\rm jet} - d \vec{e}_c \right) / R_{\rm jet}, 
    \end{aligned}
\end{equation}
where $\vec{e}_{\rm obs}$, $\vec{e}_c$, $\vec{e}_{\rm jet}$, and $\vec{e}_{\star}$ denote the directions of the observer, the donor star, the jet blob, and the vector from the donor star to the jet blob, respectively. Here, $d$ is the binary separation and $R_{\rm jet}$ represents the distance between the donor star and the jet blob. 
The Doppler factor of the jet blob is then given by 
\begin{equation}
    \delta_j = \left[\Gamma_j (1 - \beta_j \vec{e}_{\rm obs} \cdot \vec{e}_{\rm jet})\right]^{-1},
\end{equation}
where $\beta_j$ is the jet velocity $v_j$ in units of the speed of light. 

To comprehensively model both the $>0.1$~PeV and lower-energy photons, we propose three specific physical scenarios in which two distinct regions may contribute to different components of the observed radiation. In all cases, one emission region is a jet blob located at height $H_{\rm jet}$, which may form through internal shocks between jet shells with different velocities or through a termination shock caused by the interaction between the jet and the surrounding medium. The second region differs among the scenarios, as described below.

\begin{itemize}
    \item Jet-Core Model (Scenario~A): Motivated by the possible corona-jet connection in Cygnus~X-1 suggested by X-ray spectropolarimetry~\citep{Steiner-2024ApJ...969L..30S} and by theoretical studies of particle acceleration in X-ray binary coronae~\citep{Fang-2024ApJ...975L..35F}, we propose Scenario~A. In this model, in addition to the jet blob, we include an inner blob with radius $r'_{\rm in}$ located at height $H_{\rm in}$, which may originate from either the corona or internal shocks in the compact outflow, as illustrated in Fig.~\ref{fig:model-pgamma}. We assume that both blobs are co-aligned and that $H_{\rm in} \ll H_{\rm jet}$.
    \item Stellar-Wind Interaction Model (Scenario~B): This scenario is motivated by the strong stellar winds of massive donor stars in microquasar systems~\citep[e.g.,][]{Koljonen-2017MNRAS.472.2181K, MillerJones-2021Sci...371.1046M}, which provide dense targets for $pp$ interactions, while wind-interaction regions may also serve as sites for cosmic-ray acceleration~\citep[e.g.,][]{Peretti-2025AA...698A.188P,Wang-2025ApJ...989L..25W}. We assume a static external stellar-wind region where charged particles (e.g., protons and electrons) escaping from the jet blob interact with the wind material, contributing to high-energy gamma rays and neutrinos. As illustrated in Fig.~\ref{fig:model-pp1}, we assume that the external region is co-located with the jet blob and has radius $r_w=R_{\rm jet}$. The number density of this region, $n_{p,\rm ext}$, is taken to be constant and related to the stellar wind density at $R_{\rm jet}$.
    \item Extended-Jet Model (Scenario~C): Motivated by observations of extended TeV-emitting jets in microquasars such as SS~433 and V4641~Sgr~\citep{Abeysekara-2018Natur.562...82A, Alfaro-2024Natur.634..557A, LHAASO-2024arXiv241008988L} and related theoretical studies~\citep[e.g.,][]{Kimura-2020ApJ...904..188K, Carpio-2025arXiv250622550C, Ohira-2025MNRAS.541.2434O}, Scenario~C includes a large-scale external region associated with extended jet structures. In addition to the jet blob, we introduce a static external region where particles escaping from the jet blob interact with the parsec-scale environment, such as molecular clouds or the interstellar medium. As illustrated in Fig.~\ref{fig:model-pp2}, this region is modeled as a spherical shell with radius $H_{\rm ext}$, width $\Delta_{\rm ext}$, total mass $M_{p,\rm ext}$, and density $n_{p,\rm ext}$. This scenario also favors $pp$ interactions and, owing to its much larger spatial scale than Scenario~B, is expected to produce more persistent and less modulated emission.
\end{itemize}

Given the compact size of the emission region, Scenario~A naturally predicts strong variability. In contrast, Scenarios~B and~C involve more extended and static external zones and thus exhibit more stable behavior. 
Although Scenario~C includes an extended external region, our focus is on the unresolved, effectively point-like counterpart of microquasar emission rather than on modeling spatially resolved jet structures.

\subsection{Numerical Methods}

We further develop AMES to numerically calculate the multimessenger emission from microquasars, including photons and neutrinos. In this framework, the inner blob, which may represent a moving corona, is treated in the same way as the jet blob but with different parameter values. The external region is treated similarly, except that it is static and the particles injected into it originate from particles escaping from the jet blob.

We assume that the injected particle distribution for species $a$ (e.g., electrons or protons) in the blob follows a power law with a minimum Lorentz factor $\gamma'_{a,\rm min}$ and an exponential cutoff at $\gamma'_{a,\rm max}$:
\begin{equation}
    \dot{N}'_{a, \rm inj} (\gamma'_a)= \begin{cases} K'_a \gamma'^{-s_a}_a \exp{\left( - \frac{\gamma'_a}{\gamma'_{a,\rm max}} \right)}, & \gamma'_a \geq \gamma'_{a,\rm min} \\ 0, & \text { otherwise }\end{cases}
\end{equation}
where $\dot{N}'_{a, \rm inj}$ denotes the particle injection rate in the blob co-moving frame, $K'_a$ is the normalization constant, and $s_a$ is the spectral index. 

The values of $\gamma'_{\rm min}$ and $\gamma'_{\rm max}$ vary for different particle species depending on the specific physical conditions.
Considering the acceleration process, the maximum Lorentz factor can be estimated by balancing the acceleration timescale $t'_{\rm acc}$ and the cooling timescale $t'_c$ within the dynamical timescale $t'_{\rm dyn} = r'/v_j$ in the blob co-moving frame; that is, 
\begin{equation}
\label{eq:gamma_max}
    t'_{\rm acc} = \min (t'_{\rm dyn}, t'_c).
\end{equation}
The acceleration timescale at the blob co-moving frame can be written as
\begin{equation}
\label{eq:t_acc}
    t'_{\rm acc} = \eta_{\rm acc} \frac{r'_L}{c},
\end{equation}
where $r'_L = \gamma'_a m_a c^2 / (e B')$ is the gyration radius, $B'$ is the magnetic field in the blob co-moving frame, $m_a$ is the mass of a single particle $a$. Here, $\eta_{\rm acc} \approx \xi_B \beta_j^{-2}$ with $\xi_B \gtrsim 1$, hereafter referred to as the acceleration parameter, is a dimensionless parameter characterizing the specifics of the particle acceleration mechanism~\citep{Rachen-1998PhRvD..58l3005R}. 
Specifically, for electrons, both synchrotron and IC cooling can be important. 
As an example, the synchrotron cooling timescale is given by
\begin{equation}
t'_{\rm syn} = \frac{6\pi m_e c}{\sigma_T B'^2 \gamma'_e} \simeq 7.7 \times 10^4 ~{\rm s}~B'^{-2}_{1} \gamma_{e,2}'^{-1}.
\end{equation}
The maximum electron energy, $\varepsilon'_{e,\max} = \gamma'_{e, \rm max} m_e c^2$, can be determined by equating the acceleration timescale $t_{\rm acc}^\prime$ to the synchrotron cooling timescale $t'_{\rm syn}$, yielding
\begin{equation}
\varepsilon'_{e,\max} = \sqrt{\frac{6\pi e m_e^2 c^4}{\eta_{\rm acc} \sigma_T B'}} \simeq 20 ~{\rm TeV} ~\eta_{\rm acc}^{-1/2} B'^{-1/2}_1 .
\label{eq:E_e_max}
\end{equation}
For protons, radiative cooling is generally less significant. 
Their maximum energy can be estimated by equating the acceleration timescale to the dynamical timescale $t'_{\rm dyn}$, giving
\begin{equation}
\varepsilon'_{p,\max} = \frac{e B' r'}{\eta_{\rm acc} \beta_j} \simeq 300 ~{\rm TeV} ~\eta_{\rm acc}^{-1} B'_1 r'_{11} \beta_{j}^{-1}.
\label{eq:E_p_max}
\end{equation}

The magnetic field strength $B'$ in the blob co-moving frame is given by
\begin{equation}
\label{eq:B'}
    \frac{B'^2}{8\pi}
    =
    \frac{\epsilon_B}{\epsilon_a}
    \frac{3L'_{a,\rm inj}}{4\pi r'^2\beta_j c},
\end{equation}
where $\epsilon_a$ is the fractions of the blob energy carried by particle species $a$, and the magnetic energy fraction is explicitly defined as $\epsilon_B \equiv \epsilon_a B'^2 r'^2 \beta_j c / (6L'_{a,\rm inj})$.
The injection luminosity of particle species $a$ in the co-moving frame is
\begin{equation}
    L'_{a, \rm inj} = \int_{\gamma'_{a, \rm min}}^{\gamma'_{a, \rm max}}  \dot{N}'_{a, \rm inj} (\gamma'_a) (\gamma'_a - 1) m_a c^2 ~d \gamma'_a.
\end{equation}
It can also be related to the two-sided jet power $P_j$ in the lab frame as 
\begin{equation}
\label{eq:Pj}
    L'_{a, \rm inj} = \frac{2 \epsilon_a P_j}{\theta_{\rm op}^2 (H \Gamma_j / r')^2} 
    = \frac{2 \epsilon_a P_j}{\Gamma_j^2},
\end{equation}
where $r' = H \theta_{\rm op}$ is the radius of the blob in the blob co-moving frame. 
The jet power is related to the mass accretion rate by
\begin{equation}
    P_j = \eta_j\dot{M}c^2
    = \frac{\eta_j}{\eta_{\rm rad}} \dot{m}L_{\rm Edd},
\label{eq:Ledd}
\end{equation}
where $\eta_j$ is the jet-production efficiency,
$\eta_{\rm rad}$ is the radiative efficiency,
$\dot{M}_{\rm Edd}\equiv
L_{\rm Edd}/(\eta_{\rm rad}c^2)$ is the Eddington accretion rate,
and $\dot{m}\equiv\dot{M}/\dot{M}_{\rm Edd}$ is the dimensionless
accretion rate. We adopt $\eta_j=1$ and $\eta_{\rm rad}=0.1$, while
$\dot{m}$ is source dependent.
The energy fractions satisfy
$\epsilon_B+\epsilon_e+\epsilon_p< 1$, ensuring energy
conservation, where $\epsilon_e$ and $\epsilon_p$ denote the fractions
allocated to electrons and protons, respectively.

We note that the proton number density in the blob is constrained by the total injected proton luminosity. The corresponding upper limit in the blob co-moving frame is
\begin{equation}
    n'_{p, \rm ul} = \frac{3 L'_{p, \rm inj}}{4 \pi r'^2 m_p c^3 \beta_j} \simeq 5.3 \times 10^6 ~{\rm cm^{-3}} ~L'_{p, \rm inj,37} r'^{-2}_{11} \beta_j^{-1}.
\label{eq:n_p_ul}
\end{equation}
In this work, we set the proton number density entrained in the blob to be 10\% of this upper limit, i.e., $n'_{p, \rm jet} = 0.1\, n'_{p, \rm ul}$.

In our calculations, we include both leptonic and hadronic processes. Leptonic processes mainly include synchrotron radiation, IC scattering, involving both synchrotron self-Compton (SSC) and external inverse Compton (EIC) processes, and $\gamma\gamma$ annihilation. Hadronic processes mainly include proton synchrotron radiation, decay processes, $pp$ interactions, $p\gamma$ processes, and BH processes.
These lepto-hadronic processes are modeled by solving coupled transport equations for photons, electrons, neutrinos, neutrons, protons, muons, and pions in the blob co-moving frame. For a particle species $a$, the transport equation is
\begin{equation}
\begin{aligned}
\frac{\partial n_{\varepsilon'_{a}}'^{a}}{\partial t'}= & -n_{\varepsilon'_{a}}'^{a} \mathcal{A}'_{a}\left(\varepsilon'_{a}\right) \\
& +\int d \varepsilon_{a}'^{*} ~ n_{\varepsilon_{a}'^{*}}'^{a} \mathcal{B}'_{a \rightarrow a}\left(\varepsilon'_{a}, \varepsilon_{a}'^{*}\right) \\
& +\sum_b \int d \varepsilon'_{b} ~ n_{\varepsilon'_{b}}'^{b} \mathcal{C}'_{b \rightarrow a}\left(\varepsilon'_{a}, \varepsilon'_{b}\right) \\
& +\dot{n}'_{a, \rm inj}\left(\varepsilon'_{a}\right),
\end{aligned}
\end{equation}
where $n_{\varepsilon'_{a}}'^{a}$ is the differential number density of species $a$ at energy $\varepsilon'_{a}$, and $\mathcal{A}'_{a}\left(\varepsilon'_{a}\right)$ is the total interaction rate including particle escape. The term $\mathcal{B}'_{a \rightarrow a}(\varepsilon'_{a}, \varepsilon_{a}'^{*})$ describes the self-production rate of particles $a$ with energy $\varepsilon'_{a}$ generated from the same type of particles with energy $\varepsilon_{a}'^{*}$, while $\mathcal{C}'_{b \rightarrow a}(\varepsilon'_{a}, \varepsilon'_{b})$ represents the production rate of particles $a$ with energy $\varepsilon'_{a}$ originating from particles $b$ with energy $\varepsilon'_{b}$. The last term, $\dot{n}'_{a, \rm inj} \left(\varepsilon'_{a}\right) = \dot{N}'_{a, \rm inj} \left(\varepsilon'_{a}\right) / V'$, is the injection rate per unit volume, where $V' = (4/3) \pi r'^3$ is the co-moving volume of the system.

For further details on the specific coefficients $\mathcal{A}'$, $\mathcal{B}'$, and $\mathcal{C}'$ for each particle species, we refer the reader to \cite{Zhang-2023MNRAS.524...76Z} and \cite{Murase-2024PhRvD.109j3020M}. 
In this work, high-energy electrons mainly cool through adiabatic losses, synchrotron radiation, and IC scattering. In addition to primary electrons, we include secondary electrons produced by $\gamma\gamma$ annihilation, $pp$, BH, $p\gamma$, and neutron decay processes.
Protons mainly cool through adiabatic losses, synchrotron radiation, and BH processes, and can be produced through $pp$, $p\gamma$, and neutron decay channels.
Neutrons are produced via $pp$ and $p\gamma$ interactions, and their decay is included.
Photons are produced mainly by synchrotron and IC emission, as well as by $pp$ and $p\gamma$ interactions, and are attenuated through $\gamma\gamma$ absorption.
Neutrinos are primarily produced through $pp$, $p\gamma$, and neutron decay processes.
Compared with \cite{Zhang-2023MNRAS.524...76Z}, we also  solve kinetic equations for muons and pions that are produced via $pp$ and $p\gamma$ interactions, accounting for their cooling and decay as described in the following section.
For neutral particles, the escape timescale is taken to be the light-crossing time. For charged particles, we consider both advection and diffusion escape, as discussed in Sect.~\ref{sect:t_esc}.

\subsubsection{Muon/Pion Cooling}

Pions are produced through either $pp$ or $p\gamma$ interactions, with charged pions subsequently decaying into muons and neutrinos.
The production of neutrinos and photons from decays of secondary particles, such as muons and pions, can be suppressed if the dynamical time of the system is shorter than the particle decay time. This effect becomes particularly important at high energies because of relativistic time dilation.
In addition, radiative cooling processes, including synchrotron cooling of both muons and pions and IC cooling of muons, can significantly modify the resulting high-energy photon and neutrino spectra. These effects are included self-consistently in our coupled transport calculations.

We adopt a classical treatment for the synchrotron radiation of muons and pions, following equations analogous to those for electrons. The energy loss rate of particle species $a$ due to synchrotron radiation is
\begin{equation}
  \frac{d\varepsilon'_{a, \rm syn}}{dt'} = - \frac{4 \sigma_T Z^4 U'_B}{3 m_e^2 c^3} \varepsilon_a^{'2} \left(\frac{m_e}{m_a}\right)^4,
\end{equation}
where $U'_B = B'^2 / 8 \pi$ is the magnetic energy density, $Z$ is the charge number of the particle $a$, and $m_a$ is its mass. The corresponding synchrotron emissivity for the particle species $a$ is~\citep{Zhang-2021ApJ...920...55Z}
\begin{equation}
    \frac{dn'_{a, \rm syn}}{dt' d\varepsilon'_{\rm ph}
    } = \frac{1.81 \sqrt{3} Z^3 e^3 B'}{h m_a c^2 \varepsilon'_{\rm ph}} \int d \gamma'_a ~n'_a ({\gamma'_{a}}) ~ G\left(\frac{\varepsilon'_{\rm ph}}{\varepsilon'_{c}}\right),
\end{equation}
where
\begin{equation}
    G(x) = \frac{e^{-x}}{\sqrt{x^{-2/3} + ({3.62}/{\pi})^2}}, \ \text{and} \ \varepsilon'_{c} = \frac{3 Z e h B' \gamma'^2_a}{8 \pi m_a c}.
\end{equation}
Here, $\varepsilon'_{\rm ph}$ is the photon energy in the co-moving frame, and $n'_a ({\gamma'_{a}}) = d n'_a / d \gamma'_{a}$ is the differential number density of the particle $a$ in the co-moving frame.

For IC scattering of muons, we use a cross section analogous to that for electrons. Assuming isotropic particles and isotropic target photons, the IC emissivity is~\citep{Blumenthal-1970RvMP...42..237B, Zhang-2021ApJ...920...55Z}
\begin{equation}
\label{eq:emissivity_ic}
\begin{aligned}
    \frac{dn'_{a, \rm IC}}{dt' d\varepsilon'_{\rm ph}
    } & = \frac{3 \sigma_T c}{16 \pi} \left(\frac{m_e}{m_a}\right)^2  \\
    &\times \int d \gamma'_a ~n'_a({\gamma'_{a}}) \frac{1}{\gamma'^2_a}  \int d\varepsilon' ~\frac{1}{\varepsilon'} \frac{d n'_{\rm ph}}{d \varepsilon'} f(q, w),
\end{aligned}
\end{equation}
where 
\begin{equation}
    \begin{aligned}
        & f(q, w) = 2q \ln{q} + (1 + 2q) (1 - q) + \frac{1}{2} \frac{(wq)^2}{1 + wq}(1-q), \\
        & w = \frac{4 \varepsilon' \gamma'_a}{m_a c^2}, \ \text{and} \ q = \frac{\varepsilon'_{\rm ph}}{w (\varepsilon'_a - \varepsilon'_{\rm ph})}.
    \end{aligned}
\end{equation}
Here, ${d n'_{\rm ph}} / {d \varepsilon'}$ represents the differential number density of target photons, including both external photon fields and internally accumulated photons from prior processes such as synchrotron emission.
The energy loss rate of particle species $a$ due to IC scattering is
\begin{equation}
    \frac{d\varepsilon'_{a, \rm IC}}{dt'} = - V' \int d\varepsilon'_{\rm ph} ~ (\varepsilon'_{\rm ph} - \varepsilon') \frac{dn'_{a, \rm IC}}{dt' d\varepsilon'_{\rm ph}
    },
\end{equation}
where $V'$ is the volume of the system in the co-moving frame, and $dn'_{a, \rm IC} / (dt' d\varepsilon'_{\rm ph})$ represents the differential number density of IC scattered photons generated by a single particle $a$ with $\gamma'_a$, which can be obtained from Eq.~\ref{eq:emissivity_ic} by taking $n'_a({\gamma'_{a}})$ as a delta function.

Although the synchrotron emission of muons and pions and the IC emission of muons can contribute to photon production, their contributions are minor compared with other emission mechanisms. More importantly, these processes cool muons and pions, thereby suppressing the neutrino emission generated through $p\gamma$ processes. This cooling effect is particularly significant for neutrinos above TeV energies, as shown in Fig.~\ref{fig:check_muonCooling}.

\subsection{Origin of External Seed Photons}

In this work, we consider two main external photon fields. The first is the thermal emission from the donor star, which we model as a blackbody with radius $R_\star$, temperature $T_\star$, and luminosity $L_\star$. We take the stellar frame as the lab frame, since the orbital velocity of the donor star is negligible compared with the blob velocity.
As the blob moves with velocity $\beta_j$, the stellar radiation field is boosted in the blob co-moving frame. This boost is described by the Doppler factor $\delta_\star$, which is
\begin{equation}
    \delta_\star = \frac{1}{\Gamma_j (1 - \beta_j \vec{e}_\star \cdot \vec{e}_{\rm jet})}.
\end{equation}
The differential number density of the blackbody photons from the donor star, as measured at the location of the blob in the stellar (lab) frame, is given by~\citep{Zdziarski-2012MNRAS.421.2956Z}
\begin{equation}
    \frac{d n_{\rm ph,\star}}{d E_{\rm ph,\star}} = \frac{2 \pi}{c^3 h^3} \left(\frac{R_\star}{R}\right)^2 \frac{E_{\rm ph,\star}^2}{\exp{(E_{\rm ph,\star} / k T_\star)} - 1},
\label{eq:star}
\end{equation}
where $E_{\rm ph,\star} = \delta_\star \varepsilon'_{\rm ph,\star}$ is the photon energy in the lab frame, $k$ is the Boltzmann constant, and $h$ is the Planck constant. 

The second external photon field is provided by the accretion disk. The differential number density of the disk's photon field, evaluated at the blob's position in the lab frame, is
\begin{equation}
    \frac{d n_{\rm ph, disk}}{d E_{\rm ph, disk}} = \frac{L_{E_{\rm ph, disk}}}{4 \pi H^2 c E_{\rm ph, disk}},
\label{eq:disk}
\end{equation}
where $E_{\rm ph, disk} = \delta_{\rm disk}\varepsilon'_{\rm ph, disk}$ is the energy of disk photons in the lab frame, 
$\delta_{\rm disk} = \left[\Gamma_j (1 - \beta_j)\right]^{-1}$ is the related Doppler factor, and $L_{E_{\rm ph,disk}}$ is the differential isotropic-equivalent luminosity of disk photons in the lab frame. Here, $H=r'/\theta_{\rm op}$ denotes the blob location in the lab frame, and $r'$ is the blob radius in the co-moving frame.
Since disk photons are predominantly observed in the X-ray band, their number density can be estimated from the observed X-ray spectrum using $E_{\rm ph} L_{E_{\rm ph}} = 4 \pi d_L^2 E_{\rm ph} F_{E_{\rm ph}}$, where $E_{\rm ph} F_{E_{\rm ph}}$ is the observed flux at energy $E_{\rm ph}$, and $d_L$ is the distance from the source to the observer.

In addition to stellar and disk photons, other radiation fields can also serve as external seed photons. In Scenario~A, the jet blob additionally interacts with photons from the inner blob, while in Scenarios~B and~C, the external region also includes target photons from the jet blob.
We can transform these photon fields into the appropriate co-moving frame of each emission region using the fact that $d n_{\rm ph}/dE_{\rm ph}$ is a relativistic invariant~\citep{Blumenthal-1970RvMP...42..237B}.

\subsection{External Absorption}

Here, we consider two external absorption mechanisms: (i) free-free absorption, which attenuates radio photons as they propagate through the ionized stellar wind toward the observer, and (ii) external $\gamma\gamma$ absorption, which occurs when high-energy photons interact with target photons along the line-of-sight.
These external absorption processes are applied as post-processing steps outside the AMES framework.

\subsubsection{Free-Free Absorption from Winds}

We assume that the stellar wind from the donor star is smooth and spherically symmetric, following a standard velocity profile in the lab frame~\citep[e.g.,][]{Schaerer-1996AA...309..129S},
\begin{equation}
v_w(R_w) = v_{w,\infty} \left(1 - \frac{R_{0}}{R_w}\right)^{\beta_w}, \quad R_w > R_\star,
\end{equation}
where $R_w$ is the distance from the center of the donor star, $v_{w,\infty} = \sqrt{2GM_\star/R_\star}$ is the terminal wind velocity, $M_\star$ is the mass of the donor star, and $\beta_w$ parametrizes the wind acceleration. The constant $R_{0}$ is defined as
\begin{equation}
R_{0} = R_\star \left[1 - \left(\frac{v_{w,\star}}{v_{w,\infty}}\right)^{1/\beta_w}\right],
\end{equation}
where $v_{w,\star} = \sqrt{\frac{k T_\star}{\mu m_H}}$ is the initial wind velocity at the stellar surface, $T_\star$ is the temperature of the donor star, $\mu \simeq 1$ is the mean molecular weight, and $m_H$ is the hydrogen mass. Given that the blob velocity $\beta_j$ greatly exceeds the stellar wind velocity, the stellar wind frame is taken as the lab frame.
The wind number density is determined by the continuity equation,
\begin{equation}
n_w(R_w) = \frac{\dot{M}_w}{4\pi R_w^2 \mu m_H v_w(R_w)},
\end{equation}
where $\dot{M}_w$ is the stellar mass-loss rate. The local wind temperature is determined by the balance among radiative heating, cooling, and advection processes. Considering the irradiation of X-rays from the accretion disk, we adopt a characteristic wind temperature of $T_w = 10^6~\rm{K}$~\citep[e.g.,][]{Szostek-2007MNRAS.375..793S}.

For simplicity, neglecting the detailed metallicity dependence, the characteristic optical depth for free-free absorption in the stellar wind is approximated as~\citep[e.g.,][]{Murase-2024PhRvD.109j3020M}
\begin{equation}
\tau_{\rm ff} \simeq 8.5 \times 10^{-28}
~\nu_{\rm ph, 10}^{-2.1}
T_{w,4}^{-1.35}
R~ \left[n_w(R)\right]^2,
\end{equation}
where $\nu_{\rm ph} = E_{\rm ph} / h$ is the photon frequency in the lab frame, and $R$ is the distance between the donor star and the emission region. 
The corresponding suppression factor is 
\begin{equation}
f_{\rm sup, ff} = \exp(-\tau_{\rm ff}).
\end{equation}

\subsubsection{External $\gamma\gamma$ Absorption}

We consider external $\gamma\gamma$ absorption for the blob, with target photons originating from both the donor star and the accretion disk. In Scenario~A, the jet blob also interacts with photons from the inner blob, while the inner blob interacts with photons from the jet blob.

The optical depth for a gamma-ray photon of energy $E_{\rm ph}$ propagating along the observer direction $\vec{e}_{\rm obs}$ is estimated as
\begin{equation}
    \tau_{\gamma\gamma} \approx \int dR_{\gamma\gamma} \int dE_{\rm ph,t} \, \sigma_{\gamma\gamma} \, [1 - \cos\psi] \, \frac{d n_{\rm ph,t}}{d E_{\rm ph,t}}(R_{t}),
\end{equation}
where $\sigma_{\gamma\gamma}$ is the pair-production cross section~\citep{Gould-1967PhRv..155.1404G}, $R_{\gamma\gamma}$ denotes the distance measured along the line-of-sight from the gamma-ray emission site to an arbitrary interaction point $P(R_{\gamma\gamma})$ where external $\gamma\gamma$ annihilation may occur and serves as the integration variable for the external $\gamma\gamma$ absorption calculation, $R$ the distance between the emission region and the donor star, $R_t$ the distance between the target-photon source and $P(R_{\gamma\gamma})$, $\frac{d n_{\rm ph,t}}{d E_{\rm ph,t}}(R_{t})$ is the differential target photon number density at that point, and $\psi$ is the angle between the line-of-sight and the direction of the target photons.
We use the full $\gamma\gamma$ pair-production cross section, including its angular dependence, so that the difference between head-on and tail-on collisions is taken into account.

The suppression factor due to external $\gamma\gamma$ absorption is 
\begin{equation}
f_{{\rm sup}, \gamma\gamma} = \exp(-\tau_{\gamma\gamma}).
\end{equation}

\subsection{Escape Timescales of Charged Particles}
\label{sect:t_esc}

In all scenarios, we consider the escape of charged particles from two types of regions: the accelerator, corresponding to the jet blob and/or inner blob, and the reservoir, corresponding to the external region. These regions are characterized by different escape timescales.

In the accelerator, the escape timescale of charged particles in the lab frame is
\begin{equation}
    t_{\rm esc} = r / v_j,
\end{equation}
where $r$ is the blob radius in the lab frame. Note that including or neglecting this escape process has only a modest impact on the results, typically changing them by a factor of $\sim 2$.

In the external regions of Scenarios~B and~C, the injected charged particles originate from particles escaping from the jet blob. The escape fraction of charged particles from the jet blob is estimated as
\begin{equation}
\label{eq:f_esc}
    f_{\rm esc} = \min \left(1,  \frac{t_c}{t_{\rm dyn}}\right).
\end{equation}
The luminosity of escaped charged particles of species $a$ injected into the external region in the lab frame is then
\begin{equation}
    L_{a, \rm ext} = \frac{\delta_j^4 \theta_{\rm op}^2}{2} f_{\rm cov} f_{\rm esc} L'_{a, \rm inj},
\end{equation}
where $f_{\rm cov} = \Delta \Omega_{\rm ext} / (4 \pi)$ is the covering factor, $\Delta \Omega_{\rm ext}$ denotes the solid angle subtended by the external region as seen from the jet blob, and $L'_{a, \rm inj}$ is the injected luminosity of species $a$ in the blob co-moving frame.

In the reservoir, considering diffusion escape, the escape timescale for charged particles $a$ in the lab frame is
\begin{equation}
    t_{a, \rm esc} = \max(t_{a, \rm diff}, \frac{r_{\rm ext}}{c}),
\end{equation}
where $t_{a, \rm diff} = r_{\rm ext}^2 / (6 D_{a, \rm diff})$ is the diffusion timescale. The diffusion coefficient for a Kolmogorov turbulence spectrum is~\citep{Harari-2014PhRvD..89l3001H}
\begin{equation}
\begin{aligned}
    D_{a, \rm diff} & = \frac{c r_{\rm ext}}{3} \\
    & \times \left[4 \left(\frac{E_a}{E_{c, \rm diff}}\right)^{2} + 0.9 \left(\frac{E_a}{E_{c, \rm diff}}\right) + 0.23 \left(\frac{E_a}{E_{c, \rm diff}}\right)^{1/3}\right]
\end{aligned}
\end{equation}
where $E_a$ is the particle energy in the lab frame, $E_{c,\rm diff} \approx Z e B r_{\rm ext}$ is the critical energy, and $r_{\rm ext}$ is the characteristic size of the external region, corresponding to $r_w$ in Scenario~B and $\Delta_{\rm ext}$ in Scenario~C.

\section{Case Study: Cygnus~X-1 and Cygnus~X-3}
\label{sect:results}

In this section, we investigate the multimessenger emission from microquasars using Cygnus~X-1 and Cygnus~X-3 as benchmark sources. Both are persistent high-mass X-ray binaries with relatively well-characterized system parameters, and their $\gamma$-ray emission is consistent with a compact, point-like origin. More importantly, as discussed in Sect.~\ref{sect:intro}, LHAASO has recently detected $\sim100~{\rm TeV}$ photons from Cygnus~X-1 and $\sim{\rm PeV}$ photons from Cygnus~X-3~\citep{LHAASO-2024arXiv241008988L,LHAASO-2025arXiv251216638L}. These properties make them useful test cases for our model. For Cygnus~X-1, clear jet activity is observed in the hard state~\citep[e.g.,][]{Stirling-2001MNRAS.327.1273S,Zdziarski-2017MNRAS.471.3657Z}, while the jet is strongly suppressed in the soft state. However, current observations do not completely exclude a weak or intermittent soft-state outflow~\citep[e.g.,][]{Rushton-2012MNRAS.419.3194R}. We therefore present predictions for both the hard and soft states of Cygnus~X-1.

We select extensive multiwavelength observational data for these sources, covering energies from radio to PeV. The observed $0.1$-$1$~PeV photons are difficult to explain with leptonic processes alone, motivating the inclusion of hadronic processes. As described in Sect.~\ref{sect:model}, we apply three distinct scenarios with different parameter sets to fit the broadband data. The results presented here correspond to representative parameter sets that reproduce the observations, rather than statistically optimized best fits.

The general properties and observational data of Cygnus~X-1 and Cygnus~X-3 are summarized below:
\begin{enumerate}[label=(\alph*)]
    \item Cygnus~X-1 (RA = $299.6^\circ$, DEC = $35.2^\circ$) is a HMXB hosting a black hole with $M_{\rm BH} \approx 21~M_\odot$ in a 5.6-day orbit around an O-type donor star with $M_\star \approx 41~M_\odot$~\citep{MillerJones-2021Sci...371.1046M}. Although the donor star nearly fills its Roche lobe, accretion occurs via the stellar wind~\citep[e.g.,][]{Gies-2008ApJ...678.1237G}. As shown in Figs.~\ref{fig:spectra_pgamma}-\ref{fig:spectra_pp_B}, we collect non-simultaneous multiwavelength data for both the hard and soft states. These include: radio-to-millimeter data from \cite{Fender-2000MNRAS.312..853F, Pandey-2006AA...447..525P} (detected at MJD 50664-50675, 50945-50964, 52796-53392, respectively; blue points); infrared data from \cite{Mirabel-1996AA...315L.113M, Persi-1980AA....92..238P} (MJD 43808-43827; blue points); X-ray data from \cite{McConnell-2002ApJ...572..984M} (pink points); $0.1$-$100$~GeV data from \textit{Fermi}-LAT~\citep[][]{Zdziarski-2017MNRAS.471.3657Z} (MJD 54628-57705; brown); AGILE upper limits above $100$~MeV~\citep[][]{Sabatini-2013ApJ...766...83S} (MJD 54406-55121 for hard state, and MJD 55378-55647 for soft state; purple); MAGIC upper limits above $100$~GeV~\citep[][]{Fernandez-2017ICRC...35..734F} (MJD 54101-55562 for hard state; grey); and $\gtrsim 10$~TeV data from LHAASO~\citep[][]{LHAASO-2024arXiv241008988L} (MJD 58843-60340 for KM2A, and MJD 59281-60561 for WCDA; orange). For the X-ray band, we adopt the de-absorbed spectrum from \citet{Zdziarski-2017MNRAS.471.3657Z} (dotted pink line), corrected for absorption by the interstellar medium.
    
    \item Cygnus~X-3 (RA = $308.1^\circ$, DEC = $41.0^\circ$) is a wind-fed HMXB composed of a compact object and a Wolf-Rayet companion star, with an orbital period of 4.8 hours~\citep{vanKerkwijk-1992Natur.355..703V, Antokhin-2019ApJ...871..244A, Antokhin-2022ApJ...926..123A}. Although the nature of the compact object remains uncertain, a BH is favored by extensive X-ray and radio observations~\citep[e.g.,][]{Hjalmarsdotter-2009MNRAS.392..251H, Szostek-2008MNRAS.388.1001S, Koljonen-2017MNRAS.472.2181K}. In this work, we adopt the BH scenario, assuming a compact object mass of $M_{\rm BH} \approx 7~M_\odot$ and a total system mass of $19~M_\odot$, consistent with \citet{Antokhin-2022ApJ...926..123A}.
    As shown in Figs.~\ref{fig:spectra_pgamma}-\ref{fig:spectra_pp_B}, we collect non-simultaneous soft-state data from radio to PeV energies: radio-to-millimeter data from \citet{Pandey-2006AA...447..525P} (MJD 52797-53393; blue points); infrared data from \citet{Ogley-2001MNRAS.322..177O} (MJD 50613-50619; blue points); X-ray data from \citet{Szostek-2008MNRAS.388.1001S} (pink points); $0.1$-$100$~GeV data from \textit{Fermi}-LAT~\citep{LHAASO-2025arXiv251216638L} (MJD 58848-60522; brown); MAGIC upper limits above $ 100$~GeV~\citep{Abe-2026arXiv260427671A} (MJD 56600-60600; grey); and $\gtrsim 0.1$~PeV data from LHAASO~\citep[][]{LHAASO-2025arXiv251216638L} (MJD 59101-60501 for KM2A; orange). For the X-ray band, we adopt the de-absorbed spectrum from \cite{Vilhu-2023AA...674A..74V}, shown as the dotted pink line. Although \cite{Veledina-2024NatAs...8.1031V} reported that the intrinsic X-ray flux can be more than ten times higher than the observed flux, we use the originally observed value for a general example in our calculation.
\end{enumerate}
The basic parameters of Cygnus~X-1 and Cygnus~X-3 are listed in Table~\ref{tab:basic}. 
For both sources, we assume circular orbits ($e=0$), so that the distance $d$ between the donor star and the central engine is equal to the semi-major axis $a$ at all orbital phases.

For simplicity, we adopt a fixed jet geometry in all modeling scenarios, as summarized in Table~\ref{tab:fix}. 
This geometry should be regarded as an illustrative configuration, whose possible physical origin is discussed later. As a simple parametrization of a bent or intrinsically misaligned jet geometry, we set $\Delta\phi = \phi - \phi_j = -90^\circ$ as a representative choice.
We also set $\phi = 180^\circ$, which roughly corresponds to the mean flux during orbital modulation and thus serves as a proxy for the orbital-phase-averaged flux, as shown in Figs.~\ref{fig:CygX3_pgamma_varied} and~\ref{fig:flux_pgamma_varied}.
For Cygnus~X-1, the inclination inferred from the X-ray reflection spectrum, $\sim 40^\circ$, differs from the binary inclination of $27.1^\circ$~\citep{Tomsick-2018ApJ...855....3T,Parker-2015ApJ...808....9P}. We therefore adopt a representative jet polar angle of $\vartheta_j = 15^\circ$.
For Cygnus~X-3, modeling of the orbital modulation of gamma-ray emission indicates $\vartheta_j \gtrsim 25^\circ$~\citep{Zdziarski-2018MNRAS.479.4399Z, Dmytriiev-2024ApJ...972...85D}, and we adopt $\vartheta_j = 25^\circ$.

\begin{table*}
\caption{Physical parameters for Cygnus~X-1 and Cygnus~X-3 adopted in this work. See Tables.~\ref{tab:fix}-\ref{tab:pp_B} for modeling parameters.}
\label{tab:basic}
\centering
\begin{threeparttable}
\centering
\begin{minipage}{0.48\textwidth}
\centering
\begin{tabular}{cc|cc}
\hline \hline
Input Parameters & Symbol [Units] & Cyg~X-1~\tablenotemark{a} & Cyg~X-3~\tablenotemark{b} \\
\hline
Luminosity distance   & $D_L~[\mathrm{kpc}]$ & 2.2 & 9.7 \\
Orbital inclination   & $i~[^{\circ}]$ & 28 & 30 \\
Orbital period        & $P_{\mathrm{orb}}~[\mathrm{s}]$ & $4.8\times10^5$ & $1.7\times10^4$ \\
Mass of black hole    & $M_{\mathrm{BH}}~[M_\odot]$ & 21 & 7 \\
Mass of donor star    & $M_\star~[M_\odot]$ & 41 & 12 \\
Temperature of donor star & $T_\star~[\mathrm{K}]$ & $3.1\times10^4$ & $1.0\times10^5$ \\
Luminosity of donor star & $L_\star~[\mathrm{erg~s^{-1}}]$ & $1.6\times10^{39}$ & $7.7\times10^{38}$ \\
Stellar wind mass-loss rate & $\dot{M}_w~[M_\odot~\mathrm{yr^{-1}}]$ & $2.6\times10^{-6}$ & $9.6\times10^{-6}$ \\
Wind acceleration parameter & $\beta_w$ &  1.05 & 2.0 \\
Wind temperature & $T_w~[\mathrm{K}]$ & $1.0\times10^6$ & $1.0\times10^6$ \\
\hline
\end{tabular}
\end{minipage}%
\hfill
\begin{minipage}{0.48\textwidth}
\centering
\begin{tabular}{cc|cc}
\hline \hline
Derived Parameters & Symbol [Units] & Cyg~X-1 & Cyg~X-3 \\
\hline
Semi-major axis~\tablenotemark{c} & $a~[\mathrm{cm}]$ & $3.7\times10^{12}$ & $2.7\times10^{11}$ \\
Eddington luminosity & $L_{\mathrm{Edd}}~[\mathrm{erg~s^{-1}}]$ & $2.7\times10^{39}$ & $9.1\times10^{38}$ \\
Radius of donor star~\tablenotemark{d} & $R_\star~[\mathrm{cm}]$ & $1.6\times10^{12}$ & $1.0\times10^{11}$ \\
Stellar wind velocity & $v_{w, \infty}~[\mathrm{cm~s^{-1}}]$ & $8.3\times10^7$ & $1.7 \times 10^8$ \\
\hline
\end{tabular}
\end{minipage}
\vspace{5pt}
\tablenotetext{a}{Parameters from \cite{Ziolkowski-2005MNRAS.358..851Z,Szostek-2007MNRAS.375..793S,MillerJones-2021Sci...371.1046M}.}
\tablenotetext{b}{Parameters from \cite{Koljonen-2017MNRAS.472.2181K,Reid-2023ApJ...959...85R,Antokhin-2022ApJ...926..123A}.}
\tablenotetext{c}{Assuming a circular orbit, the semi-major axis is given by 
$a = \left(\frac{G P_{\mathrm{orb}}^2 (M_{\mathrm{BH}} + M_\star)}{4\pi^2}\right)^{1/3}$.}
\tablenotetext{d}{The stellar radius $R_\star$ is derived from $L_\star = 4\pi R_\star^2 \sigma_\mathrm{SB} T_\star^4$.}
\end{threeparttable}
\end{table*}

\begin{table*}
\caption{Geometry parameters used in this work, unless otherwise specified. }
\label{tab:fix}
\centering
\begin{threeparttable}
\centering
\begin{tabular}{cc|ccc} 
\hline \hline
Geometry Parameters & Symbol [Units] & Cygnus~X-1 (hard) &  Cygnus~X-1 (soft) & Cygnus~X-3 \\
\hline
Jet polar angle  &  $\vartheta_j~\rm [deg]$              & $15$         & $15$ & $25$ \\
Jet azimuthal angle  &  $\phi_{j}~\rm [deg]$            & $-90$           & $-90$ & $-90$ \\
Azimuthal angle of donor star  &  $\phi~\rm [deg]$           & $180$            & $180$ & $180$ \\
\hline
\end{tabular}
\end{threeparttable}
\end{table*}

In our calculations, we adopt a total evolution time of $10^6$~s for Scenarios~A and~B. The soft state is generally less stable and shorter-lived than the hard state, and Cygnus~X-1 and Cygnus~X-3 exhibit different variability patterns. Nevertheless, this duration is sufficient for the systems considered here. Cygnus~X-1 can remain in stable hard or soft states for more than 300 hours~\citep{Grinberg-2013AA...554A..88G}, and Cygnus~X-3 has been observed in a prolonged ($\sim 10-30$ days) hypersoft state~\citep{Koljonen-2018AA...612A..27K}. These timescales significantly exceed the dynamical timescales in Scenarios~A and~B, allowing the systems to reach steady states. Therefore, uncertainties in the precise duration of each state should have a negligible effect on our results.
For Scenario~C, which represents more persistent large-scale emission, we set the evolution time to the system age: $10^6~$yr for Cygnus~X-1~\citep{Ziolkowski-2005MNRAS.358..851Z} and $10^5~$yr for Cygnus~X-3~\citep{Meynet-2005AA...429..581M}. 
Since these ages are much longer than the relevant dynamical timescales, uncertainties in the precise system ages should also have little impact on our results. 
The orbital periods of both sources are much shorter than the adopted evolution times, so we neglect orbital motion in our calculations.

\subsection{Optical Depth and Production of High-Energy Photons}

\begin{figure*}[ht!]
	\centering
	\subfigure[]{
		\includegraphics[scale=0.3]{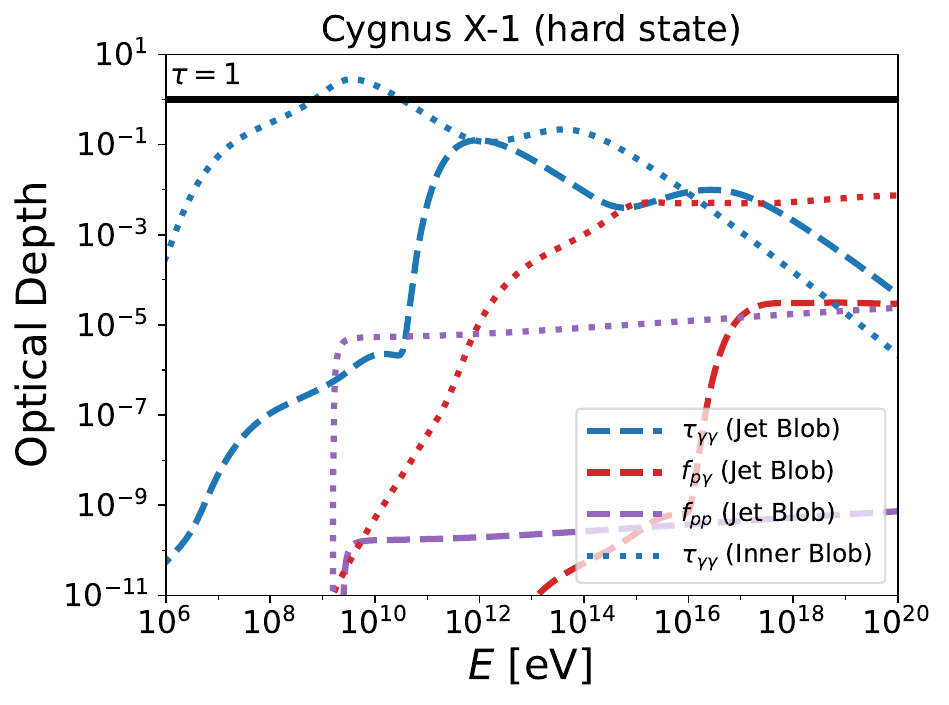}
	}%
    \subfigure[]{
		\includegraphics[scale=0.3]{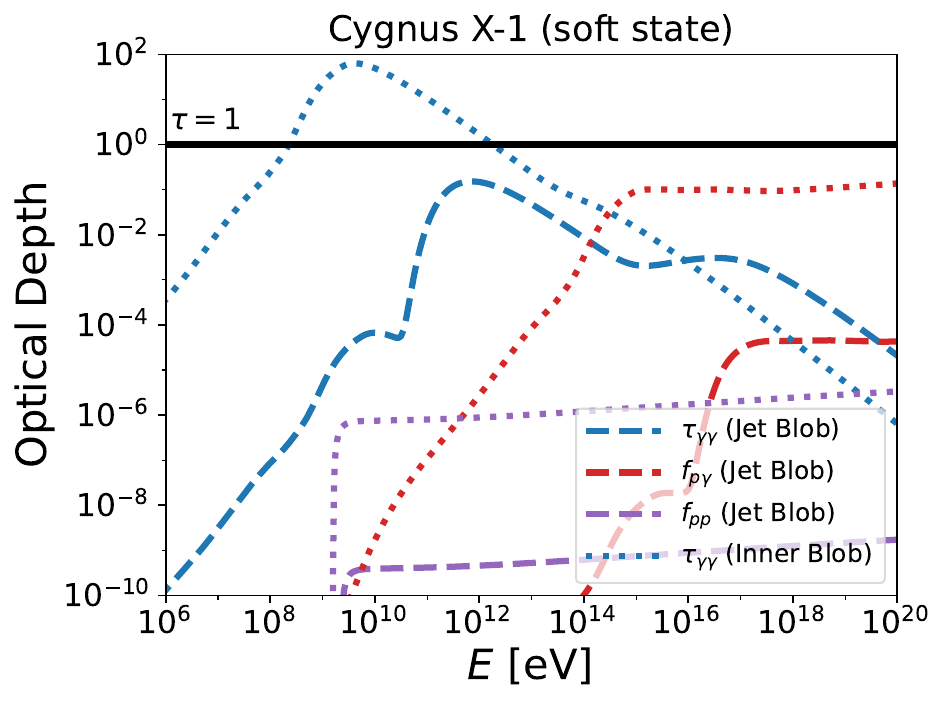}
	}%
	\subfigure[]{
		\includegraphics[scale=0.3]{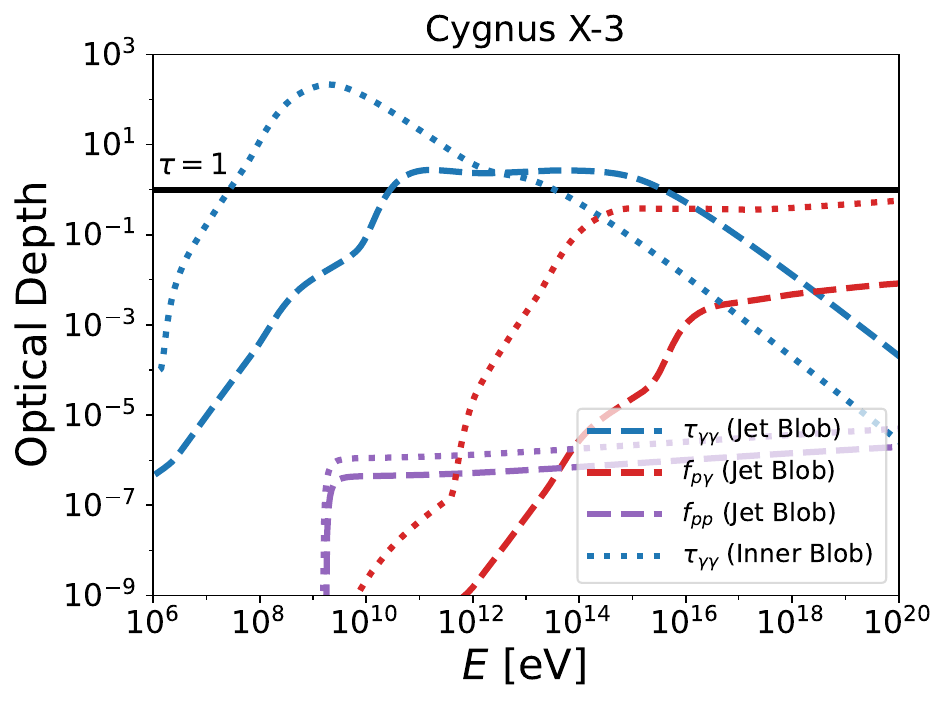}
	}%
    
	\subfigure[]{
		\includegraphics[scale=0.3]{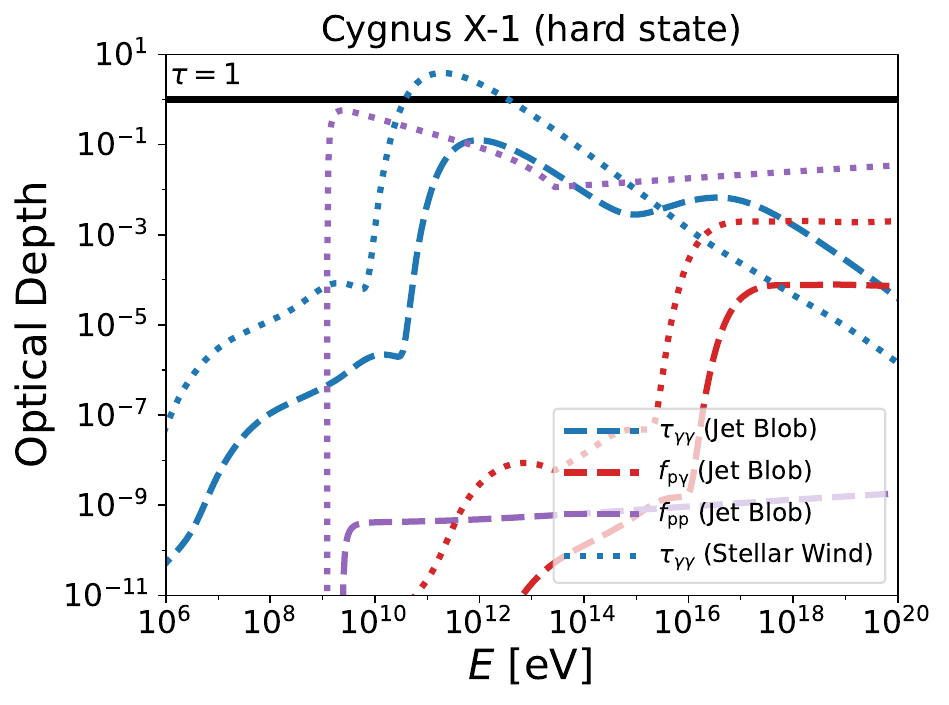}
	}%
    \subfigure[]{
		\includegraphics[scale=0.3]{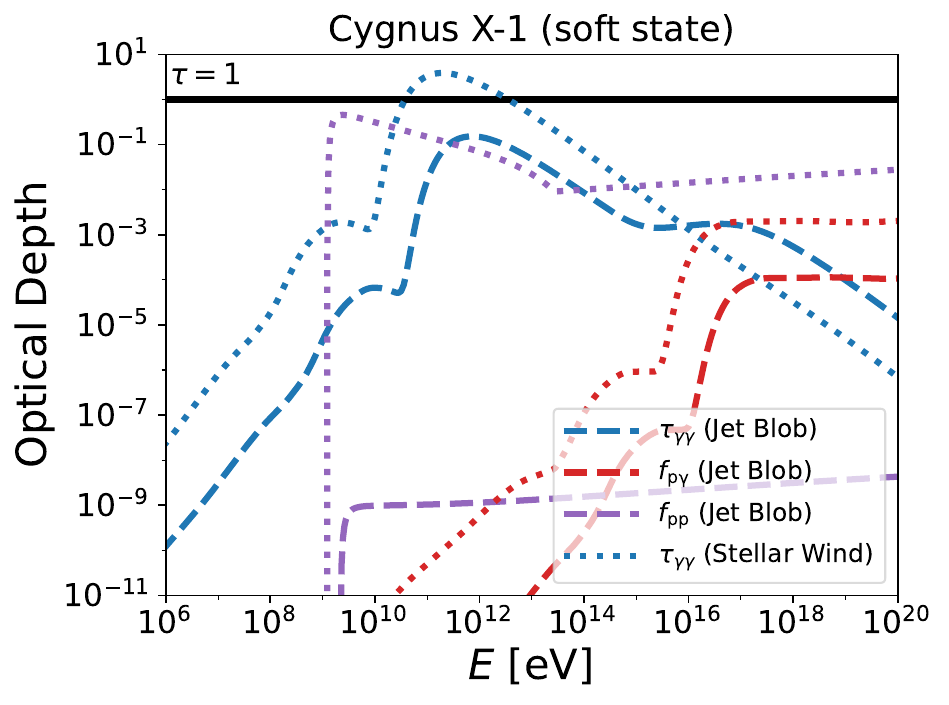}
	}%
	\subfigure[]{
		\includegraphics[scale=0.3]{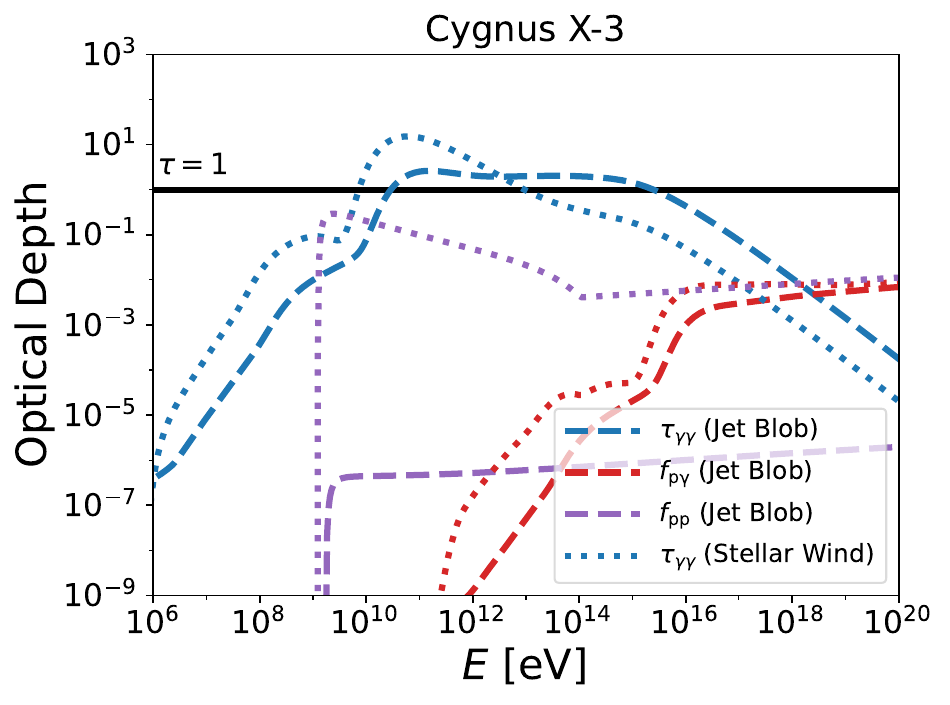}
	}%

	\subfigure[]{
		\includegraphics[scale=0.3]{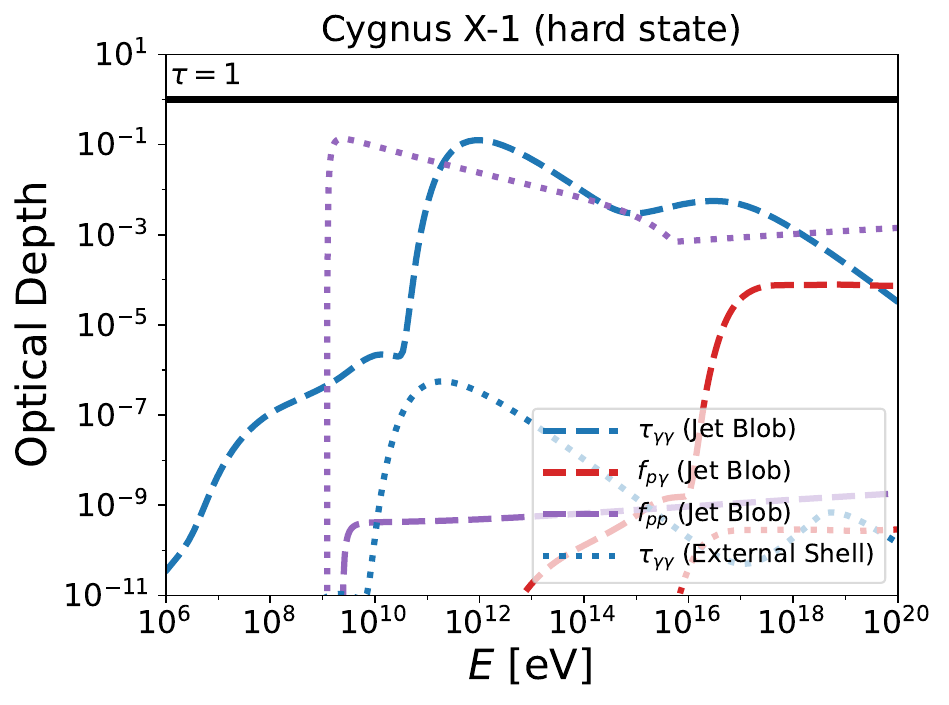}
	}%
    \hspace{5cm}
	\subfigure[]{
		\includegraphics[scale=0.3]{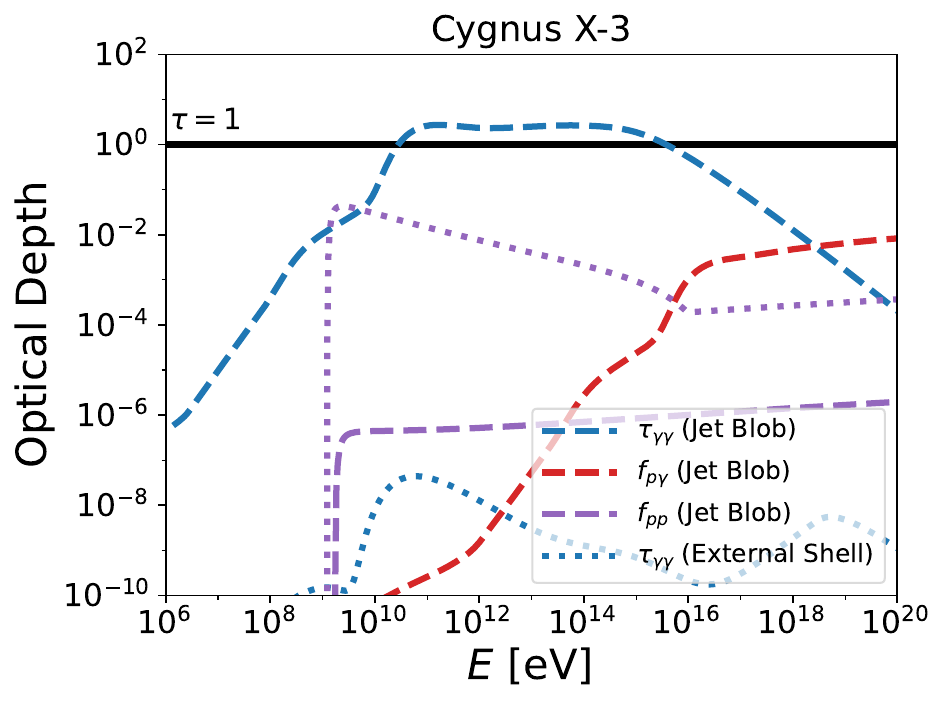}
	}%
		
	\caption{Optical depths for $\gamma\gamma$ annihilation (blue), $p\gamma$ interactions (red), and $pp$ interactions (purple) in the observer frame. The first, second, and third rows correspond to Scenarios~A, B, and C, respectively. Different line styles correspond to different emission zones, as indicated in the legend.}
	\label{fig:optical_depth}
\end{figure*}

In our model, high-energy photons are primarily produced via the $p\gamma$ and $pp$ processes. Fig.~\ref{fig:optical_depth} shows the corresponding interaction efficiencies for each scenario and source, illustrating the efficiency of high-energy photon production.

Within both the jet blob and the inner blob, photons above $0.1$~PeV are dominated by the $p\gamma$ process due to dense target photon fields.
For a target photon field with $dn_{\rm ph, t}/dE_{\rm ph, t} \propto E_{\rm ph, t}^{-\alpha}$ and $\alpha \gtrsim 1$, such as the external seed photons from the accretion disk, the $p\gamma$ interaction is dominated by the $\Delta-$resonance. The $p\gamma$ interaction efficiency can be estimated as~\citep{Murase-2016PhRvL.116g1101M}:
\begin{equation}
\label{eq:f_pgamma1}
    \begin{aligned}
        f_{p\gamma} & \approx \frac{t_{\rm dyn}}{t_{p\gamma}} \approx \eta_{p\gamma} r \beta_j^{-1} n_{\rm ph,t} \hat{\sigma}_{p\gamma} \left(\frac{E_{p}}{E_{{\rm ch}, p\gamma}}\right)^{\alpha-1} \\
        & \simeq 8.8 \times 10^{-7} ~ \eta_{p\gamma} r_{9} \beta^{-1}_{j,-0.1} n_{\rm ph, t, 13} \left(\frac{E_{p}}{E_{{\rm ch}, p\gamma}}\right)^{\alpha-1},
    \end{aligned}
\end{equation}
where $\eta_{p\gamma} \approx 2 / (1 + \alpha)$, $t_{p\gamma}$ is the energy-loss timescale of a proton with energy $E_p$ through the $p\gamma$ process in the lab frame, $n_{\rm ph,t}$ is the target photon number density in the lab frame, and $\hat{\sigma}_{p\gamma} = \kappa_{p\gamma} \sigma_{p\gamma} \sim 0.7 \times 10^{-28} ~\rm cm^2$ is the product of the inelasticity and the cross section for $p\gamma$ process. The characteristic proton energy is $E_{{\rm ch}, p\gamma} = 0.5 \Gamma_j^2 m_p c^2 \bar{\varepsilon}_\Delta / E_{\rm ph, t} \simeq 141 {~\rm TeV} (E_{\rm ph, t} / 1~\rm keV)^{-1} \Gamma_{j,0}^2$, where $\bar{\varepsilon}_\Delta \sim 0.3~$GeV and $E_{\rm ph, t}$ is the target photon energy in the observer frame. 
For stellar photons, which follow a blackbody distribution and do not satisfy the simple power-law approximation with $\alpha \gtrsim 1$, higher resonances and multipion production make $f_{p\gamma}$ approximately constant. 
At high energies, the corresponding interaction efficiency can be estimated as
~\citep{Dermer-2014JHEAp...3...29D}:
\begin{equation}
\label{eq:f_pgamma2}
    \begin{aligned}
        f_{p\gamma} & \approx \frac{t_{\rm dyn}}{t_{p\gamma}} \approx r \beta_j^{-1} n_{\rm ph,t} \hat{\sigma}_{p\gamma} \left[1 - \frac{1 + \ln y_u}{y_u}\right] \\
        & \simeq 8.8 \times 10^{-7} ~ r_{9} \beta^{-1}_{j,-0.1} n_{\rm ph,t, 13} \left[1 - \frac{1 + \ln y_u}{y_u}\right],
    \end{aligned}
\end{equation}
where $y_u = \left(4 \Gamma_j \gamma'_p E_{\rm ph, t} / \bar{\varepsilon}_{{\rm thr},p\gamma
}\right)^2$, $\bar{\varepsilon}_{{\rm thr},p\gamma
} \simeq 400 \Gamma_j m_e c^2$, $E_{\rm ph, t} \simeq 2.82 k T_\star \simeq 24.3 ~{\rm eV}~ T_{\star, 5}$ is the peak energy of the blackbody target photons in the observer frame, and $\gamma'_p$ is the proton Lorentz factor in the blob co-moving frame.

The typical observed photon flux produced via the $p\gamma$ process is
\begin{equation}
\label{eq:pgamma}
E_{\rm ph} F_{E_{\rm ph}} \approx \frac{1}{2} f_{p\gamma} \, E_p F_{E_p},
\end{equation}
where $E_p F_{E_p}$ is the proton flux at energy $E_p$ in the observer frame, and the typical photon energy is $E_{\rm ph} \sim 0.1 \, E_p$. 
Combining Eqs.~\ref{eq:f_pgamma1}-\ref{eq:pgamma}, we find that the blob can generate photons with a flux of $\sim 10^{-13}~\mathrm{erg~cm^{-2}~s^{-1}}$ around 10~TeV (PeV), given sufficient X-ray (UV) photon density.

However, in $p\gamma$ scenarios, the same photon field can also attenuate the escaping gamma-rays. The optical depth for internal $\gamma\gamma$ annihilation can be estimated as~\citep[e.g.,][]{Murase-2016PhRvL.116g1101M}
\begin{equation}
\begin{aligned}
    \tau_{\gamma\gamma} & \approx \eta_{\gamma\gamma} r \beta_j^{-1} n_{\rm ph,t} \sigma_T \\
    & \simeq 8.4 \times 10^{-3} ~\eta_{\gamma\gamma} r_9 \beta_{j, -0.1}^{-1} n_{\rm ph,t, 13} ,
\end{aligned}
\end{equation}
where $\sigma_T \approx 6.65 \times 10^{-25}~\mathrm{cm^2}$ is the Thomson cross section, $\eta_{\gamma\gamma} \approx 7/[6\alpha^{5/3}(1+\alpha)]$ for $1 < \alpha < 7$~\citep{Svensson-1987MNRAS.227..403S}, and $n_{\rm ph,t}$ is the number density of target photons with energy $E_{\rm ph,t}$ in the observer frame. The corresponding typical gamma-ray energy is $E_{\rm ph} \approx \Gamma_j^2 m_e^2 c^4 E_{\rm ph,t}^{-1} \simeq 26~{\rm GeV} \left(E_{\rm ph,t} / {10 ~\rm eV}\right)^{-1} \Gamma_{j,0}^2$. As shown by the blue lines in Fig.~\ref{fig:optical_depth}, photons with energies $\gtrsim10$~TeV are not significantly attenuated by $\gamma\gamma$ annihilation, with $\tau_{\gamma\gamma}<1$.

In the external region, photons above $0.1$~PeV are mainly produced through the $pp$ process, where escaping protons interact with dense ambient gas and efficiently convert their kinetic energy into high-energy photons. The $pp$ interaction efficiency can be estimated as~\citep{Kelner-2006PhRvD..74c4018K,Murase-2020PhRvL.125a1101M}:
\begin{equation}
\label{eq:f_pp}
\begin{aligned}
    f_{pp} & \approx \frac{t_{\rm esc}}{t_{pp}} \approx c n_{p} \kappa_{pp} \sigma_{pp} \times \max{(t_{\rm diff}, \frac{r_{\rm ext}}{c})}, \\
    & \simeq 6.0 \times 10^{-4} ~n_{p, 9} t_{\rm esc, 3}
\end{aligned}
\end{equation}
where $t_{pp}$ is the interaction timescale of the $pp$ process in the lab frame, $n_p$ is ambient proton number density in the lab frame, $\kappa_{pp} \approx 0.5$ is the inelasticity coefficient, and $\sigma_{pp} \sim 4 \times 10^{-26} ~\rm cm^2$ is the inelastic $pp$ cross section.

The typical observed photon flux produced via the $pp$ process is 
\begin{equation}
\label{eq:pp}
E_{\rm ph} F_{E_{\rm ph}} \approx \frac{1}{3} f_{pp} \, E_p F_{E_p},
\end{equation}
where the typical photon energy is $E_{\rm ph} \sim 0.08 \, E_p$. 
Combining Eqs.~\ref{eq:f_pp}-\ref{eq:pp}, we find that the external region can produce photons with a flux of $\sim10^{-13}~{\rm erg~cm^{-2}~s^{-1}}$ at $\sim10$~TeV and can extend to PeV energies with fluxes of $\sim10^{-14}-10^{-15}~{\rm erg~cm^{-2}~s^{-1}}$.

Overall, both analytical estimates and numerical calculations indicate that microquasars can produce detectable photons above $0.1$~PeV, even after accounting for $\gamma\gamma$ annihilation. The $\gamma\gamma$ absorption is most important in the GeV band for the inner and jet blobs, where interactions with target photons are efficient. In contrast, its effect is negligible in the external region.

\subsection{Results for Scenario~A}

\begin{table*}
\caption{Fitting parameter values for the jet blob shared by all scenarios. }
\centering
\begin{threeparttable}
\centering
\begin{tabular}{cc|ccc} 
\hline \hline
Jet Blob Parameters  &  Symbol [Units]   & Cygnus~X-1 (hard) & Cygnus~X-1 (soft) & Cygnus~X-3 \\
\hline
Emission region location  & $H_{\rm jet} ~\rm [cm]$                   &        $1.0 \times 10^{13}$          &  $1.0 \times 10^{13}$   & $3.0 \times 10^{11}$   \\
Half-opening angle~\tablenotemark{a} & $\theta_{\rm op} ~\rm [deg]$ & $5$ & $5$ & $15$ \\
Emission region radius  & $r'_{\rm jet} ~\rm [cm]$                   &        $8.7 \times 10^{11}$            &  $8.7 \times 10^{11}$   & $7.9 \times 10^{10}$               \\
Lorentz factor (velocity) & $\Gamma_{j} (\beta_j)$     &              $1.80 ~(0.83)$    &         $1.40 ~(0.70)$        &    $1.15 ~(0.49)$   \\
Acceleration parameter & $\eta_{\rm acc}$           &           $1.0 ~\beta_j^{-2}$         &     $1.0 ~\beta_j^{-2}$        &    $1.0 ~\beta_j^{-2}$   \\
$p$ fraction~\tablenotemark{b} & ${{\epsilon}_p}$ &          $0.2$         &       $0.2$         &   $0.2$ \\
$p$ spectral index  & $s_p$           &          $2.0$          &        $2.0$       &  $2.0$   \\
$p$ Minimum Lorentz factor  & $\gamma'_{p, \rm min}$           &          $1.0$          &        $1.0$       &  $1.0$   \\
$e^-$ spectral index  & $s_e$           &          $2.8$          &        $2.8$       &  $2.8$   \\
$e^-$ Minimum Lorentz factor & $\gamma'_{e, \rm min}$           &        $1.0 \times 10^2$            &      $1.0 \times 10^2$        &   $1.6 \times 10^3$  \\
Stellar wind number density~\tablenotemark{c}  & $n_{w}~\rm [cm^{-3}]$                       & $9.8 \times 10^{8}$ & $9.8 \times 10^{8}$ & $2.1 \times 10^{12}$ \\
Ambient number density~\tablenotemark{d}  & $n'_{p, \rm jet}~\rm [cm^{-3}]$                    &       $1.4 \times 10^4$             & $2.7 \times 10^4$    &  $9.5 \times 10^7$              \\
\hline
\end{tabular}
\tablenotetext{a}{The adopted jet half-opening angles are representative values guided by observational constraints. For Cygnus~X-3, we use a value consistent with $\theta_{\rm op}\lesssim15^\circ$~\citep{Veledina-2024NatAs...8.1031V}. For Cygnus~X-1, although the resolved radio jet suggests $\theta_{\rm op}\lesssim2^\circ$~\citep{Stirling-2001MNRAS.327.1273S}, we adopt $\theta_{\rm op}=5^\circ$ as a fiducial value to allow for uncertainties in extrapolating this constraint to the inner emission region.}
\tablenotetext{b}{Motivated by observational constraints on the accretion and jet properties of the two sources, we adopt the source-dependent dimensionless accretion rates $\dot{m}=5\times10^{-3}$ for Cygnus~X-1~\citep{Prabu-2025arXiv251209645P} and $\dot{m}=0.2$ for Cygnus~X-3~\citep{Zdziarski-2012MNRAS.421.2956Z, Veledina-2024NatAs...8.1031V} with $\eta_j=1$ and $\eta_{\rm rad}=0.1$. These values are fiducial estimates intended to characterize relatively active instantaneous states rather than long-term averages. The corresponding proton injection luminosities are calculated using Eqs.~\ref{eq:Pj} and~\ref{eq:Ledd}.}
\tablenotetext{c}{The wind density at the center of the jet blob is given by $n_{w} = {\dot{M}_w} / {\left(4\pi R_{\rm jet}^2 m_p v_w \right)}$, where $R_{\rm jet}$ is the distance between the donor star and the jet blob.}
\tablenotetext{d}{We set the proton number density entrained
in the jet to be $10\%$ of the upper limits calculated from Eq.~\ref{eq:n_p_ul}.}
\end{threeparttable}
\label{tab:jet_blob}
\end{table*}

In Scenario~A, we model the emission using a jet blob and an inner blob with parameters listed in Tables~\ref{tab:jet_blob} and~\ref{tab:pgamma}. 
Table~\ref{tab:jet_blob} summarizes the jet-blob parameters shared by all scenarios. For Cygnus~X-1, we adopt a jet-blob height of $10^{13}~\mathrm{cm}$ in the lab frame, consistent with estimates based on gamma-ray modulation~\citep{Zanin-2016AA...596A..55Z}. For Cygnus~X-3, the corresponding height is $3\times10^{11}~\mathrm{cm}$, in agreement with predictions from gamma-ray modulation studies~\citep{Zdziarski-2018MNRAS.479.4399Z}.
The injected particle distributions are primarily determined by the microphysical parameters, including $\eta_{\rm acc}$, $\tilde{\epsilon}_p$, $\tilde{\epsilon}_e$, and $\tilde{\epsilon}_B$. The maximum Lorentz factors of electrons and protons are calculated self-consistently using Eqs.~\ref{eq:gamma_max}-\ref{eq:E_p_max}.

If the inner blob is associated with the corona, its magnetic field can be compared with observational constraints from X-ray studies.
As shown in Table~\ref{tab:pgamma}, the magnetic field for Cygnus~X-1 is $\approx 6.4 \times 10^{3}$ G ($\approx 5.2 \times 10^{3}$ G) for the hard (soft) state, consistent with the observationally inferred upper limits of $\sim 10^5$~G ($\sim 10^6$~G) for the hard (soft) states~\citep{DelSanto-2013MNRAS.430..209D}. 
For Cygnus~X-3, we obtain $B'_{\rm in} \approx 8.9 \times 10^4~{\rm G}$, lower than the equipartition coronal field of $\sim 5 \times 10^6~{\rm G}$ estimated by \citet{Cerutti-2011A&A...529A.120C}.
These estimates are model dependent and should be regarded as indicative.
Magnetic one-photon pair creation is unimportant when $B' \lesssim 6.3 \times 10^{5}~{\rm G}~\Gamma_{\rm in} E_{\rm ph,14}^{-1}$~\citep{Murase-2012ApJ...745L..16M}. For the inner blob considered here ($\Gamma_{\rm in} \simeq 1$ and $E_{\rm ph} \sim 100$ TeV), the adopted magnetic fields are below this limit. Therefore, this attenuation process does not alter our conclusion, although it may become important for more strongly magnetized or higher-energy emission regions in other models.

Fig.~\ref{fig:spectra_pgamma} presents the resulting neutrino (black lines) and photon (red lines) spectra. The top panels show the broadband emission from radio to PeV energies, while the bottom panels focus on the gamma-ray range. In this scenario, photons above $0.1$~PeV are predominantly produced via the $p\gamma$ process, whereas GeV photons arise mainly from EIC scattering. In the energy range between 0.1 and 10~TeV, the photon spectrum exhibits a pronounced valley caused by the combined effects of the $p\gamma$ interactions and $\gamma\gamma$ annihilation. Recent MAGIC observations provide only upper limits~\citep[][]{Abe-2026arXiv260427671A}. We also show the sensitivity of the Cherenkov Telescope Array Observatory (CTAO) Northern Array~\citep[][]{Abe-2026arXiv260427671A} and Large Array of imaging atmospheric Cherenkov Telescope (LACT)~\citep[][]{Zhang-2025arXiv250920694Z} for a 50-hour exposure. Detecting a signal with CTAO and/or LACT would likely require a longer observation time.

The radio emission originates from synchrotron radiation but is significantly suppressed by free-free absorption. The light-red curves represent spectra without free-free absorption and external $\gamma\gamma$ annihilation, while the red curves include these processes. The suppression of radio flux in the latter case suggests that the observed radio emission may originate from regions at larger heights~\citep[$\sim 10^{14}~\mathrm{cm}$; e.g.,][]{Zdziarski-2012MNRAS.422.1750Z,Zdziarski-2018MNRAS.479.4399Z,Tetarenko-2019MNRAS.484.2987T}, where the stellar wind density is lower.

\subsubsection*{(a) Cygnus~X-1}

For Cygnus~X-1, Figs.~\ref{fig:spectra_pgamma} (a) and~(d) present the model fitting results for the hard state, while Figs.~\ref{fig:spectra_pgamma} (b) and~(e) show those for the soft state. The dotted pink lines show the de-absorbed X-ray emission from the accretion disk, which differs between the two states. The dotted orange lines represent the blackbody radiation from the donor star, which is the same for both states. Observationally, only GeV upper limits are available in the soft state, whereas GeV emission is detected by \textit{Fermi}-LAT in the hard state.

For both states, the LHAASO data ($\gtrsim 10$~TeV) can be reproduced by the $p\gamma$ process occurring in the inner blob, where the target photons originate from the X-ray radiation of the accretion disk surrounding the central engine. 
We also find that the inner blob can accelerate protons (electrons) up to $\sim 100$~TeV ($\sim 500$~GeV).

\subsubsection*{(b) Cygnus~X-3}

\begin{figure*}[ht!]
	\centering
	\subfigure[]{
		\includegraphics[scale=0.3]{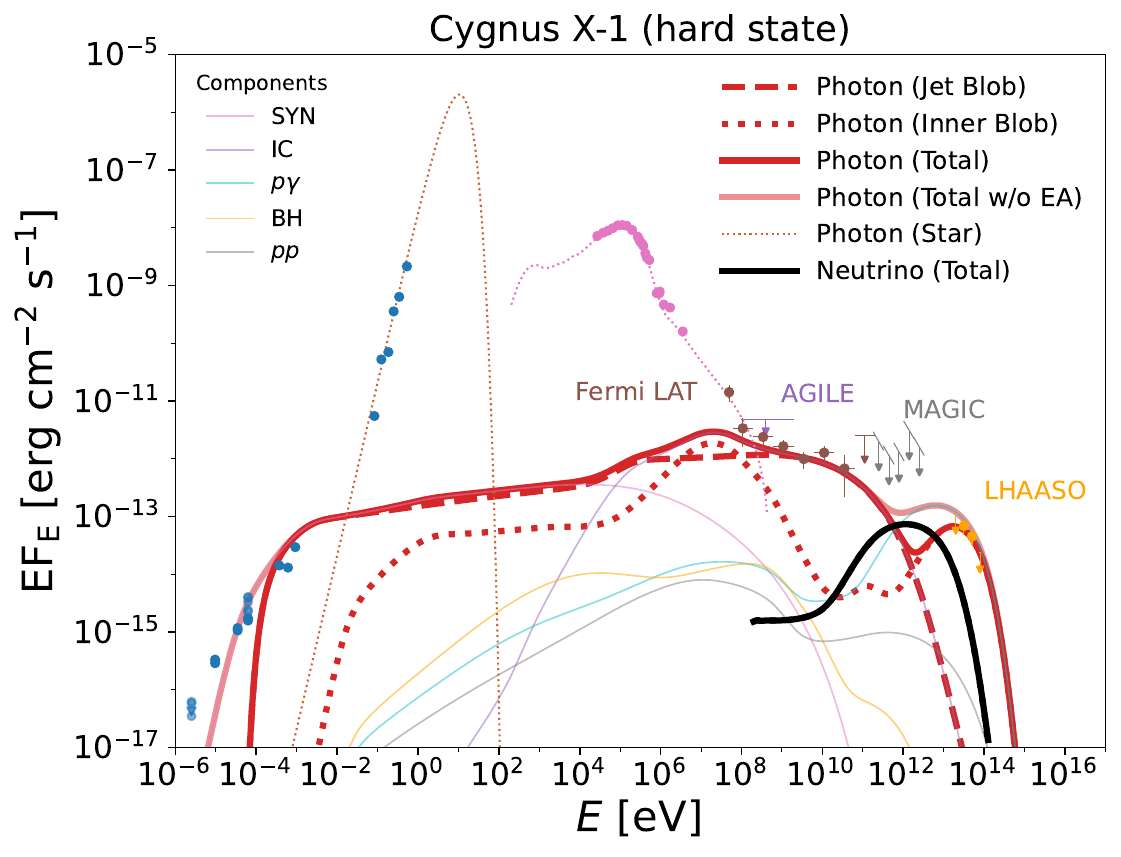}
	}%
        \subfigure[]{
		\includegraphics[scale=0.3]{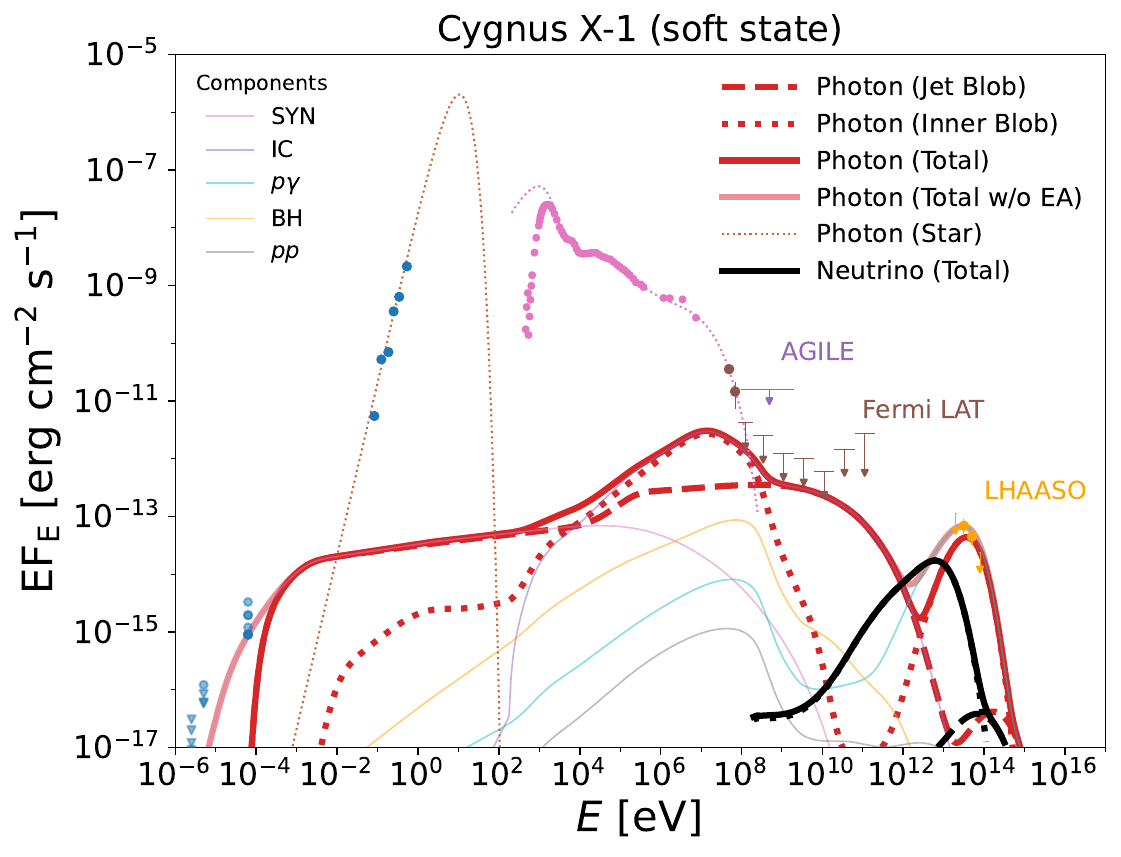}
	}%
        \subfigure[]{
		\includegraphics[scale=0.3]{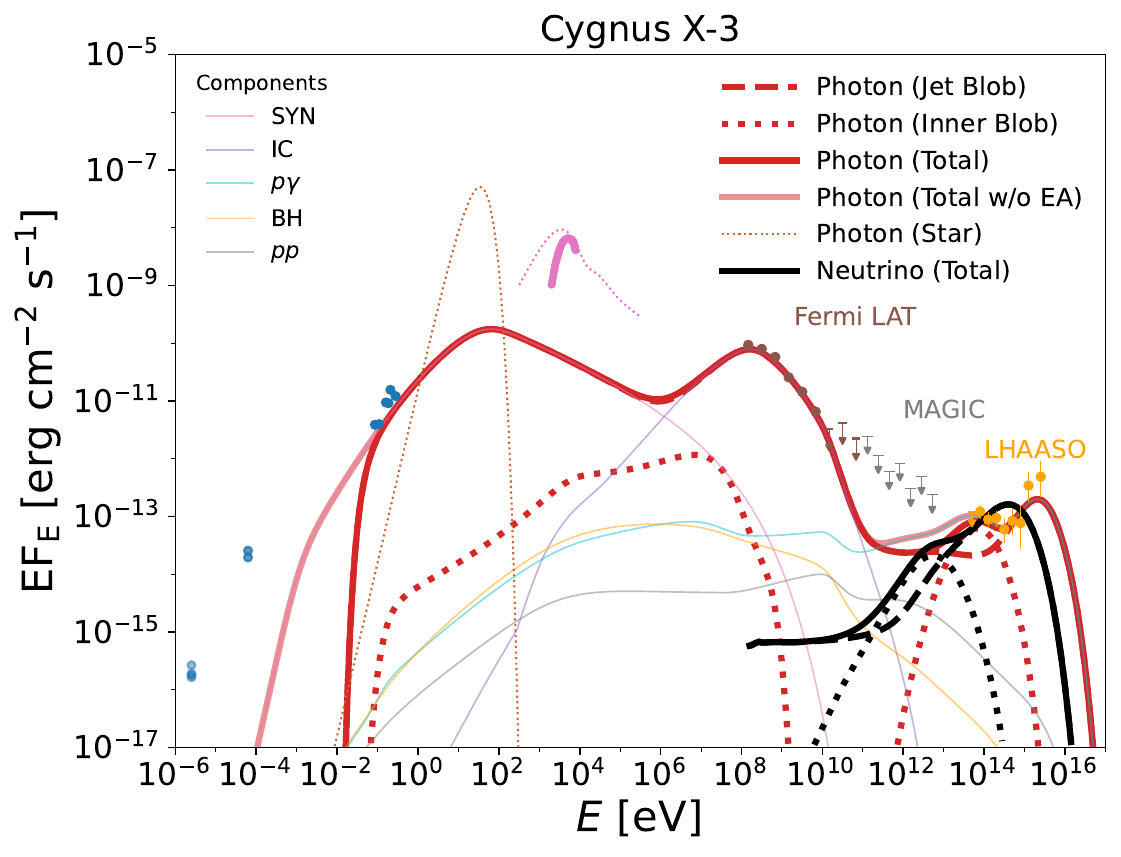}
	}%
    
	\subfigure[]{
		\includegraphics[scale=0.3]{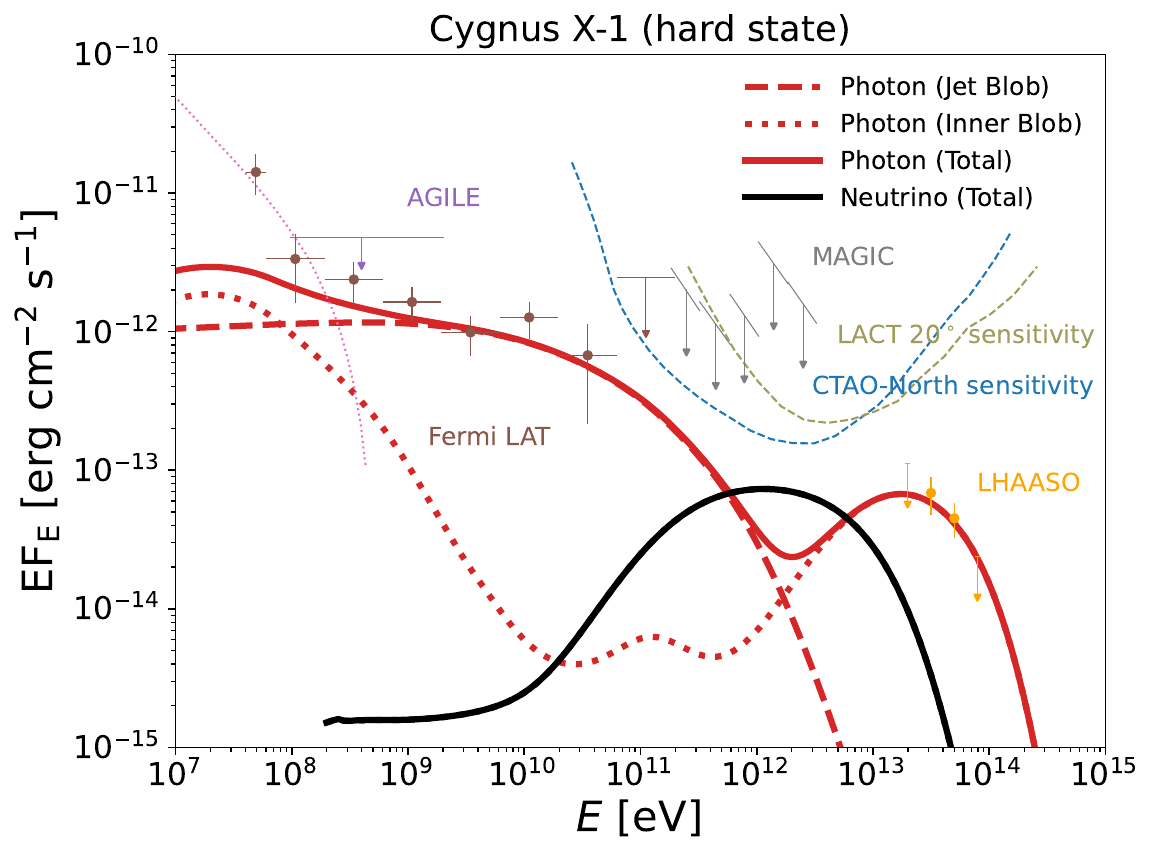}
	}%
	\subfigure[]{
		\includegraphics[scale=0.3]{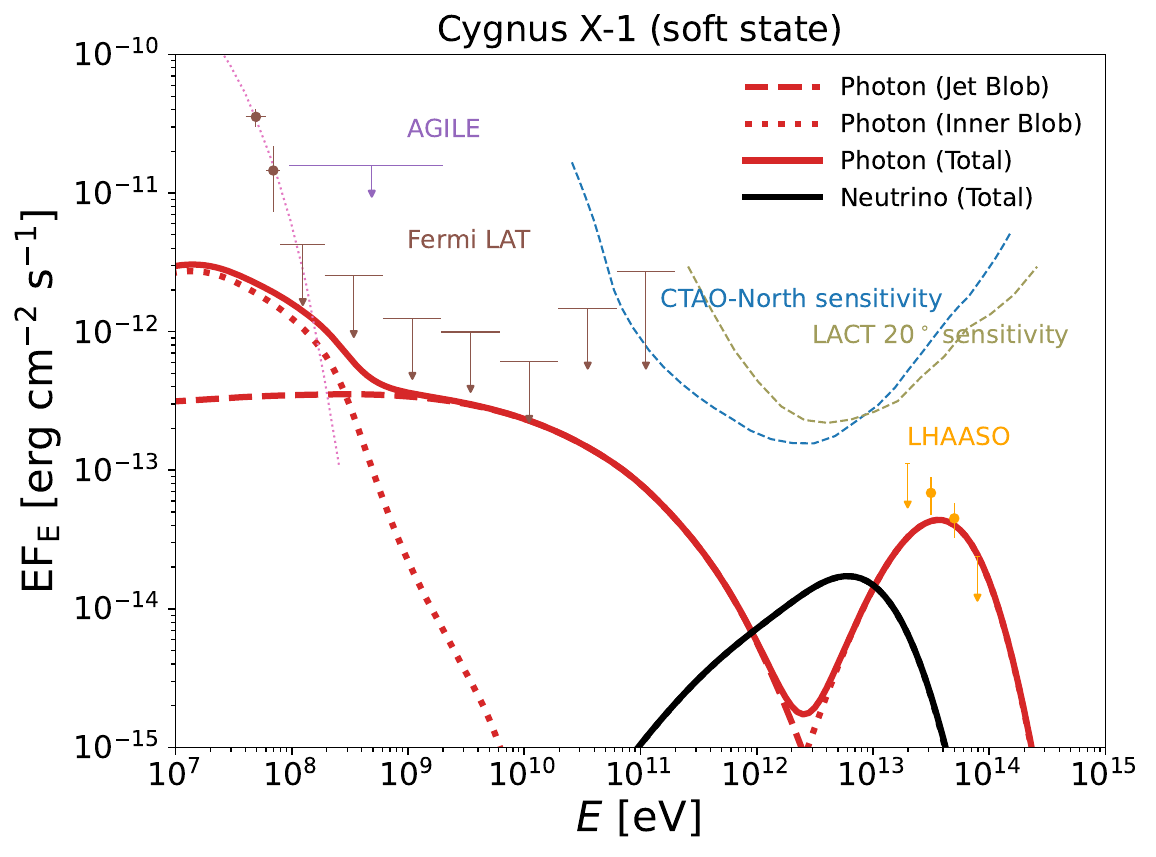}
	}%
	\subfigure[]{
		\includegraphics[scale=0.3]{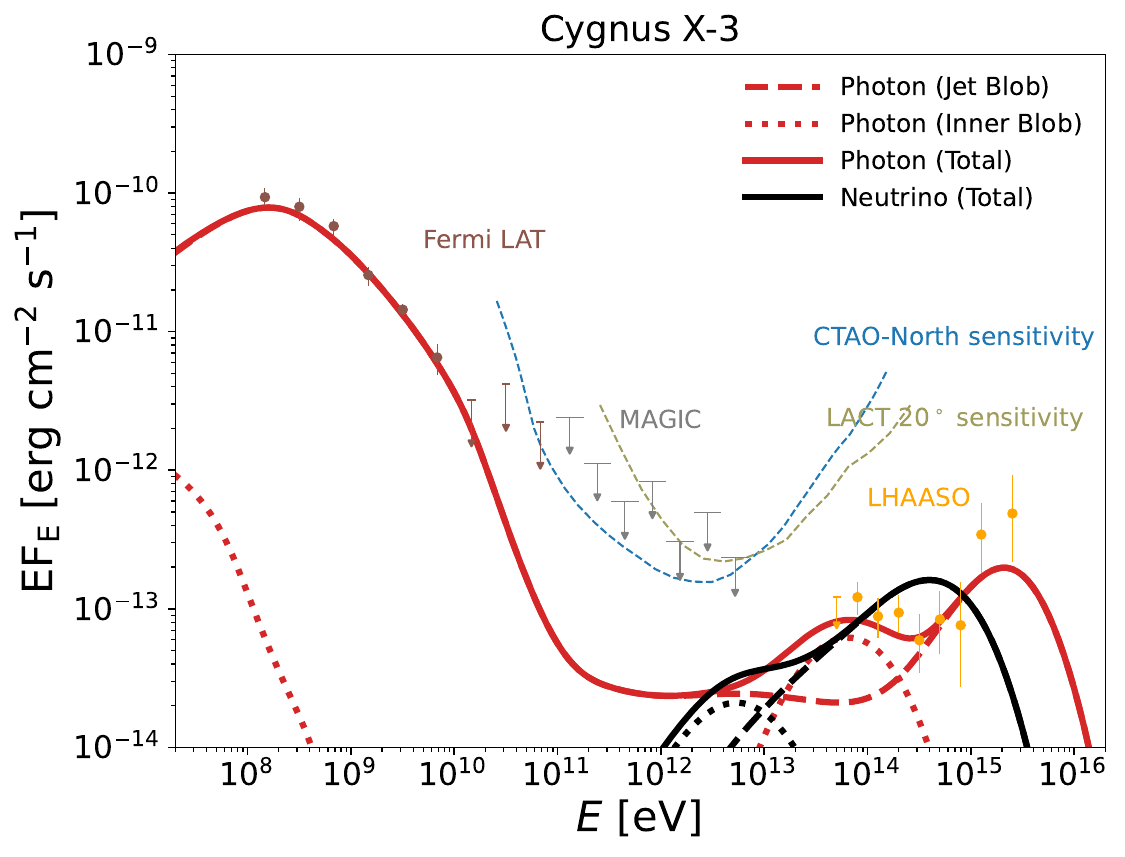}
	}%
		
	\caption{Photon (red) and neutrino (black) spectra obtained from Scenario~A. The contributions to the photon and neutrino fluxes are indicated by dashed lines for the jet blob and dotted lines for the inner blob. The solid lines represent the total fluxes, summing the inner and jet blob contributions. Fainter red solid lines show photon spectra calculated without the external absorption (i.e., without including free-free absorption and external $\gamma\gamma$ annihilation). Additionally, the light thin lines represent various emission components (electron synchrotron, electron IC, $p\gamma$, BH, and $pp$) summed across both regions, as specified in the legend. The dotted orange lines represent the blackbody radiation from the donor star, while the dotted pink lines indicate the de-absorbed X-ray emission from the accretion disk. The bottom row provides a zoomed-in view of the top row. The dashed blue and green-yellow lines show the differential sensitivities for 50-hour exposures of the CTAO Northern Array~\footnote{https://www.ctao.org/for-scientists/performance/} and the LACT full array at a zenith angle of $20^\circ$~\citep[][]{Zhang-2025arXiv250920694Z}, respectively.
    The first, second, and third columns correspond to model fits for the hard state of Cygnus~X-1, the soft state of Cygnus~X-1, and Cygnus~X-3, respectively. Note that the observational data at different energy ranges are not simultaneous.
	}
	\label{fig:spectra_pgamma}
\end{figure*}

For Cygnus~X-3, as shown in Figs.~\ref{fig:spectra_pgamma} (c) and (f), our model reproduces the LHAASO data above $0.1$~PeV, and the predicted spectrum exhibits a two-peak structure. The first peak arises from $p\gamma$ interactions in the inner blob, with target photons provided by the X-ray radiation of the accretion disk, while the second peak originates from $p\gamma$ interactions in the jet blob, where the target photons come from the blackbody radiation of the donor star. 
Here, we use the observed X-ray radiation without accounting for the possibly ten-times-higher intrinsic X-ray radiation~\citep{Veledina-2024NatAs...8.1031V}. This serves as a general example, and consequently, the X-ray photon density in the jet blob is insufficient to generate the required $\sim 100$~TeV photons, which explains the difference from the modeling in \citet{LHAASO-2025arXiv251216638L}.
We find that the jet blob can accelerate protons (electrons) up to $\sim 50$~PeV ($\sim 1$~TeV), whereas the inner blob can accelerate protons (electrons) up to $\sim 500$~TeV ($\sim 100$~GeV).

\begin{table*}[ht]
\caption{Fitting parameter values used for the jet blob and inner blob in Scenario~A. The remaining jet-blob parameters are summarized in Table~\ref{tab:jet_blob}.}
\label{tab:pgamma}
\centering
\begin{threeparttable}
\centering
\begin{minipage}{0.48\textwidth}
\centering
\begin{tabular}{cc|ccc} 
\hline \hline
Jet Blob Parameters  &  Symbol [Units]   & Cygnus~X-1 (hard) & Cygnus~X-1 (soft) & Cygnus~X-3 \\
\hline
{Magnetic field fraction} & ${{\epsilon}_B}$ &         $7.1 \times 10^{-4}$        &     $3.5 \times 10^{-4}$          &  $2.1 \times 10^{-3}$   \\
Magnetic field & $B'_{\rm jet}$~[G] &         $4.3$        &     $4.2$          &  $6.1 \times 10^{2}$   \\
$e^-$ Fraction & ${{\epsilon}_e}$ &          $3.0 \times 10^{-3}$         &       $4.5 \times 10^{-4}$         &   $1.7 \times 10^{-3}$ \\
$p$ injection luminosity  & $L'_{p, \rm jet}~\rm [erg~s^{-1}]$                         &     $1.7 \times 10^{37}$               &    $2.7 \times 10^{37}$       &  $5.5 \times 10^{38}$        \\
$p$ Maximum Lorentz factor  & $\gamma'_{p, \rm max}$           &       $9.9 \times 10^5$             &    $8.3 \times 10^5$           &  $7.5 \times 10^6$   \\
$e^-$ Maximum Lorentz factor  & $\gamma'_{ e, \rm max}$           &       $7.4 \times 10^7$             &    $6.3 \times 10^7$           &  $2.3 \times 10^6$   \\
\hline
\end{tabular}
\end{minipage}%
\hfill
\begin{minipage}{0.48\textwidth}
\centering
\begin{tabular}{cc|ccc} 
\hline \hline
Inner Blob Parameters  &  Symbol [Units]   & Cygnus~X-1 (hard) & Cygnus~X-1 (soft) & Cygnus~X-3 \\
\hline
Emission region location  & $H_{\rm in} ~\rm [cm]$                   &          $2.0 \times 10^8$          &         $2.0 \times 10^8$    & $6.7 \times 10^7$       \\
Half-opening angle~\tablenotemark{a} & $\theta_{\rm op}^{\rm in} ~\rm [deg]$ & $60$ & $60$ & $60$ \\
Emission region radius~\tablenotemark{b} & $r_{\rm in} ~\rm [cm]$                   &        $1.9 \times 10^8$            &     $1.9 \times 10^8$         & $6.4 \times 10^7$      \\
Lorentz factor (velocity) & $\Gamma_{\rm in} (\beta_{\rm in})$     &              $1.05 ~(0.30)$    &         $1.05 ~(0.30)$        &    $1.05 ~(0.30)$   \\
Acceleration parameter & $\eta_{\rm acc}^{\rm in}$           &           $1.0 ~\beta_j^{-2}$         &     $1.0 ~\beta_j^{-2}$        &    $1.0 ~\beta_j^{-2}$   \\
{Magnetic field fraction~\tablenotemark{c}} & ${{\epsilon}_B^{\rm in}}$ &         $9.0 \times 10^{-6}$        &     $6.0 \times 10^{-6}$          &  $1.5 \times 10^{-5}$   \\
Magnetic field & $B'_{\rm in}$~[G] &         $6.4 \times 10^{3}$        &     $5.2 \times 10^{3}$          &  $8.9 \times 10^{4}$   \\
$p$ fraction~\tablenotemark{c} & $\epsilon_p^{\rm in}$ &          $2.5 \times 10^{-2}$         &       $3.5 \times 10^{-3}$         &   $1.3 \times 10^{-4}$ \\
$e^-$ Fraction~\tablenotemark{c} & $\epsilon_e^{\rm in}$ &          $1.0 \times 10^{-5}$         &       $1.0 \times 10^{-5}$         &   $1.0 \times 10^{-5}$ \\
$p$ injection luminosity & $L'_{p, \rm in}~\rm [erg~s^{-1}]$                         &       $6.0 \times 10^{36}$             &  $8.5 \times 10^{35}$        & $4.3 \times 10^{35}$          \\
$p$ spectral index  & $s_p^{\rm in}$           &          $2.0$          &        $2.0$       &  $2.0$   \\
$p$ Minimum Lorentz factor  & $\gamma_{p, \rm min}'^{~\rm in}$           &          $1.0$          &        $1.0$       &  $1.0$   \\
$p$ Maximum Lorentz factor & $\gamma_{p, \rm max}'^{~\rm in}$           &          $1.2 \times 10^5$          &    $9.5 \times 10^4$           &  $5.5 \times 10^5$   \\
$e^-$ spectral index  & $s_e^{\rm in}$           &          $2.8$          &        $2.8$       &  $2.8$   \\
$e^-$ Minimum Lorentz factor & $\gamma_{ e, \rm min}'^{~\rm in}$           &        $1.0 \times 10^2$            &      $1.0 \times 10^2$        &   $1.6 \times 10^3$  \\
$e^-$ Maximum Lorentz factor & $\gamma_{ e, \rm max}'^{~\rm in}$           &          $1.2 \times 10^6$          &    $1.1 \times 10^6$           &  $1.9 \times 10^5$   \\
Ambient number density~\tablenotemark{d} & $n'_{p, \rm in}~\rm [cm^{-3}]$                    &          $3.0 \times 10^{11}$          & $4.2 \times 10^{10}$ & $1.8 \times 10^{11}$ \\
\hline
\end{tabular}
\tablenotetext{a}{For simplicity, the half-opening angle of the inner blob is estimated as $1/\Gamma_j^{\rm in}$.}
\tablenotetext{b}{We adopt an inner radius of $r'_{\rm in} = 30 R_s$, where $R_s = 2GM_{\rm BH}/c^2$ is the Schwarzschild radius of black hole.}
\tablenotetext{c}{For simplicity, we adopt the same blob power as the jet blob.}
\tablenotetext{d}{Similar to the jet blob, we set $n'_{p, \rm in} = 0.1 n'_{p, \rm ul}$.}
\end{minipage}
\vspace{5pt}
\end{threeparttable}
\end{table*}

\subsection{Results for Scenario~B}

\begin{table*}[ht]
\caption{Similar to Table~\ref{tab:pgamma}, fitting parameter values used for the jet blob and the stellar-wind region in Scenario~B. }
\label{tab:pp_A}
\centering
\begin{threeparttable}
\centering
\begin{minipage}{0.48\textwidth}
\centering
\begin{tabular}{cc|ccc} 
\hline \hline
Jet Blob Parameters  &  Symbol [Units]   & Cygnus~X-1 (hard) & Cygnus~X-1 (soft) & Cygnus~X-3 \\
\hline
{Magnetic field fraction} & ${{\epsilon}_B}$ &         $3.3 \times 10^{-4}$        &     $3.5 \times 10^{-4}$          &  $1.5 \times 10^{-3}$   \\
Magnetic field & $B'_{\rm jet}$~[G] &         $2.9$        &     $4.2$          &  $5.1 \times 10^{2}$   \\
{$e^-$ Fraction} & ${{\epsilon}_e}$ &          $3.0 \times 10^{-3}$         &       $2.5 \times 10^{-4}$         &   $1.5 \times 10^{-3}$ \\
$p$ injection luminosity  & $L'_{p, \rm jet}~\rm [erg~s^{-1}]$                         &     $1.7 \times 10^{37}$               &    $2.7 \times 10^{37}$       &  $5.5 \times 10^{38}$        \\
$p$ Maximum Lorentz factor  & $\gamma'_{p, \rm max}$           &       $6.8 \times 10^5$             &    $8.3 \times 10^5$           &  $6.4 \times 10^6$   \\
$e^-$ Maximum Lorentz factor  & $\gamma'_{ e, \rm max}$           &       $5.7 \times 10^7$             &    $4.0 \times 10^7$           &  $2.5 \times 10^6$   \\
\hline
\end{tabular}
\end{minipage}%
\hfill
\begin{minipage}{0.48\textwidth}
\centering
\begin{tabular}{cc|ccc} 
\hline \hline
Stellar Wind Parameters &  Symbol [Units]   & Cygnus~X-1 (hard) & Cygnus~X-1 (soft) & Cygnus~X-3 \\
\hline
Emission region Radius~\tablenotemark{a} & $r_{w} ~\rm [cm]$                    &     $1.1 \times 10^{13}$  &   $1.1 \times 10^{13}$    & $4.0 \times 10^{11}$      \\
Magnetic field~\tablenotemark{b} & $B_{w} ~\rm [G]$  &     $4.1 \times 10^{-2}$     &  $4.1 \times 10^{-2}$   &  $4.3$   \\
$p$ escaping luminosity & $L_{p, \rm ext}~\rm [erg~s^{-1}]$                                &    $4.5 \times 10^{35}$   &  $5.9 \times 10^{35}$   &  $5.4 \times 10^{37}$        \\
$e^-$ escaping luminosity & $L_{e, \rm ext}~\rm [erg~s^{-1}]$                                &    $6.8 \times 10^{33}$   &  $7.3 \times 10^{32}$   &  $1.6 \times 10^{32}$        \\
Ambient number density~\tablenotemark{c}  & $n_{p, \rm ext}~\rm [cm^{-3}]$                       & $2.5 \times 10^{10}$ & $2.0 \times 10^{10}$ & $2.1 \times 10^{11}$ \\
\hline
\end{tabular}
\tablenotetext{a}{The radius of the stellar-wind region is estimated as the distance from the donor star to the center of the jet blob.
}
\tablenotetext{b}{We assume the magnetic field in the stellar-wind region is given by $B_w = \sqrt{ \sigma_w \dot{M}_w v_w^2 / [2 r_{w}^2 c (1 + \sigma_w)] }$, with the magnetization parameter $ \sigma_w = 0.01 $~\citep[e.g.,][]{Kennel-1984ApJ...283..710K,Shibata-2003MNRAS.346..841S}.}
\tablenotetext{c}{In Scenario~B, we allow source-dependent,
order-of-magnitude variations around the fiducial stellar-wind proton
densities listed in Table~\ref{tab:jet_blob} to account for uncertainties
in the wind properties.}
\end{minipage}
\vspace{5pt}
\end{threeparttable}
\end{table*}

\begin{figure*}[ht!]
	\centering
	\subfigure[]{
		\includegraphics[scale=0.3]{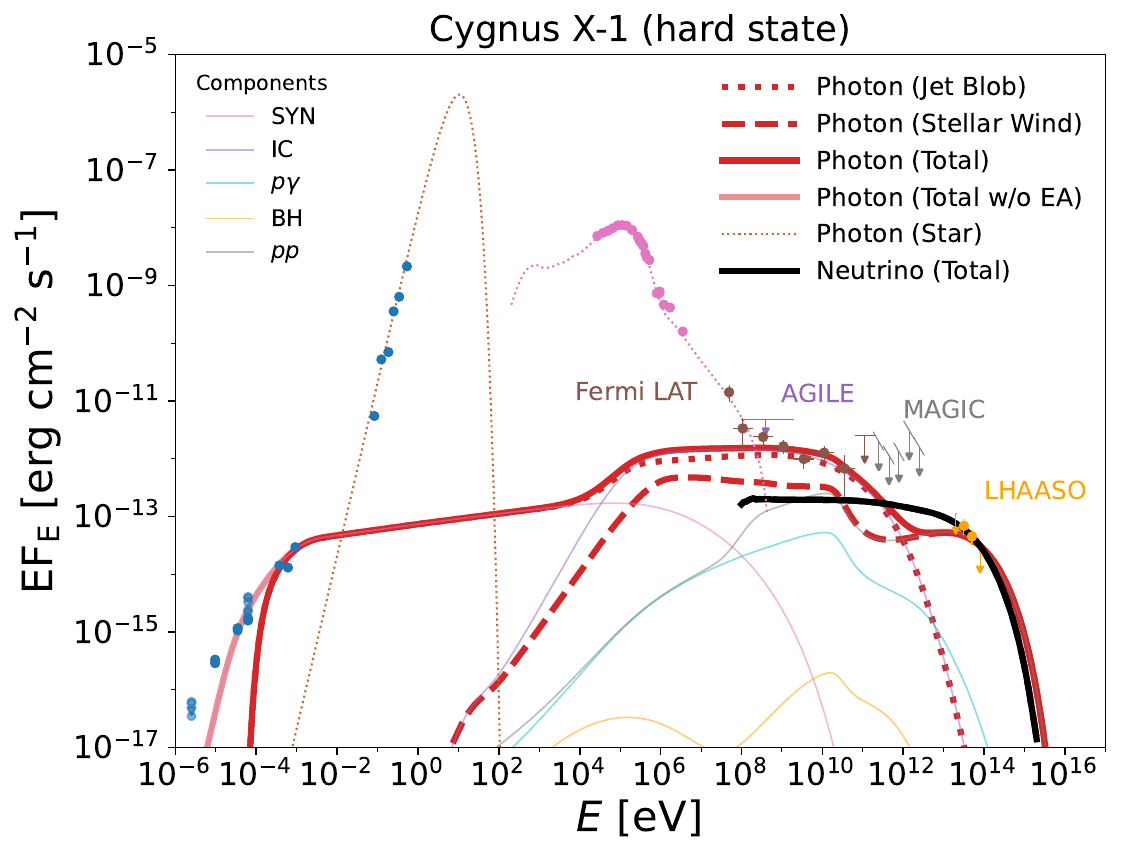}
	}%
        \subfigure[]{
		\includegraphics[scale=0.3]{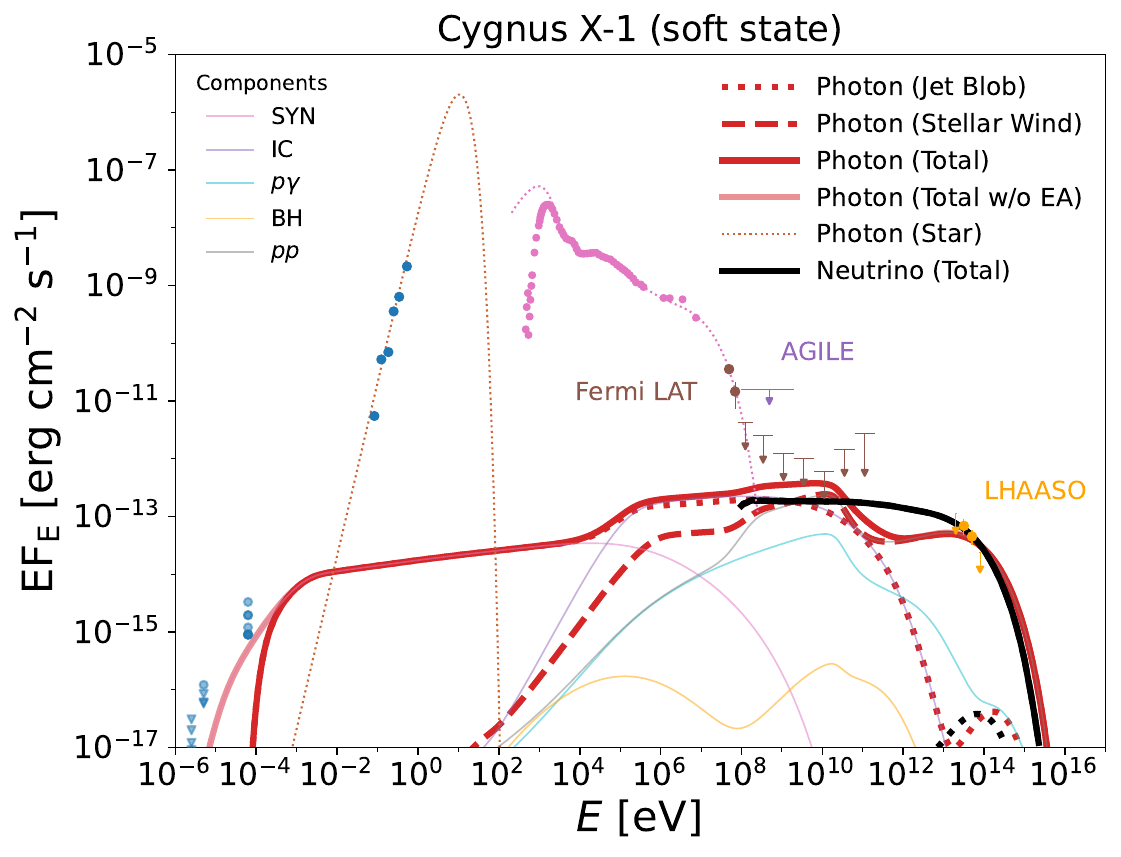}
	}%
        \subfigure[]{
		\includegraphics[scale=0.3]{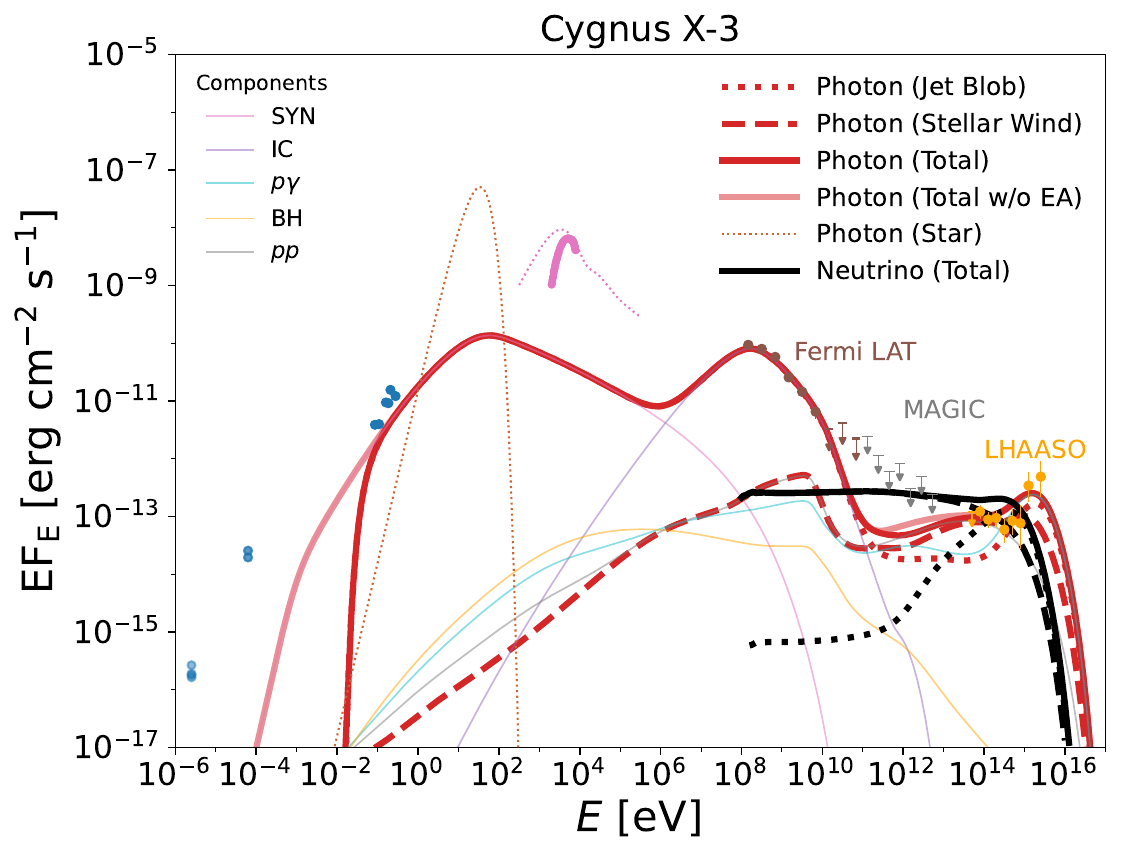}
	}%
    
	\subfigure[]{
		\includegraphics[scale=0.3]{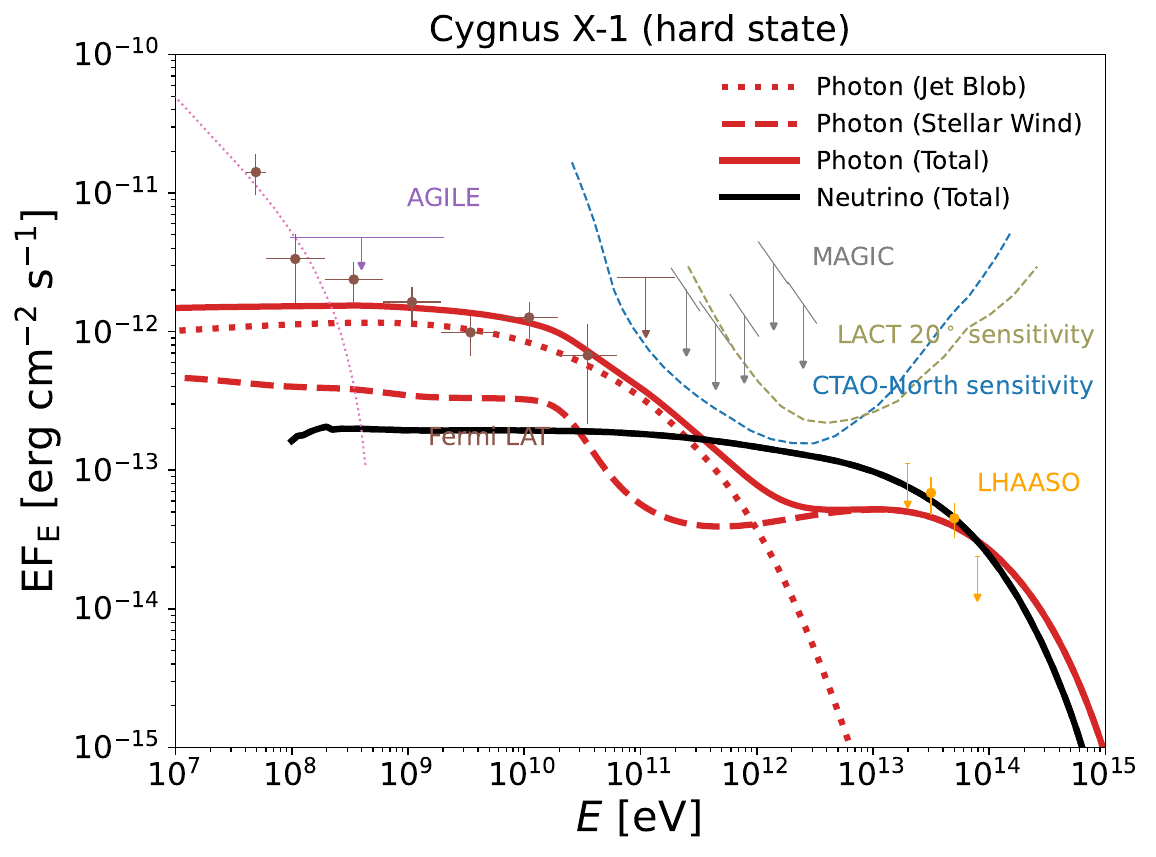}
	}%
	\subfigure[]{
		\includegraphics[scale=0.3]{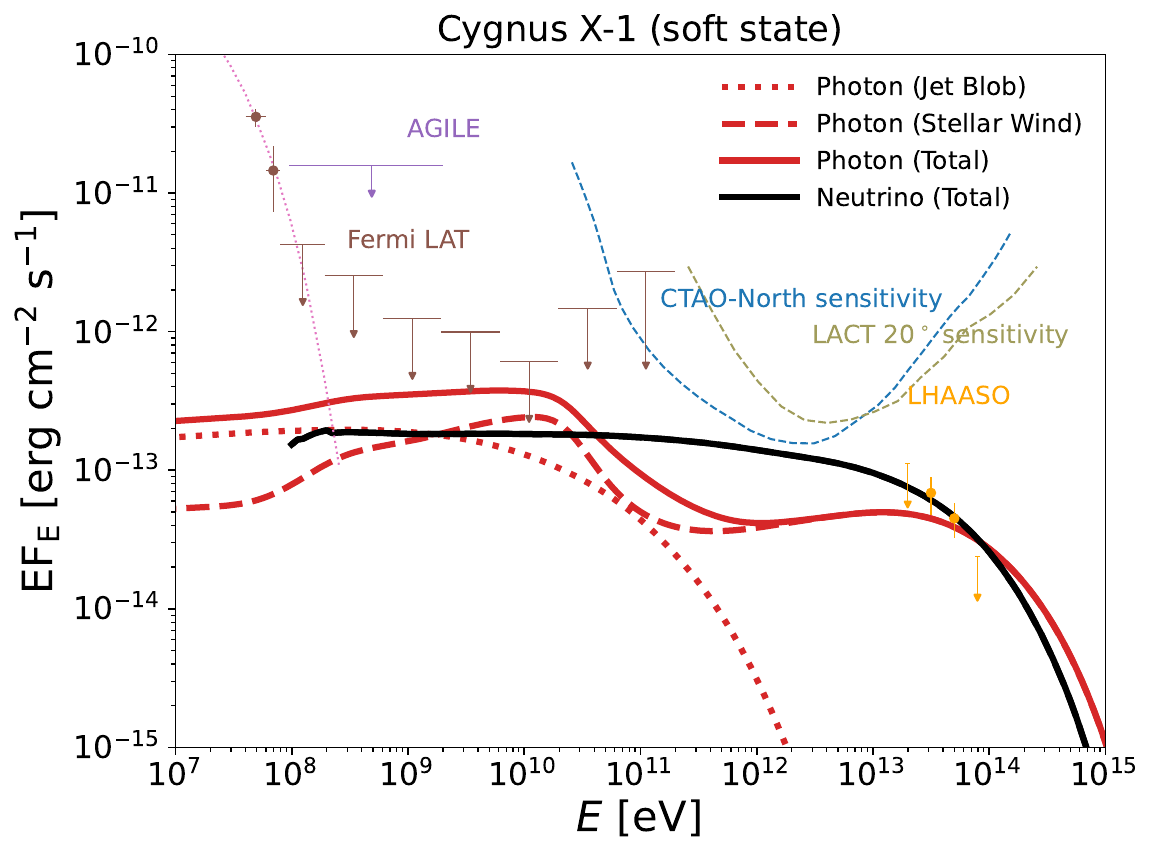}
	}%
	\subfigure[]{
		\includegraphics[scale=0.3]{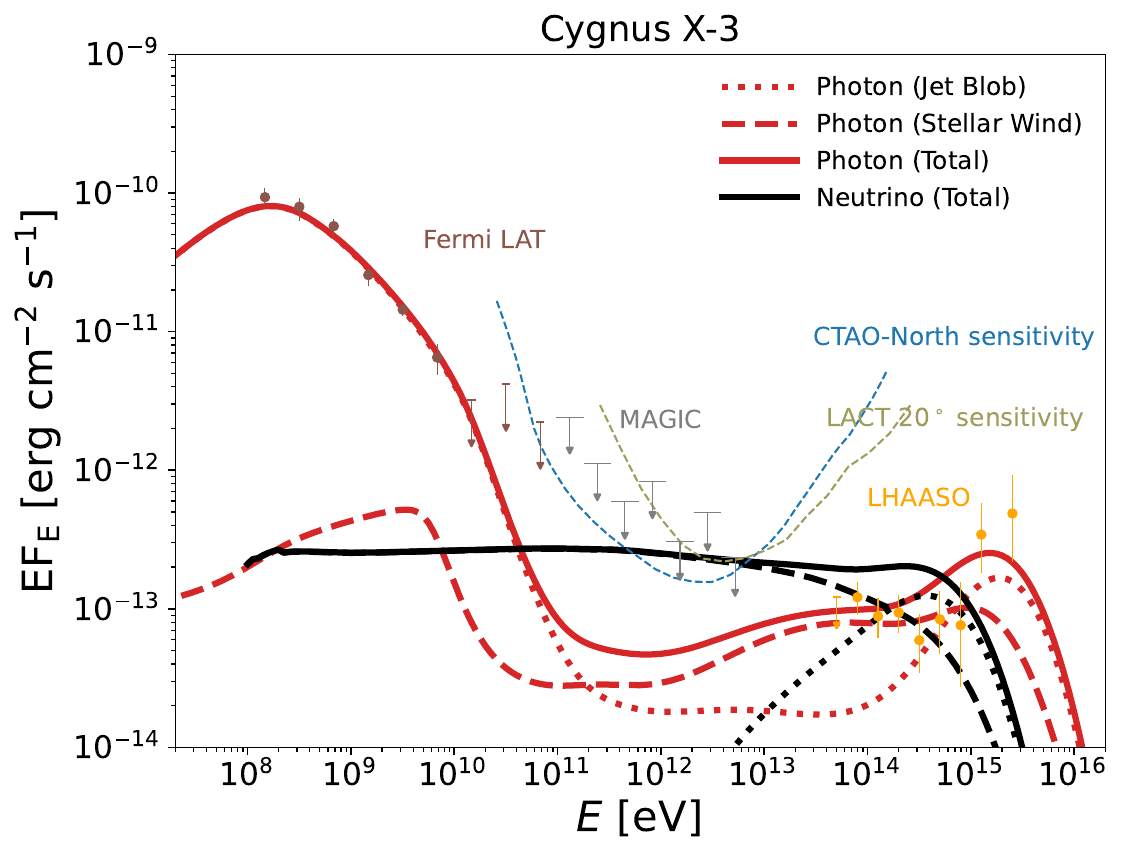}
	}%
		
	\caption{Similar to Fig.~\ref{fig:spectra_pgamma}, but for Scenario~B. The contributions to the photon (red) and neutrino (black) fluxes are indicated by dotted lines for the jet blob and dashed lines for the stellar-wind region.
	}
	\label{fig:spectra_pp_A}
\end{figure*}

The fitting parameters for Scenario~B are listed in Table~\ref{tab:pp_A}. The jet blob has the same location and radius as in Scenario~A, while some microphysical parameters are allowed to vary. The external stellar-wind region is approximated as having a uniform proton density and magnetic field. 
The fiducial wind proton density listed in Table~\ref{tab:jet_blob} is evaluated at the center of the jet blob and is comparable to the average density of the stellar-wind region when the wind density profile is taken into account. To reproduce the observed data, we adopt ambient number densities of 20 and 25 times the fiducial value for the soft and hard states of Cygnus~X-1, respectively, and 0.1 times the fiducial value for Cygnus~X-3. Such variations are plausible given the uncertainties in the wind properties and possible deviations from a smooth, spherically symmetric wind~\citep[e.g.,][]{Mivskovivcova_2016A&A...590A.114M, Koljonen-2017MNRAS.472.2181K, Hirsch_2019A&A...626A..64H, Vilhu_2021A&A...649A.176V}.

Similar to Fig.~\ref{fig:spectra_pgamma}, Fig.~\ref{fig:spectra_pp_A} presents the photon (red lines) and neutrino (black lines) spectra for Scenario~B. 
In the $0.1$-$10$~TeV range, the photon spectrum shows a mild suppression due to the spectral shape of the $pp$ component together with $\gamma\gamma$ annihilation. The predicted flux in this band is generally below the 50-hour sensitivities of CTAO and LACT, indicating that a longer exposure would likely be needed for detection. As in Scenario~A, GeV photons mainly originate from EIC scattering, while the free-free absorption of the radio emission suggests an origin at larger jet heights, around $\sim10^{14}~\mathrm{cm}$.

\subsubsection*{(a) Cygnus~X-1}

For both states of Cygnus~X-1, the GeV emission is attributed to the EIC process occurring in both the jet blob and the stellar-wind region, as shown in as shown in Figs.~\ref{fig:spectra_pp_A} (a), (b), (d), and~(e).
The LHAASO data are well reproduced by the $pp$ process occurring in the stellar-wind region, where escaping protons interact with ambient protons in the stellar wind.
Although our predicted $\sim 80$~TeV flux slightly exceeds the 95\% confidence upper limit reported by LHAASO, this discrepancy can be alleviated if the proton distribution exhibits a cutoff steeper than a simple exponential form. Such super-exponential cutoffs can naturally arise in acceleration scenarios involving magnetic reconnection~\citep[e.g.,][]{Werner-2016ApJ...816L...8W,Kagan-2018MNRAS.476.3902K}.

Here, we set the ambient proton number density in the stellar-wind region to be around ten times larger than the value at the center of the jet blob. This density requirement can be relaxed by increasing the proton luminosity of the jet blob.
This scenario also suggests that protons (electrons) can be accelerated up to energies of $\sim 500~\mathrm{TeV}$ ($\sim 10$~TeV), potentially contributing to the population of Galactic high-energy cosmic rays.

\subsubsection*{(b) Cygnus~X-3}

For Cygnus~X-3, as shown in Figs.~\ref{fig:spectra_pp_A} (c) and~(f), Scenario~B exhibits a plateau-plus-peak structure above $10$~TeV, rather than the two-peak feature seen in Scenario~A. The plateau is mainly produced by $pp$ interactions in the stellar-wind region, while the PeV peak receives contributions from both $p\gamma$ interactions in the jet blob and $pp$ interactions in the stellar-wind region.
The predicted PeV-band spectrum shows some tension with the detailed shape of the LHAASO data, being somewhat high at sub-PeV and lower-PeV energies but below the highest-energy point. This tension can be alleviated by adjusting the proton luminosity and/or the cutoff shape of the proton distribution. We adopt the current parameter values as a representative example to maintain a consistent proton fraction $\epsilon_p$ across all modeling cases.

Here, the value of $n_{p,\rm ext}$ in the stellar-wind region is assumed to be roughly ten times lower than that at the center of the jet blob, and it can be increased if a lower proton luminosity is adopted for the jet blob. 
Similar to Scenario~A, Scenario~B is capable of accelerating protons (electrons) up to $\sim 50~\mathrm{PeV}$ ($\sim 1$~TeV).

\subsection{Results for Scenario~C}

\begin{table*}[ht]
\caption{Similar to Table~\ref{tab:pp_A}, fitting parameter values used for the jet blob and the external shell in Scenario~C. }
\label{tab:pp_B}
\centering
\begin{threeparttable}
\centering
\begin{minipage}{0.48\textwidth}
\centering
\begin{tabular}{cc|ccc} 
\hline \hline
Jet Blob Parameters  &  Symbol [Units]   & Cygnus~X-1 (hard) &  Cygnus~X-3 \\
\hline
{Magnetic field fraction} & ${{\epsilon}_B}$ &         $8.3 \times 10^{-4}$        &  $2.1 \times 10^{-3}$   \\
Magnetic field & $B'_{\rm jet}$~[G] &         $4.7$        &     $6.1 \times 10^{2}$   \\
$e^-$ Fraction & $\epsilon_e$ &          $1.5 \times 10^{-3}$         &   $1.7 \times 10^{-3}$ \\
$p$ injection luminosity  & $L'_{p, \rm jet}~\rm [erg~s^{-1}]$                         &     $1.7 \times 10^{37}$               &  $5.5 \times 10^{38}$        \\
$p$ Maximum Lorentz factor  & $\gamma'_{p, \rm max}$           &       $1.1 \times 10^6$            &  $7.5 \times 10^6$   \\
$e^-$ Maximum Lorentz factor  & $\gamma'_{ e, \rm max}$           &       $4.5 \times 10^7$            &  $2.3 \times 10^6$   \\
\hline
\end{tabular}
\end{minipage}%
\hfill
\begin{minipage}{0.48\textwidth}
\centering
\begin{tabular}{cc|cc} 
\hline \hline
External Shell Parameters & Symbol [Units]   & Cygnus~X-1 (hard) & Cygnus~X-3 \\
\hline
Emission region location  &  $H_{\rm ext} ~\rm [cm]$                         &         $2.5 \times 10^{19}$ & $4.6 \times 10^{19}$       \\
Emission region width & $\Delta_{\rm ext} ~\rm [cm]$                    &     $8.2 \times 10^{18}$  & $1.5 \times 10^{19}$      \\
Magnetic field & $B_{\rm ext} ~\rm [\mu G]$  &     $10$     &  $10$   \\
$p$ escaping luminosity & $L_{p, \rm ext}~\rm [erg~s^{-1}]$                                &    $4.5 \times 10^{35}$    &  $5.4 \times 10^{37}$        \\
$e^-$ escaping luminosity & $L_{e, \rm ext}~\rm [erg~s^{-1}]$                                &    $3.4 \times 10^{33}$  &  $1.6 \times 10^{32}$        \\
Ambient number density~\tablenotemark{a} & $n_{p, \rm ext}~\rm [cm^{-3}]$                       & $1.3 \times 10^3$ & $1.8 \times 10^2$ \\
Total Ambient Mass~\tablenotemark{b} & $M_{p, \rm ext}~\rm [M_\odot]$                       & $6.9 \times 10^{4}$ & $6.3 \times 10^{4}$ \\
\hline
\end{tabular}
\tablenotetext{a}{We estimate the ambient proton number density contributed by the interstellar medium or a molecular cloud in Scenario~C.}
\tablenotetext{b}{The total ambient mass is calculated as $M_{p,\rm ext} = V_{\rm ext} m_p n_{p,\rm ext}$, with the volume given by $V_{\rm ext} = 4 \pi H_{\rm ext}^2 \Delta_{\rm ext}$.}
\end{minipage}
\vspace{5pt}
\end{threeparttable}
\end{table*}

\begin{figure*}[ht!]
	\centering
	\subfigure[]{
		\includegraphics[scale=0.3]{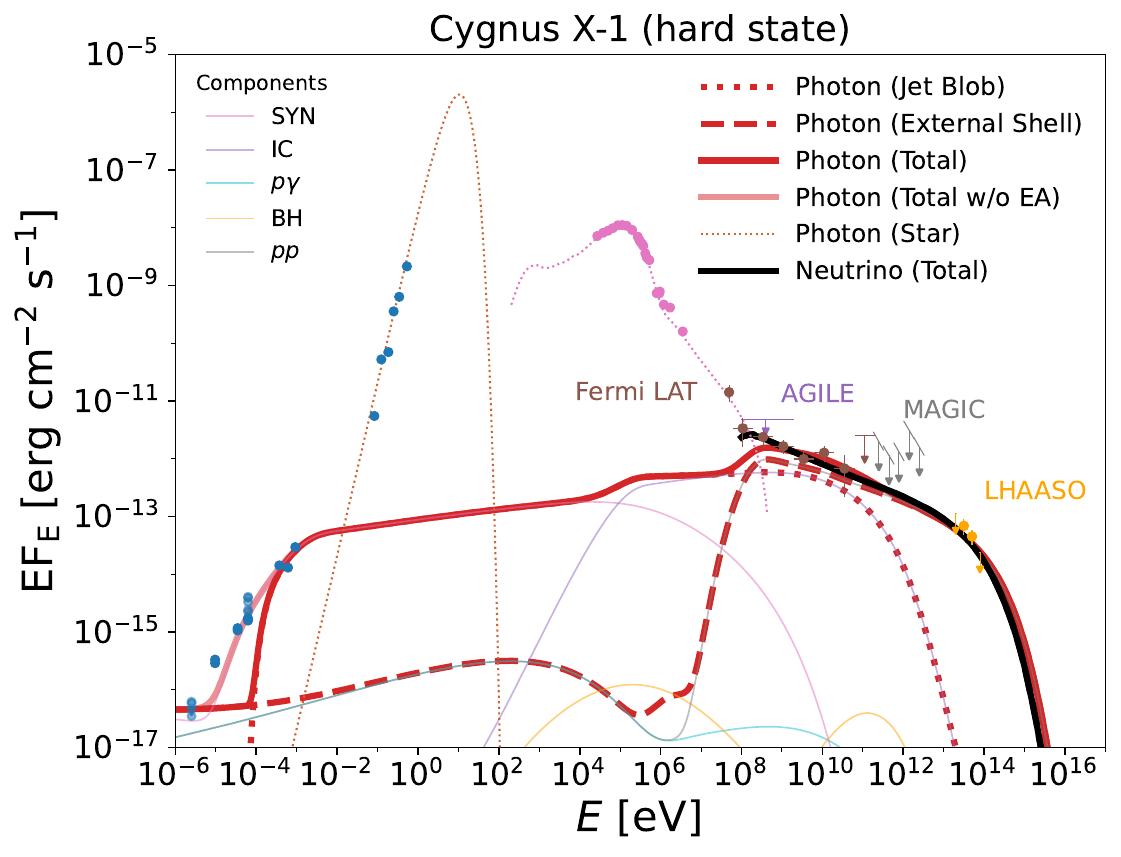}
	}%
	\subfigure[]{
		\includegraphics[scale=0.3]{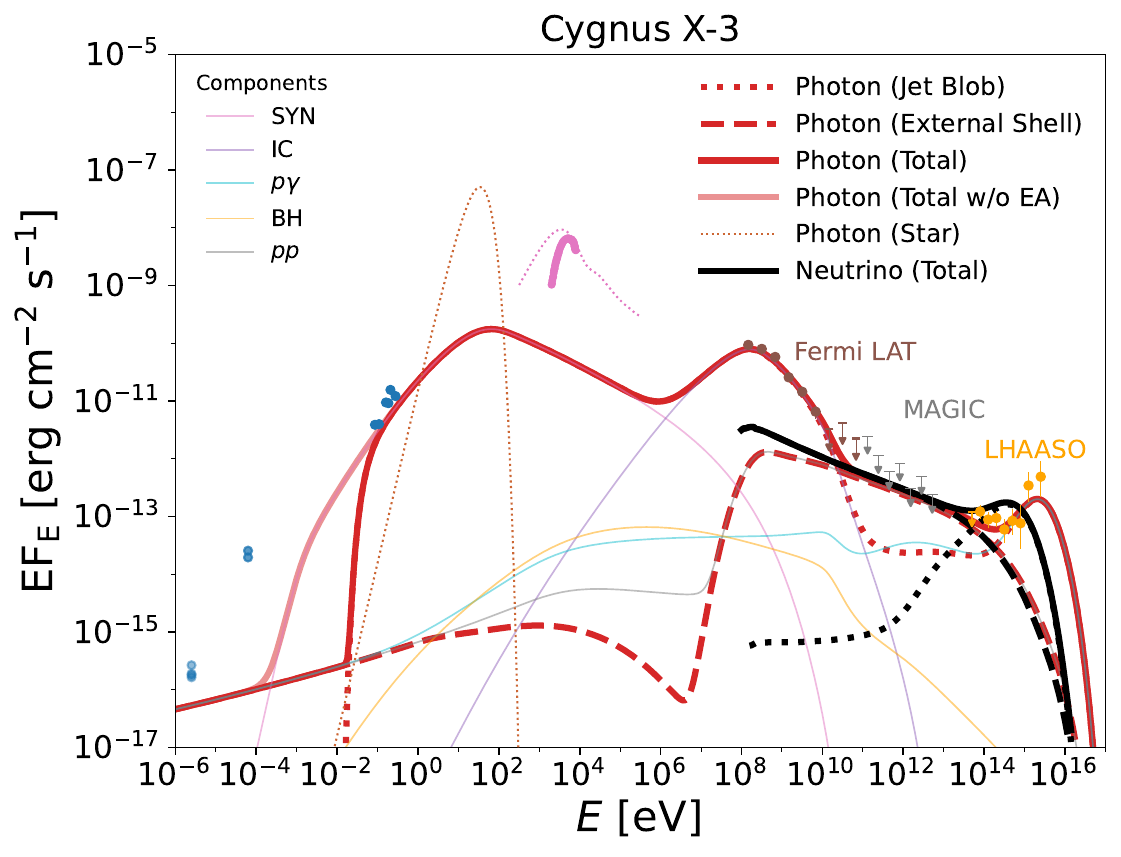}
	}%

        \subfigure[]{
		\includegraphics[scale=0.3]{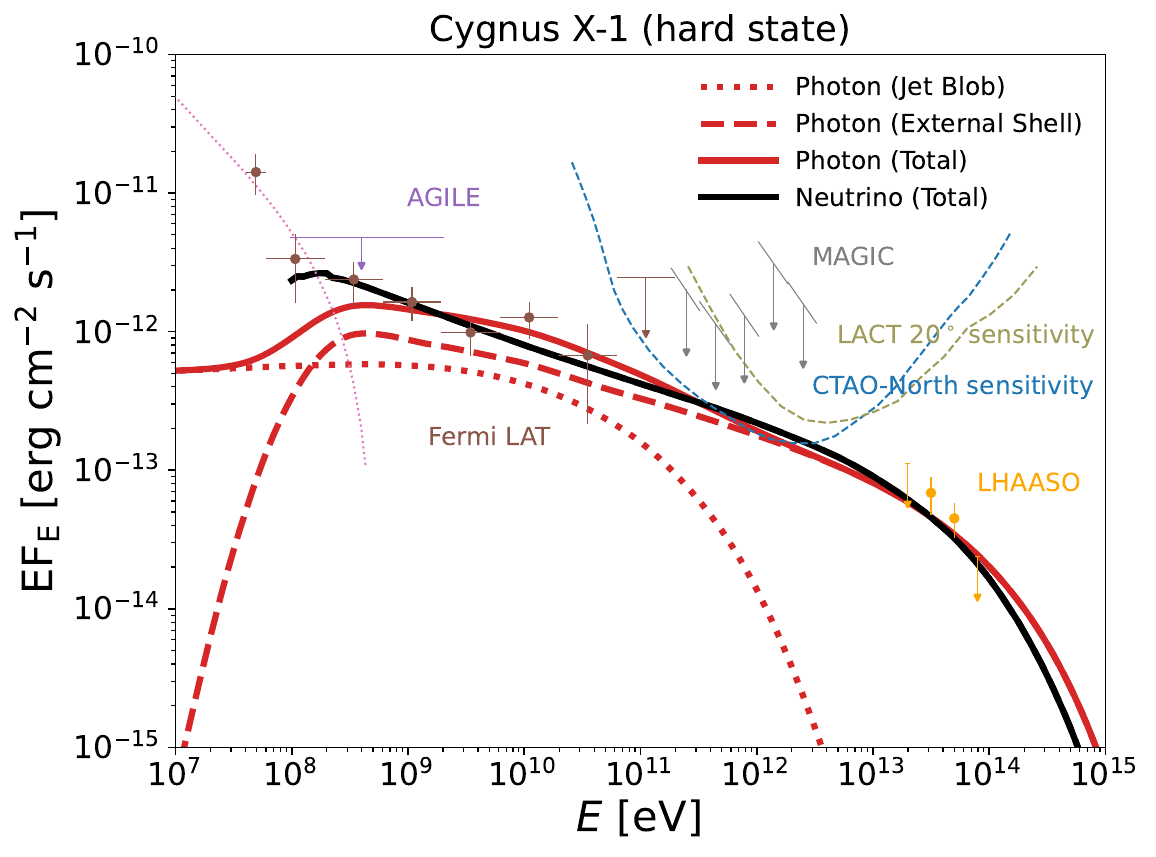}
	}%
	\subfigure[]{
		\includegraphics[scale=0.3]{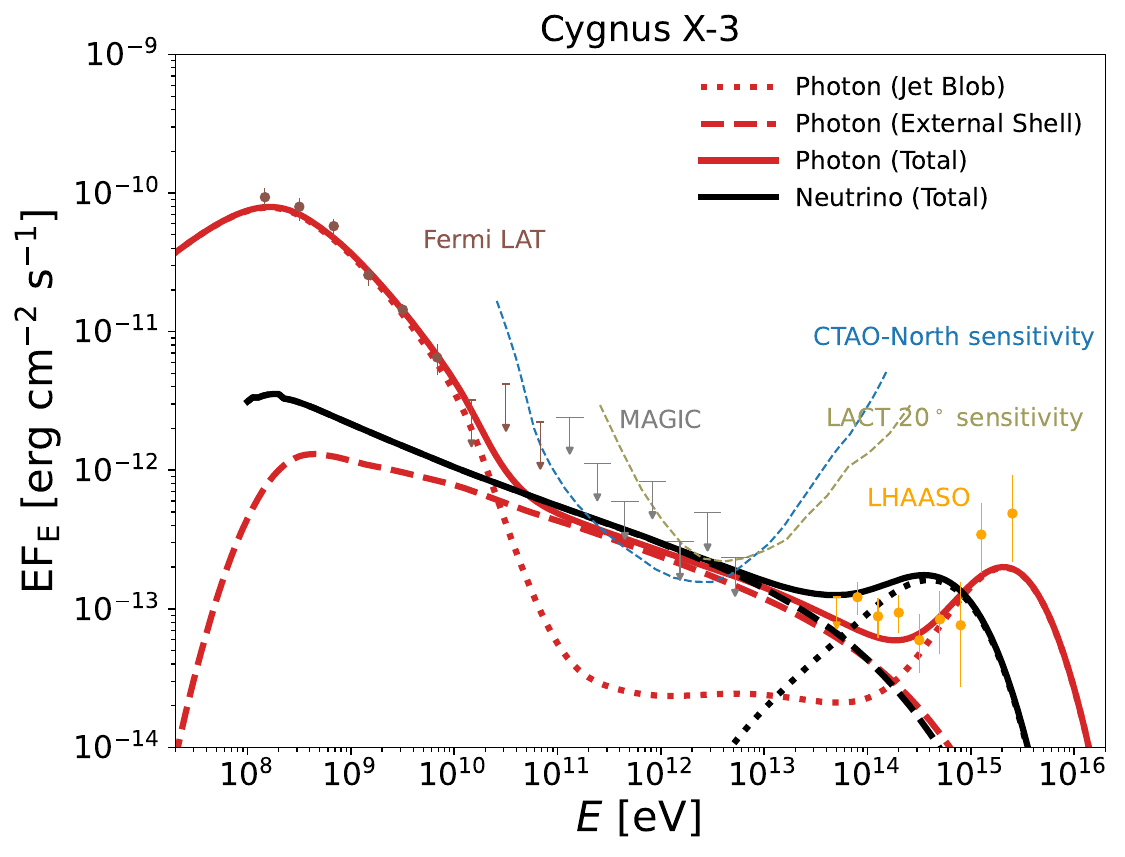}
	}%
		
	\caption{Similar to Fig.~\ref{fig:spectra_pgamma}, but for Scenario~C. The contributions to the photon (red) and neutrino (black) fluxes are indicated by dotted lines for the jet blob and dashed lines for the external shell.
	}
	\label{fig:spectra_pp_B}
\end{figure*}

The parameters for Scenario~C are listed in Table~\ref{tab:pp_B}. Here, the external component represents parsec-scale emission from particles escaping the jet blob and interacting with the surrounding interstellar medium or molecular clouds. This large-scale origin also implies less variable emission than in Scenario~B, where the external emission is produced in a nearby stellar-wind region. 
We neglect cooling and interactions during particle transport from the orbital-separation-scale jet blob to parsec-scale distances. This approximation is motivated by our escape prescription in Eq.~\ref{eq:f_esc}, where only particles with sufficiently long cooling timescales can escape efficiently. In the parameter range considered here, synchrotron and IC losses are negligible because of the relatively low magnetic field strength. Electron synchrotron and IC cooling are also subdominant below $10^{16}~\mathrm{eV}$, as shown in Fig.~\ref{fig:t_cool} (g, h). Therefore, neglecting intermediate cooling provides a reasonable approximation.

The assumed angular sizes for Cygnus~X-1 ($\approx 0.21^\circ$) and Cygnus~X-3 ($\lesssim 0.09^\circ$) are consistent with the LHAASO constraints that characterize them as point-like sources ($\lesssim 0.22^\circ$ for Cygnus~X-1, \citealt{LHAASO-2024arXiv241008988L}; and $\lesssim 0.1^\circ$ for Cygnus~X-3, \citealt{LHAASO-2025arXiv251216638L}). To produce sufficient gamma-rays through the $pp$ process, a dense and massive ambient medium is required. Observed giant molecular clouds in the Cygnus~X region, with average densities of $\sim 10^3~{\rm cm^{-3}}$ and masses of $\sim 10^5~M_\odot$, satisfy these conditions~\citep{Schneider-2006AA...458..855S}.
We adopt an idealized large-scale shell geometry for simplicity. In reality, molecular clouds at parsec scales are likely inhomogeneously distributed and have covering factors below unity. Since the gamma-ray emission in this shell is dominated by $pp$ interactions, the resulting flux is mainly controlled by the total target mass. A smaller covering factor can be compensated by a higher local gas density, reflecting the degeneracy between the target density and cloud volume. Therefore, the inferred densities should be interpreted as average values under these simplified geometric assumptions.

Similar to Fig.~\ref{fig:spectra_pgamma}, Fig.~\ref{fig:spectra_pp_B} shows the photon (red) and neutrino (black) spectra for Scenario~C. Since the external emission is produced in a large-scale shell, $\gamma\gamma$ annihilation is negligible, and the photon spectrum maintains an approximately power-law shape without clear suppression in the $0.1$-$10$~TeV range. MAGIC observations still provide only upper limits~\citep{Fernandez-2017ICRC...35..734F,Abe-2026arXiv260427671A}, while CTAO and/or LACT may be able to detect such a signal with an exposure time of $50$~hours, which would help test the viability of this scenario. Although the external shell is not affected by free-free absorption, its magnetic field and low-energy electron population are insufficient to explain the observed radio emission. The observable radio emission may therefore still originate from regions at larger jet heights, around $\sim 10^{14}~\mathrm{cm}$.

\subsubsection*{(a) Cygnus~X-1}

For Cygnus~X-1, we focus on the hard state, which is the most stable and commonly observed spectral state of this source~\citep{Grinberg-2013AA...554A..88G}. As listed in Table~\ref{tab:pp_B}, the required ambient density and total mass are $1.3 \times 10^3~\mathrm{cm^{-3}}$ and $6.9 \times 10^4~M_\odot$, respectively.
The value of $n_{p,\rm ext}$ adopted here is higher than the value of $10~\mathrm{cm^{-3}}$ used by \citet{Ohira-2025MNRAS.541.2434O}, mainly because that work adopted a much larger proton luminosity ($10^{39}~{\rm erg~s^{-1}}$), which is inconsistent with constraints on the current instantaneous kinetic power of the jet~\citep{Prabu-2025arXiv251209645P}.

Figs.~\ref{fig:spectra_pp_B} (a) and~(c) show the resulting photon and neutrino spectra. The GeV-to-PeV emission is dominated by $pp$ interactions in the large-scale external shell, reproducing both the LHAASO and \textit{Fermi}-LAT data. In this case, protons (electrons) can still be accelerated up to energies of $\sim 1~\mathrm{PeV}$ ($\sim 10~$TeV).

\subsubsection*{(b) Cygnus~X-3}

For Cygnus~X-3, Table~\ref{tab:pp_B} gives an ambient density of $1.8 \times 10^2~{\rm cm^{-3}}$ and a total mass of $6.3 \times 10^4~M_\odot$. As shown in Figs.~\ref{fig:spectra_pp_B} (b) and~(d), the spectrum above $10$~TeV exhibits a power-law-plus-peak structure. The power-law component originates from $pp$ interactions in the large-scale external shell, while the peak is mainly produced by $p\gamma$ interactions in the jet blob. Scenario~C can accelerate protons and electrons up to $\sim 50$~PeV and $\sim 1$~TeV, respectively.

\subsection{Constraints on the Magnetic Field}

Based on the detection of $>\mathrm{TeV}$ photons, and assuming that they are produced through $p\gamma$ or $pp$ interactions, the magnetic field in the emission region can be constrained using Eq.~\ref{eq:E_p_max}. As an illustrative example, we consider Cygnus~X-3. In all the scenarios considered here, the $\approx4~\mathrm{PeV}$ photons are produced through the $p\gamma$ process in the jet blob, with the target photon field provided by the donor star.

The proton energy required to generate such high-energy photons is $E_{p, \rm max} \sim 10~\mathrm{PeV}$. In addition, the observed orbital modulation implies that the emission region is located at a distance smaller than or comparable to the semi-major axis, i.e., $H_{\rm jet} \lesssim a$.
Using Eq.~\ref{eq:E_p_max}, the lower limit on the magnetic field in the co-moving frame of the jet blob is
\begin{equation}
B'_{\rm jet} \gtrsim \frac{\eta_{\rm acc} \beta_j \Gamma_j E_{p,\rm max}}{e a}
\simeq 330~{\rm G}~\eta_{\rm acc} \beta_j \Gamma_j E_{p,\rm max,16} a^{-1}_{11}.
\end{equation}
Adopting the parameters of Cygnus~X-3, we obtain a required magnetic field of $B'_{\rm jet} \gtrsim 260~\mathrm{G}$ in the co-moving frame of the jet blob.

The magnetic field is also constrained from above by the jet energy budget. According to Eqs.~\ref{eq:B'} and~\ref{eq:Pj}, requiring $\epsilon_B < 1$ and $P_j < L_{\rm Edd}$ gives
\begin{equation}
\begin{aligned}
    B'_{\rm jet} & \lesssim \sqrt{\frac{12 \eta_j \dot{m} L_{\rm Edd}}{\eta_{\rm rad} r_{\rm jet}'^2 \Gamma_j^2 \beta_j c}} \\
    & \simeq 6.4 \times 10^3~{\rm G}~L_{\rm Edd,39}^{1/2} r_{\rm jet,11}'^{-1/2} \Gamma_j^{-1} \beta_j^{-1/2} \eta_{\rm rad}^{-1/2} \eta_j^{1/2} \dot{m}^{1/2}.
\end{aligned}
\end{equation}
For Cygnus~X-3, this constraint gives $B'_{\rm jet} \lesssim 1.3 \times 10^4~\mathrm{G}$.

Lower-energy photons produced by leptonic processes can further constrain the magnetic field. In our scenario, radio-IR photons are produced by synchrotron emission, while GeV photons are produced through EIC scattering of the donor-star photon field. These two components jointly impose an upper limit on the magnetic field strength.
The EIC flux, $F_{\rm EIC}$, and the synchrotron flux, $F_{\rm SYN}$, can be approximately written as
\begin{equation}
F_{\rm EIC} \sim \frac{Y L_e}{(1 + Y)\, 4 \pi D_L^2},
\end{equation}
and
\begin{equation}
F_{\rm SYN} \sim \frac{L_e}{(1 + Y)\, 4 \pi D_L^2},
\end{equation}
where $Y \approx t'_{\rm syn} / t'_{\rm IC}$ is the total Compton $Y$-parameter. The synchrotron and inverse Compton cooling timescales in the co-moving frame are $t'_{\rm syn} \propto U_B'^{-1} \gamma_e'^{-1}$ and $t'_{\rm IC} \propto U_{\rm ph}'^{-1} \gamma_e'^{-1}$, respectively. Here, $U'_{\rm ph}$ is the energy density of the target photon field, $U'_B = B_{\rm jet}'^2 / (8\pi)$ is the magnetic energy density, and $L_e$ is the electron luminosity in the lab frame.
Combining these relations, the magnetic field can be estimated as
\begin{equation}
\label{eq:B_up}
\begin{aligned}
B'_{\rm jet} & \sim \sqrt{8\pi \, \frac{F_{\rm SYN}}{F_{\rm EIC}} \, U'_{\rm ph}} \\
& \sim 10^2~{\rm G} \, F_{{\rm SYN}, -11}^{1/2} \, F_{\rm EIC, -10}^{-1/2} \, U_{\rm ph, 4}'^{1/2}.
\end{aligned}
\end{equation}
This relation is valid only if the synchrotron and EIC fluxes used in the estimate are produced by the same electron population.
For Cygnus~X-3, the observed GeV flux is $F_{\rm EIC} \approx 9.3 \times 10^{-11}~\mathrm{erg~cm^{-2}~s^{-1}}$ at $E_{\rm ph, EIC} \approx 1.5 \times 10^8~\mathrm{eV}$, which is predominantly produced by electrons with Lorentz factor $\gamma_{e,\rm min}$. The target photon energy density is estimated to be $U'_{\rm ph} \approx 6.7 \times 10^3~\mathrm{erg~cm^{-3}}$.
Previous studies indicate that the observed IR emission of Cygnus~X-3 is likely dominated by the stellar wind rather than the jet blob~\citep[e.g.,][]{Antokhin-2022ApJ...926..123A}, and therefore provides an upper limit on the synchrotron contribution from the jet blob. Adopting $F_{\rm SYN} \approx 4.0 \times 10^{-12}~\mathrm{erg~cm^{-2}~s^{-1}}$ at $E_{\rm ph, SYN} \approx 0.11~\mathrm{eV}$, we account for the scaling needed to convert the synchrotron flux to the photon energy corresponding to electrons with Lorentz factor $\gamma_{e,\rm min}$, assuming a spectral dependence $F_{\rm SYN} \propto E_{\rm ph}^{-\beta_{\rm SYN}}$ with $\beta_{\rm SYN} \sim 1/2-4/3$.
Using Eq.~\ref{eq:B_up}, we obtain an approximate upper limit of $B'_{\rm jet} \lesssim 500~\mathrm{G}$, broadly consistent with our numerical calculations.

The above estimates indicate that a magnetic field of order $\sim 10^2~\mathrm{G}$ in the jet blob is required to explain the PeV photons detected from Cygnus~X-3. This value is somewhat higher than the magnetic fields usually inferred from previous modeling of the GeV emission, $B \lesssim 100~\mathrm{G}$~\citep[e.g.,][]{Zdziarski-2012MNRAS.421.2956Z, Zdziarski-2018MNRAS.479.4399Z, Dmytriiev-2024ApJ...972...85D}. However, these constraints are model dependent because they rely on the assumed jet geometry, the relative contributions of the SSC and EIC components, and the detailed shape of the orbitally modulated GeV light curve. These previous studies argued that a sufficiently strong magnetic field would enhance the SSC component relative to the EIC component, potentially reducing the strong orbital modulation associated with anisotropic EIC emission. This argument, however, does not necessarily apply if the jet direction varies with orbital phase~\citep[e.g.,][]{Dmytriiev-2024ApJ...972...85D}. In that case, the Doppler factor also changes along the orbit, so phase-dependent Doppler boosting can produce orbital modulation even for an SSC-dominated GeV component. In the representative parameter sets adopted in this work, the GeV emission remains dominated by the EIC component, although the relative importance of EIC and SSC depends on the model parameters. Therefore, the GeV constraints on the magnetic field are sensitive to the assumed geometry, and a field strength of a few hundred gauss in the jet blob is allowed.

\section{Energy-Dependent Orbital Modulation of GeV and PeV Emission}
\label{sect:modulation}

In the previous sections, we adopted a fixed system geometry. Here, we include the effects of orbital motion to investigate the orbital modulation in high-energy bands.
For a distant observer, the orbital motion can be described as a variation in the donor-star azimuthal angle, $\phi$.
Since the GeV and PeV components considered here are mainly produced in the jet blob, we focus on the jet-blob emission of Cygnus~X-3 and adopt the same parameters as in Scenario~A.
Given the uncertainties in the orbital-modulation model, our calculations should be regarded as a plausible illustrative scenario rather than a definitive prediction.
Once the system reaches a steady state, radiation is produced continuously. The orbital phase dependence arises from the changing binary geometry and the corresponding phase-dependent interaction conditions.

There are two primary mechanisms that can lead to orbital modulation of high-energy emission.
In the first scenario, the jet orientation remains fixed relative to the observer~\citep[e.g.,][]{Dubus-2010MNRAS.404L..55D, Zdziarski-2018MNRAS.479.4399Z}. The modulation then arises from changes in the distance between the donor star and the jet blob, which modify the local photon field and hence the IC and $p\gamma$ emission.
In the second scenario, variations in the jet azimuthal angle, $\phi_j$, change the Doppler factor and produce observable flux variability~\citep[e.g.,][]{Dubus-2010AA...516A..18D,Dmytriiev-2024ApJ...972...85D}.
In both scenarios, strong $\gamma\gamma$ annihilation can produce energy-dependent modulation patterns, with a more pronounced effect in the GeV band than in the PeV band due to the higher photon-photon opacity at lower energies.
Both models have their respective advantages and limitations~\citep{Dmytriiev-2024ApJ...972...85D}. 

\begin{figure*}[ht!]
\centering

\subfigure[]{
    \includegraphics[width=\textwidth]{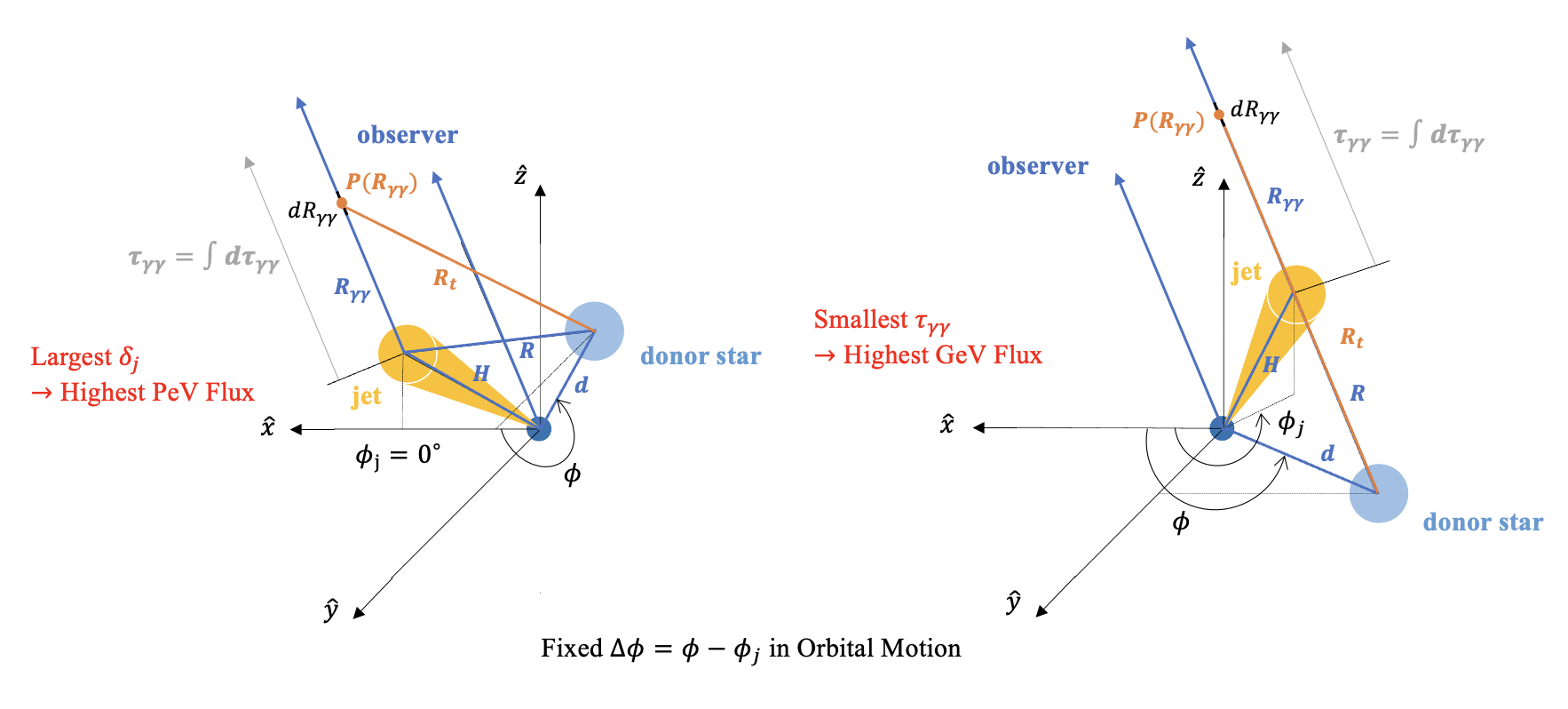}
}

\vspace{1em}

\subfigure[]{
    \includegraphics[width=0.3\textwidth]{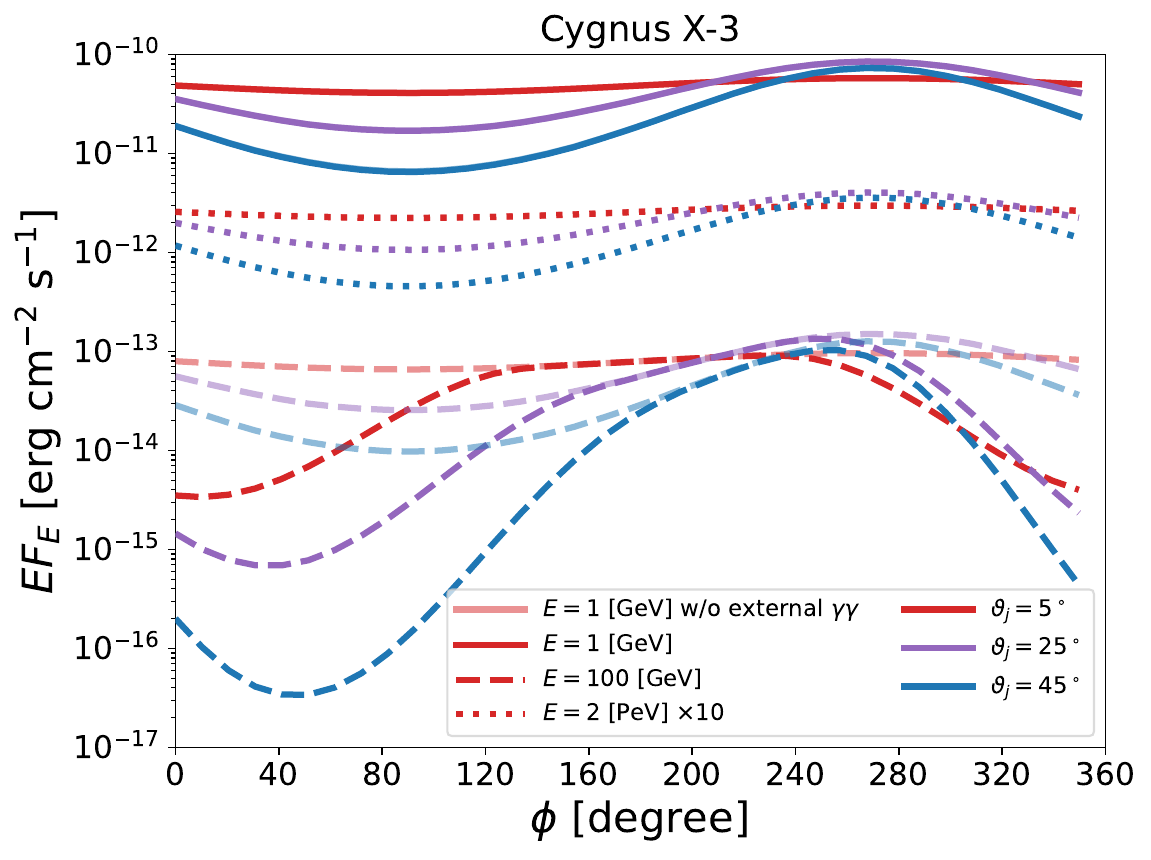}
}
\hfill
\subfigure[]{
    \includegraphics[width=0.3\textwidth]{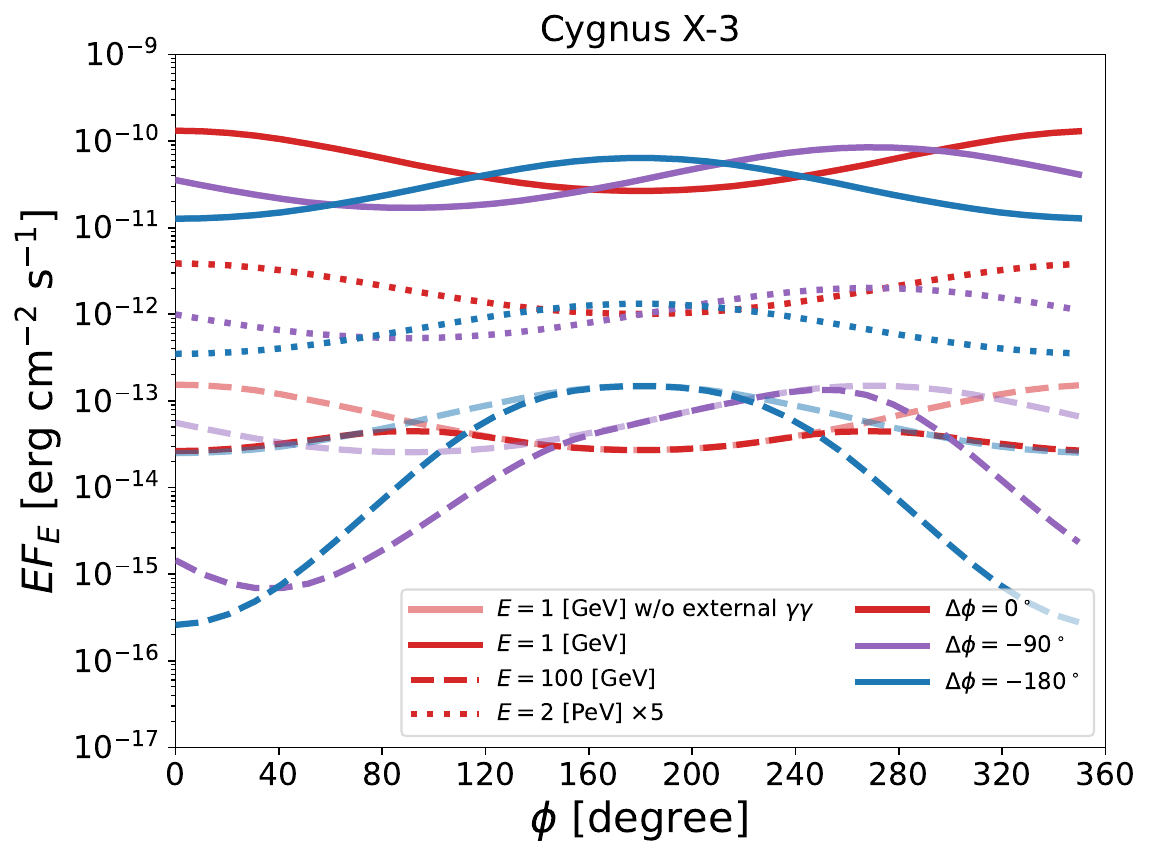}
}
\hfill
\subfigure[]{
    \includegraphics[width=0.3\textwidth]{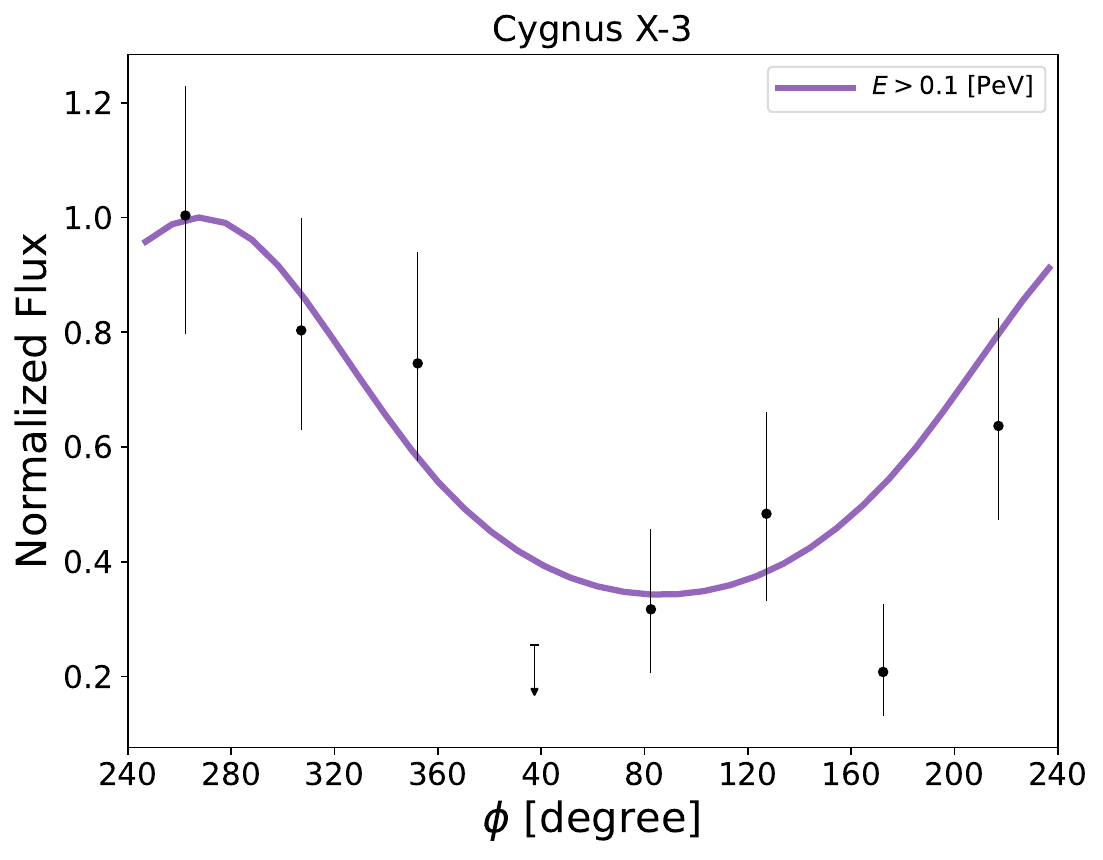}
}
    
\caption{\textbf{Panel (a)} provides a schematic illustration of the adopted deflected or misaligned jet geometry. The values of $\Delta\phi$ and $\vartheta_j$ are assumed to remain constant throughout the orbital motion. External $\gamma\gamma$ optical depth is accumulated along the gamma-ray trajectory, with $P(R_{\gamma\gamma})$ marking an arbitrary point along the path. The maximum Doppler factor and the weakest external $\gamma\gamma$ annihilation occur at different orbital phases, resulting in energy-dependent modulation.
\textbf{Panels (b) and (c)} show the fluxes at selected energies for the jet blob with varying $\vartheta_j$, $\Delta \phi$, and orbital phase $\phi$. Different line styles correspond to different energy bands.
Panel (b) adopts $\Delta \phi = -90^{\circ}$, whereas Panel (c) adopts $\vartheta_j = 25^{\circ}$. In Panel (b), different colors indicate varying $\vartheta_j$, while in Panel (c), colors represent different $\Delta \phi$, as detailed in the legend. Fainter colors correspond to results calculated without external $\gamma\gamma$ annihilation.
\textbf{Panel (d)} shows the flux above 0.1~PeV as a function of $\phi$ for Scenario~A, for both the jet blob and the inner blob, with $\vartheta_j = 25^{\circ}$ and $\Delta \phi = -90^{\circ}$, using the same parameters as listed in Tables~\ref{tab:fix}-\ref{tab:pgamma}. The black points represent the orbital light curve of Cygnus~X-3 measured by LHAASO ($\geq 0.1$~PeV)~\citep{LHAASO-2025arXiv251216638L}.}
\label{fig:CygX3_pgamma_varied}
\end{figure*}

Here, we adopt the latter geometric scenario as an illustrative example. Stellar-wind ram pressure can, in principle, bend jets in high-mass X-ray binaries, as suggested by recent radio observations of Cygnus~X-1~\citep{Prabu-2025arXiv251209645P}. For Cygnus~X-3, however, a large deflection is unlikely to be caused by the stellar wind alone, since standard estimates yield a bending angle much smaller than that required by some orbital-modulation models~\citep{Dmytriiev-2024ApJ...972...85D}. The apparent jet orientation may instead reflect an intrinsically misaligned launching geometry, such as a tilted inner accretion flow or funnel~\citep{Yang-2023MNRAS.526L...1Y}, while the dense Wolf-Rayet wind and rapid orbital motion may further modify the outflow through recollimation, asymmetry, or Coriolis-like trailing distortions.
For simplicity, we keep $\Delta\phi = \phi - \phi_j$ and $\vartheta_j$ fixed throughout the orbit.
In addition to the internal $\gamma\gamma$ annihilation automatically computed by AMES, we include external $\gamma\gamma$ annihilation for photons propagating along the line of sight and interacting with target photon fields.
The modulation is then driven by the phase dependence of both the Doppler factor and external $\gamma\gamma$ annihilation. Since the maximum Doppler factor and the weakest external $\gamma\gamma$ annihilation occur at different orbital phases, the resulting modulation is energy dependent, as illustrated in Fig.~\ref{fig:CygX3_pgamma_varied} (a).

\citet{LHAASO-2025arXiv251216638L} reported possible differences in the orbital modulation of the PeV, GeV, and X-ray bands observed by LHAASO, \textit{Fermi}-LAT, and MAXI. Specifically, the minimum of the \textit{Fermi}-LAT light curve precedes that of the LHAASO light curve by approximately $90^\circ$ in phase, while the maximum of the MAXI light curve lies between the minima of the LHAASO and \textit{Fermi}-LAT data.

In Figs.~\ref{fig:CygX3_pgamma_varied} (b) and~(c), we show examples with $\Delta\phi = 0^\circ$, $-90^\circ$, and $-180^\circ$, and with $\vartheta_j = 5^\circ$, $25^\circ$, and $45^\circ$ for Cygnus~X-3.
Although $\Delta\phi = 0^\circ$ is unlikely in a simple wind-jet interaction scenario where the average jet axis is aligned with the orbital axis, it could arise from jet precession and/or jet inclination.
While LHAASO measures emission above $0.1$~PeV and \textit{Fermi}-LAT covers $0.1$-$100$~GeV, we present the predicted orbital modulation at $2$~PeV, $100$~GeV, and $1$~GeV as illustrative examples, all dominated by jet-blob emission. For the case with $\Delta\phi = -90^\circ$ and $\vartheta_j = 25^\circ$, the 100~GeV and 2~PeV light curves show a phase shift of approximately $90^\circ$, broadly consistent with the observational trend.

For comparison with the LHAASO orbital light curve, Fig.~\ref{fig:CygX3_pgamma_varied} (d) shows the predicted flux of Cygnus~X-3 integrated above $0.1$~PeV at each orbital phase. The calculation uses the same parameters as in Figs.~\ref{fig:spectra_pgamma} (c) and~(f), and the predicted light curve is normalized to a peak value of $\sim 9.2 \times 10^{-16}\ \mathrm{cm^{-2}\ s^{-1}}$. The model reproduces the overall PeV modulation pattern and is broadly consistent with the LHAASO measurements~\citep{LHAASO-2025arXiv251216638L}.

Our results show that, if $\gamma\gamma$ annihilation is sufficiently strong, the GeV and PeV bands can exhibit distinct modulation patterns. The inverse correlation between the X-ray and PeV/GeV fluxes can also be explained by the adopted jet geometry. When the donor star obscures the accretion disk and suppresses the X-ray flux, the jet is oriented closer to the line of sight. This orientation maximizes the Doppler factor of the jet blob and enhances the observed PeV and GeV fluxes.

However, the model faces two main challenges. First, it predicts similar modulation behavior at $1$~GeV and $2$~PeV, suggesting that either stronger external annihilation or additional mechanisms may be needed. Second, while the Doppler-induced flux variation in the PeV band is comparable to the observed modulation, the absorption-induced variation predicted at $100$~GeV can exceed the observed value by up to an order of magnitude. A more detailed treatment of external $\gamma\gamma$ annihilation and additional emission or absorption effects may therefore be required to reproduce the GeV-band observations.

Overall, this section provides a possible qualitative explanation for the energy-dependent modulation by combining phase-dependent Doppler boosting with external $\gamma\gamma$ annihilation.
Although the model captures some aspects of the observed phase differences between bands, it does not fully reproduce the data, and more detailed phase-dependent calculations are required.

\section{Predicted Expected Number of Neutrinos}
\label{sect:neutrino}

In our calculations, we include muon and pion cooling through synchrotron radiation for both muons and pions, as well as IC scattering for muons. These processes were omitted in previous studies, and their inclusion reduces the predicted neutrino flux, suggesting that earlier estimates may have been overestimated. Therefore, muon and pion cooling is important for interpreting the gamma-ray-to-neutrino flux ratio and for using multimessenger observations to diagnose the physical conditions of the emission region~\citep[e.g.,][]{Murase-2013PhRvL.111m1102M,Murase-2016PhRvD..94j3006M}.

\begin{figure*}[ht!]
	\centering
	\subfigure[]{
		\includegraphics[scale=0.3]{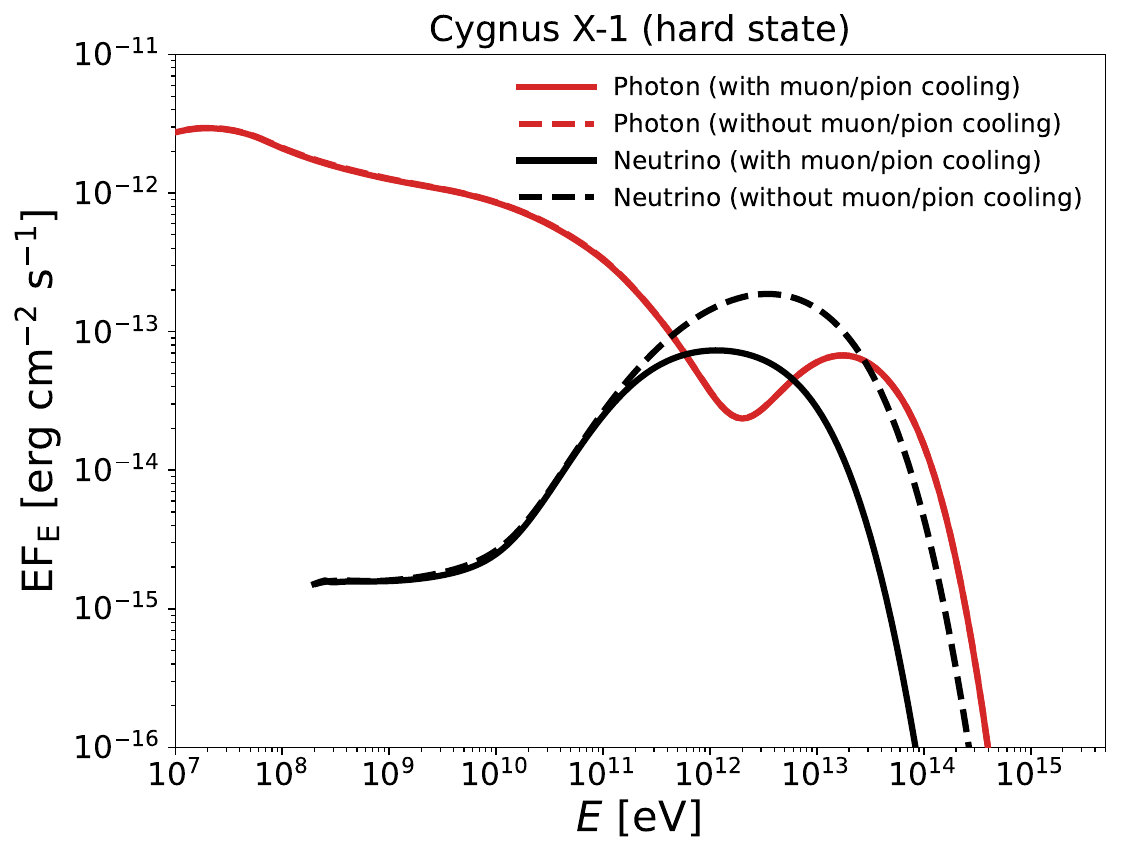}
	}%
    \subfigure[]{
		\includegraphics[scale=0.3]{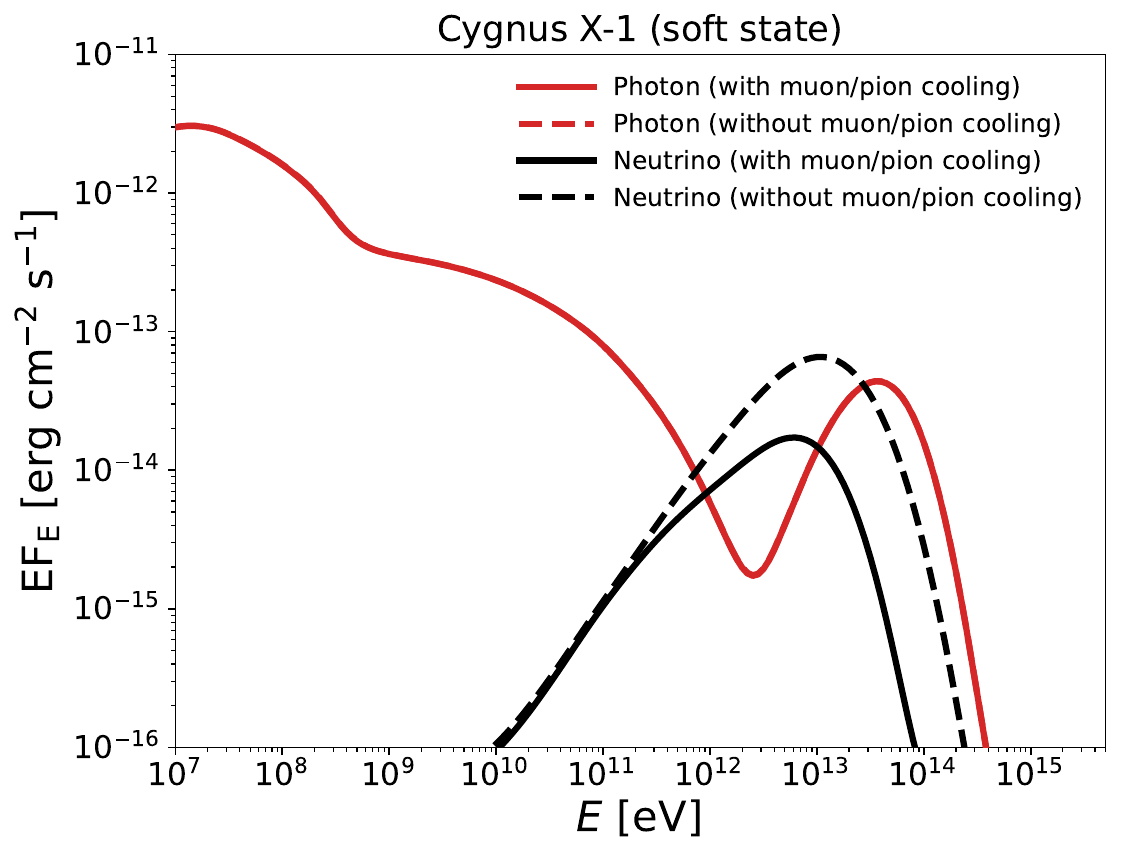}
	}%
	\subfigure[]{
		\includegraphics[scale=0.3]{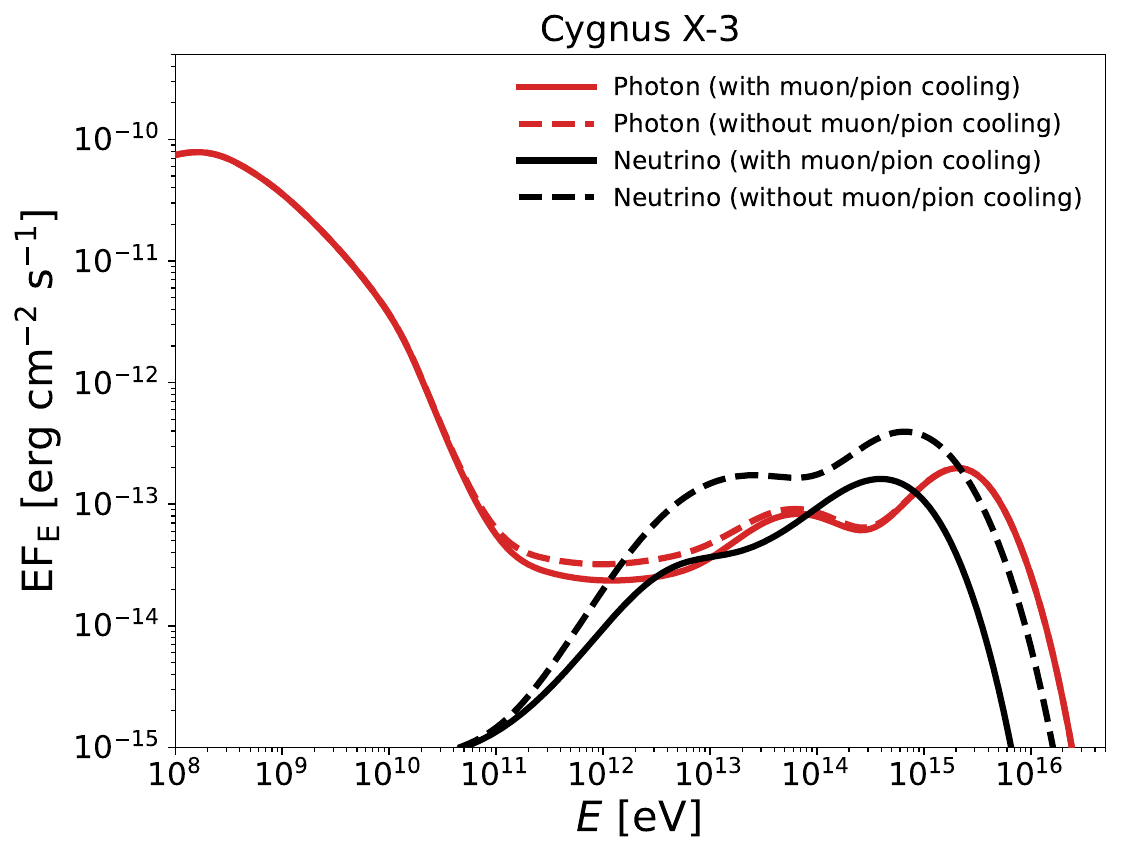}
	}%

    \subfigure[]{
		\includegraphics[scale=0.3]{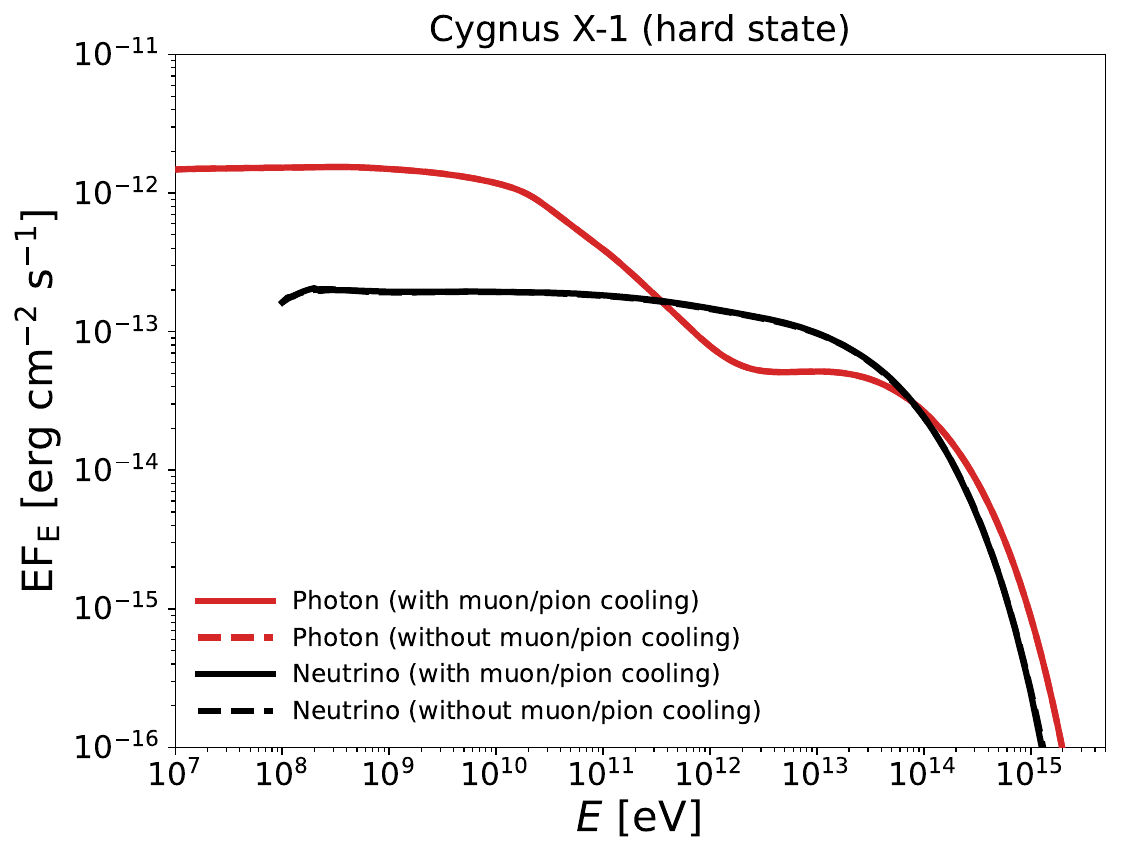}
	}%
    \subfigure[]{
		\includegraphics[scale=0.3]{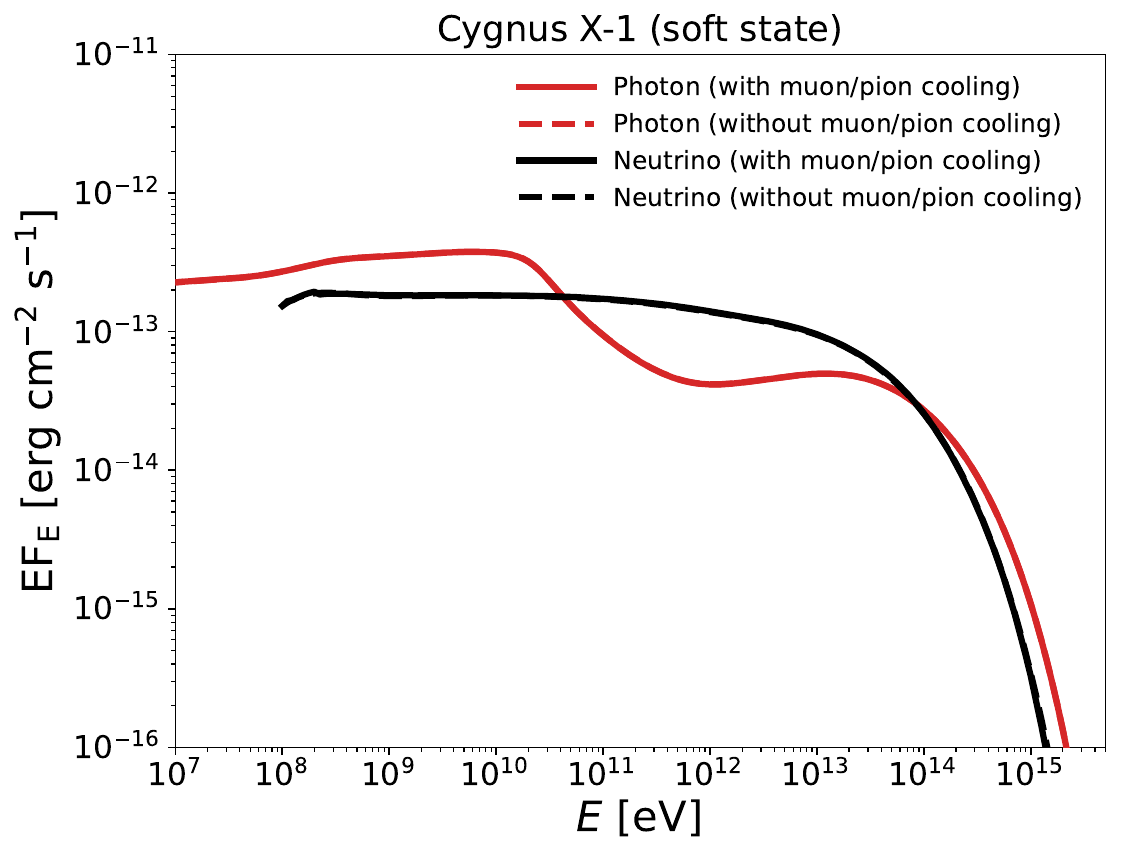}
	}%
	\subfigure[]{
		\includegraphics[scale=0.3]{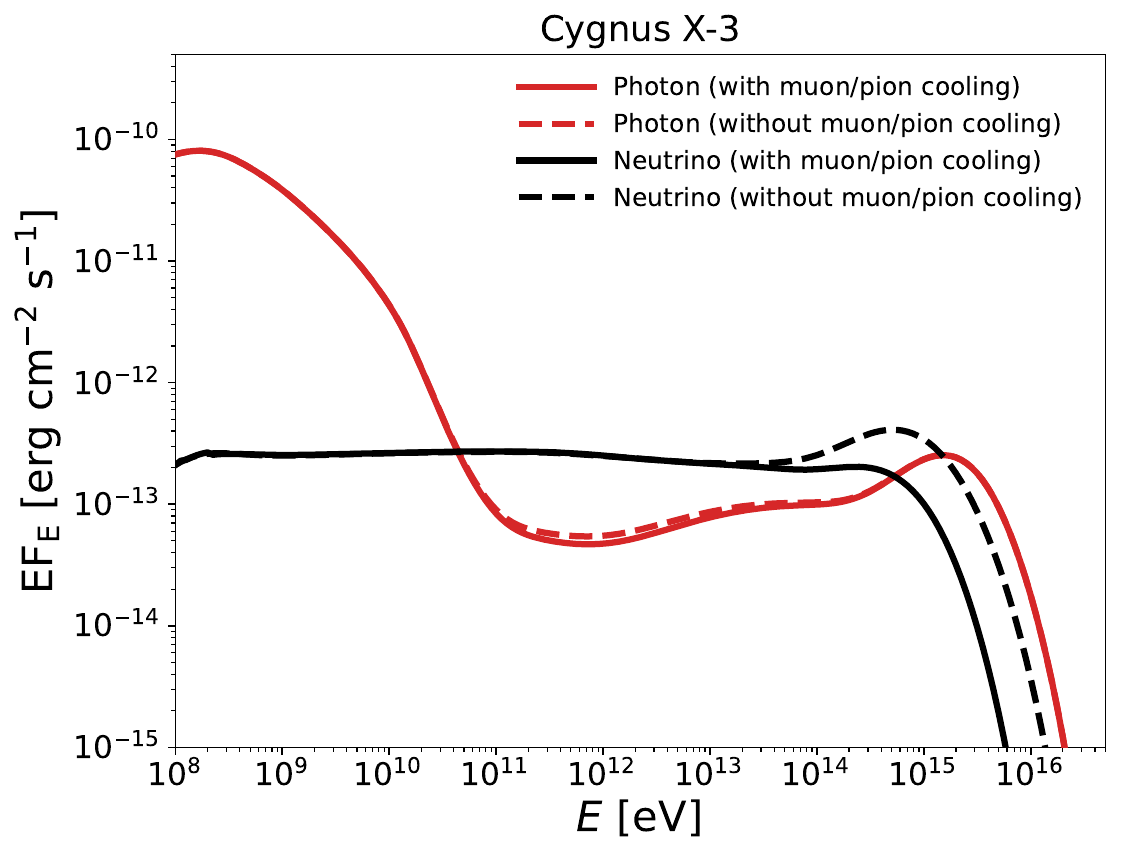}
	}%

    \subfigure[]{
		\includegraphics[scale=0.3]{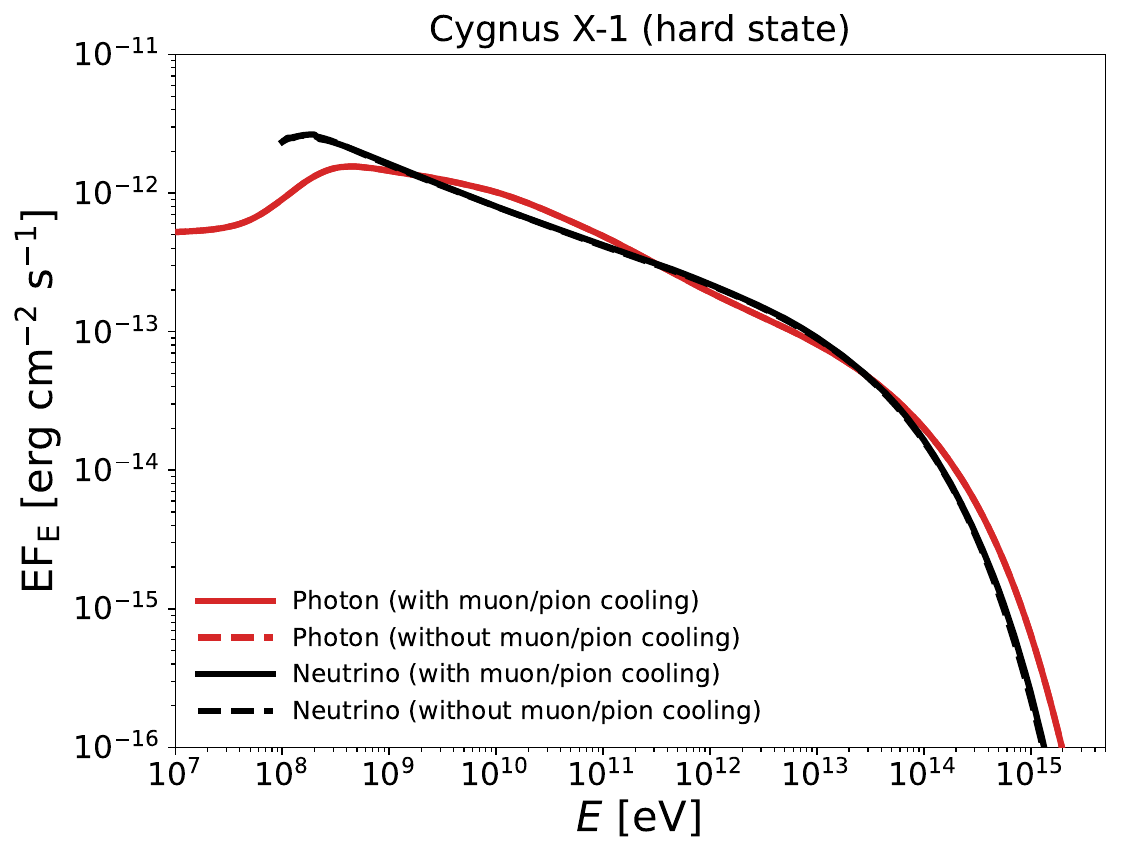}
	}%
    \hspace{5.9cm}
	\subfigure[]{
		\includegraphics[scale=0.3]{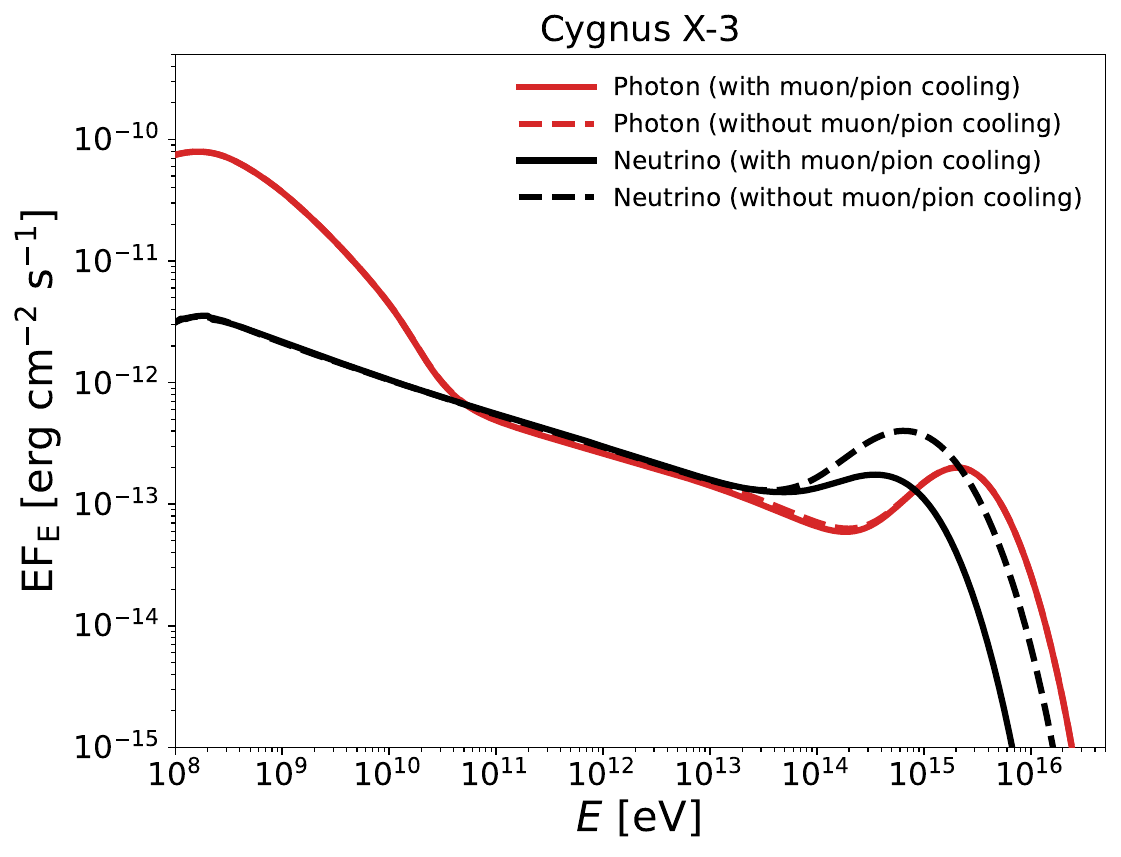}
	}%

	\caption{Photon (red) and neutrino (black) spectra are shown for Scenarios A (first row), B (second row), and C (third row). Solid lines represent results including muon and pion cooling, whereas dashed lines show results without these cooling processes.
	}
	\label{fig:check_muonCooling}
\end{figure*}

Using the same fitting parameters listed in Tables~\ref{tab:jet_blob}-\ref{tab:pp_B}, we find that IC cooling of muons is negligible, whereas synchrotron cooling of muons and pions significantly suppresses the high-energy neutrino flux. The finite lifetimes of muons and pions also contribute to this suppression.
Fig.~\ref{fig:check_muonCooling} shows this comparison. Solid lines include muon and pion cooling, while dashed lines neglect these effects. The suppression is significant in the jet and inner blobs, reducing the neutrino spectra by a factor of $\sim 5$. In the external region, these effects are minor, and the neutrino spectra remain largely unchanged because of the longer dynamical timescales and weaker magnetic fields.

Including muon and pion cooling, we calculate the expected number of neutrinos, $N_{\rm exp,\nu}$, predicted by each scenario for different detectors. The expected number is given by~\citep[e.g.,][]{Murase-2016PhRvD..94j3006M}
\begin{equation}
    N_{\rm exp, \nu} (>E) = \Delta t_{\rm int} \int_{E}^{E_{\rm max}} \phi_E~ A_{\rm eff}~ dE,
\end{equation}
where $E$ is the neutrino energy in the observer frame, $E_{\rm max}$ is the upper integration bound, $\phi_E$ is the predicted differential neutrino flux in units of $\rm eV^{-1}~cm^{-2}~s^{-1}$, $A_{\rm eff}$ is the detector effective area which depends on energy and source direction, and $\Delta t_{\rm int}$ is the net integration time. The net integration time can be shorter than the total observing time after accounting for the observational duty cycle. Based on archival data, Cygnus~X-1 spent 75.6\% (11.6\%) of the time between MJD 50087 and 55200 in its hard (soft) state~\citep{Grinberg-2013AA...554A..88G}, while Cygnus~X-3 was in a soft state for 43.7\% of the period between MJD 58848 and MJD 60522~\citep{LHAASO-2025arXiv251216638L}. Therefore, the total observing time required to achieve a given $\Delta t_{\rm int}$ is approximately $1.3 \Delta t_{\rm int}$ ($8.6 \Delta t_{\rm int}$) for Cygnus~X-1 in the hard (soft) state, and $2.3 \Delta t_{\rm int}$ for Cygnus~X-3 in the soft state.

\begin{figure*}[ht!]
	\centering
	\subfigure[]{
		\includegraphics[scale=0.35]{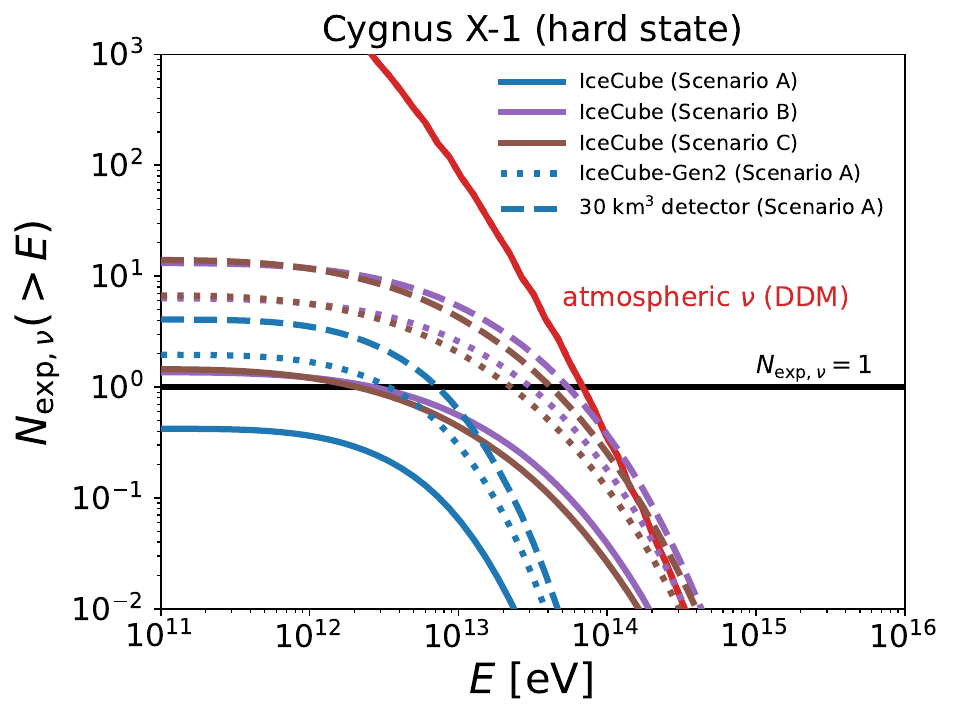}
	}%
	\subfigure[]{
		\includegraphics[scale=0.35]{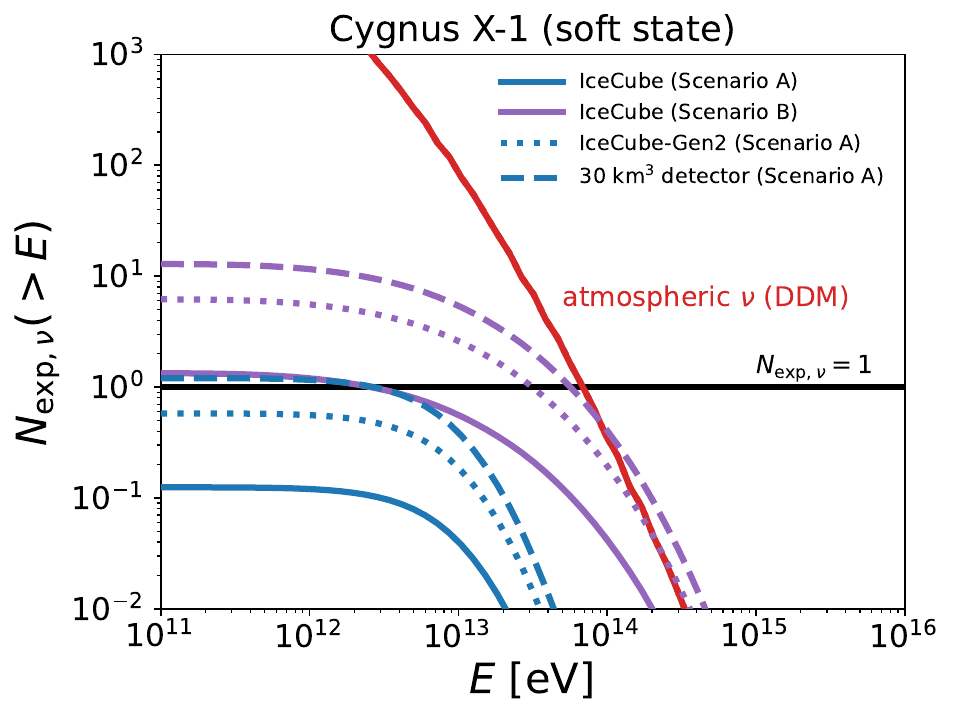}
	}%
	\subfigure[]{
		\includegraphics[scale=0.35]{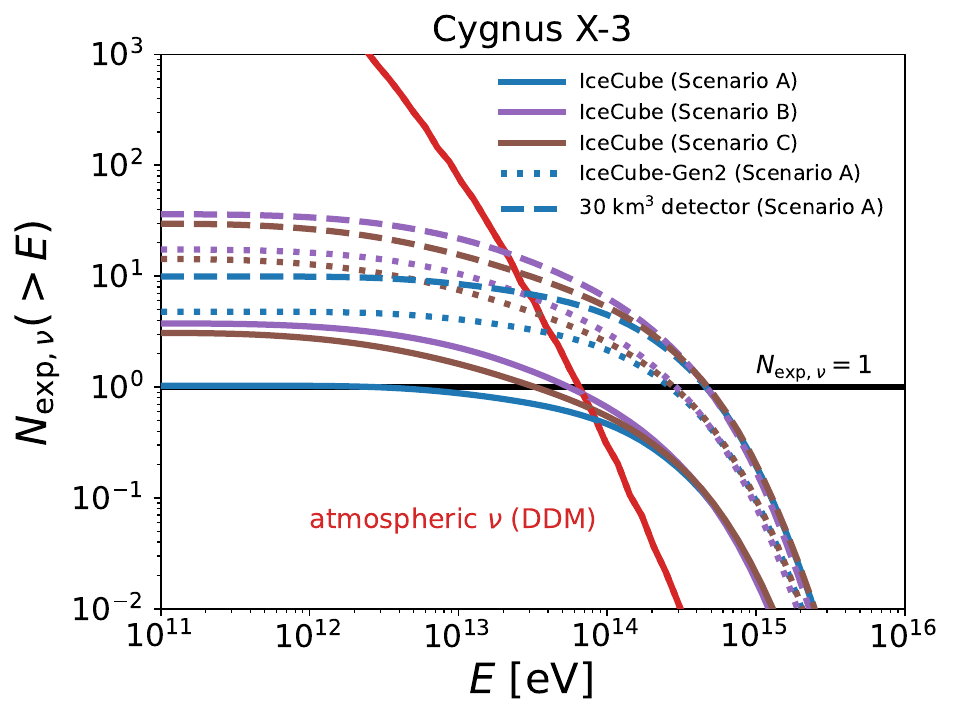}
	}%

	\caption{Expected number of neutrinos for $\Delta t_{\rm int} = 20~{\rm yr}$. Colors represent different scenarios, as indicated in the legend. Solid lines indicate the expected numbers for IceCube, dotted lines show those for IceCube-Gen2, and dashed lines represent those for a 30 km$^3$ detector. The red solid line shows the atmospheric neutrino background predicted by DDM. The horizontal black line marks $N_{\rm exp,\nu}=1$.
	}
	\label{fig:N_exp_neutrino}
\end{figure*}

Fig.~\ref{fig:N_exp_neutrino} shows the expected number of neutrinos detected by IceCube, IceCube-Gen2, and a 30 km$^3$ detector for $\Delta t_{\rm int} = 20~{\rm yr}$. We use the IceCube effective area, $A_{\rm eff}$, from the all-sky point-source data collected between 2008 and 2018~\citep{IceCube-2021arXiv210109836I}. For IceCube-Gen2, the effective area is estimated by scaling the IceCube value by a factor of $10^{2/3}$, based on its approximately ten-times-larger detector volume~\citep{Aartsen-2021JPhG...48f0501A}.
For a 30 km$^3$ detector, such as HUNT~\citep{Huang:2023mzt}, the effective area is similarly estimated by scaling the IceCube value by a factor of $30^{2/3}$. 
We calculate the expected atmospheric neutrino background using the Data-driven muon-calibrated neutrino flux (DDM) model~\citep{Yanez-2023PhRvD.107l3037Y}, shown as the red solid lines in Fig.~\ref{fig:N_exp_neutrino}.
Following previous studies~\citep[e.g.,][]{Murase-2013PhRvL.111m1102M}, the angular window for the background is defined as 
\begin{equation}
    \Delta \Omega = \pi \max{(\theta_{\rm res}^2, \theta_k^2)},
\end{equation}
where the angular resolution is taken as $\theta_{\rm res} = 0.5^\circ$ ($0.1^\circ$) for IceCube (IceCube-Gen2 and 30 km$^3$ detector), and the kinematic angle is approximated as $\theta_k \simeq 1.5^\circ \left( E / 1~\mathrm{TeV} \right)^{-1/2}$.

Fig.~\ref{fig:N_exp_neutrino} presents the results for the three scenarios, with different colors denoting Scenario~A, B, and C.
For a source to be considered detectable, the expected number of source neutrinos, $N_{\rm exp,\nu}$, should exceed both the atmospheric background and a threshold value, typically taken as $N_{\rm thr}=1$.
For an integration time of $\Delta t_{\rm int}=20~{\rm yr}$, a significant neutrino detection from either Cygnus~X-1 or Cygnus~X-3 with IceCube is unlikely. This result is consistent with the absence of a significant neutrino excess in previous IceCube searches of Galactic X-ray binaries~\citep{Abbasi_2022ApJ...930L..24A}.
For Cygnus~X-1, detection remains difficult even with IceCube-Gen2, because the expected signal is obscured by the atmospheric neutrino background. For Cygnus~X-3, the expected number of signal events in IceCube-Gen2 can exceed the atmospheric background near $100~{\rm TeV}$, although the small event number makes a statistically significant detection challenging.
Even with a $30~{\rm km}^{3}$ detector, detecting neutrinos from Cygnus~X-1 remains difficult. Such a detector could, however, detect neutrinos from Cygnus~X-3 in the energy range $0.1$-$1~{\rm PeV}$. After accounting for the source duty cycle, the required calendar observation time is approximately $46~{\rm yr}$.

\section{Summary and Discussion}
\label{sect:summary}
In this work, we utilized AMES to investigate multimessenger emission from microquasars. 
Motivated by recent LHAASO detections of photons above $0.1$~PeV from Cygnus~X-1 and Cygnus~X-3~\citep{LHAASO-2024arXiv241008988L,LHAASO-2025arXiv251216638L}, we modeled their multiwavelength emission from radio to PeV energies with three distinct physical scenarios. Scenario~A represents the jet-core model, comprising a jet blob produced by the terminal shock or internal shocks, and an inner blob originating from the corona or internal shocks in the compact outflow. 
In contrast, Scenarios~B and~C consist of a primary jet blob as well as a static stellar-wind region or large-scale external shell, within which escaping charged particles interact with the surrounding environment.

We found that the multiwavelength data for both Cygnus~X-1 and Cygnus~X-3 could be reproduced in all scenarios, except for the radio band, which was suppressed by free-free absorption, suggesting that the radio-emitting region likely lay at larger heights ($\sim 10^{14}$~cm). For Cygnus~X-1, we adopted a lower Lorentz factor in the soft state than in the hard state, consistent with theoretical expectations.
Although idealized, our results illustrated a possible scenario in which a “GeV-dark” jet could exist during the soft state and potentially contribute to emission in the TeV-PeV energy range.
For Cygnus~X-3, we also adopted a relatively low Lorentz factor for the soft state. A larger jet polar angle was required to fit the data, implying a more inclined jet orientation. We further found that explaining the observed PeV emission requires a magnetic field of order $\sim 10^2~\mathrm{G}$ in the jet blob.

Our modeling also provided insights into the particle acceleration mechanism, characterized by the acceleration parameter $\eta_{\rm acc}$, which set the maximum proton Lorentz factor $\gamma'_{p,\rm max}$. For both Cygnus~X-1 and Cygnus~X-3, $\eta_{\rm acc} = 1.0~\beta_j^{-2}$ reproduced the observations, suggesting efficient particle acceleration in microquasars. However, the weaker magnetic field required for Cygnus~X-1 by the lower-energy data resulted in a lower maximum proton energy, preventing the production of photons above $0.1$~PeV. For Cygnus~X-3, in contrast, protons could be accelerated to energies of around $10$~PeV and produce PeV photons. Leptonic processes, such as inverse Compton scattering, could not explain the PeV emission because electron cooling limited the maximum electron energy to below $0.1$~PeV.

In all scenarios, the GeV emission from Cygnus~X-3 was dominated by leptonic processes within the jet blob. For Cygnus~X-1, the dominant mechanism varied among scenarios: in Scenario~A, the GeV emission was dominated by leptonic processes in the jet blob; in Scenario~B, it received contributions from leptonic processes in both radiation zones; and in Scenario~C, it was dominated by hadronic processes in the external region.
However, at lower energies ($E \lesssim 100$~GeV), the emission did not exhibit sufficiently distinct behavior to differentiate between these scenarios.

The model predictions for different scenarios diverged at higher energies.
For Cygnus~X-1, the TeV gamma-ray data were explained either by the $p\gamma$ process in the inner blob (Scenario~A), where the target photons originated from the X-ray radiation of the accretion disk, or by the $pp$ process in the external region (Scenarios~B and~C).
For Cygnus~X-3, Scenario~A produced a two-peak structure, while Scenarios~B and~C produced plateau/power-law-plus-peak structures. The lower-energy peak (plateau/power-law component) was produced by the $p\gamma$ ($pp$) process in the inner blob (external region) for Scenario~A (Scenarios~B and~C). The higher-energy peak was mainly associated with $p\gamma$ interactions in the jet blob, where the target photons originated from the blackbody radiation of the donor star.
This suggested another possible explanation for the absence of a prominent PeV feature in Cygnus~X-1. In addition to its smaller $\gamma'_{p,\rm max}$, the lack of such a feature was likely related to the lower target photon density caused by the wider binary separation. However, PeV photon production in Cygnus~X-1 could not be ruled out if the power of its central engine were to increase.

Although all scenarios reproduced the current data, they differed in their predictions at other energies, particularly in the $0.1$-$10$~TeV band, where only upper limits are currently available. Scenario~A predicted a deep valley, Scenario~B a mild suppression, and Scenario~C a featureless power-law spectrum. Additional observations in this range are crucial for distinguishing between the possible emission origins.
Future instruments such as CTAO~\citep[e.g.,][]{Acharya-2013APh....43....3A} and the Large Array of Imaging Atmospheric Cherenkov Telescopes (LACT)~\citep[e.g.,][]{Zhang-2025arXiv250920694Z} will be pivotal. With an expected sensitivity of $\sim 10^{-13}$~erg~cm$^{-2}$~s$^{-1}$ in the $0.1$-$100$~TeV range for a 50-hour exposure, these facilities may have difficulty detecting the signals predicted in Scenarios~A and~B, but they could detect the signal predicted in Scenario~C.

Around 10~TeV, different emission mechanisms dominated in Scenarios~A, B, and C, offering a potential way to distinguish between these scenarios through future variability observations.
Scenario~A was expected to produce strong flux variability, whereas Scenarios~B and~C were expected to exhibit weaker variability because their emission originated from larger spatial regions. In Scenario~A, the emission arose from the inner blob, which had short dynamical timescales of $10^{-3}$-$10^{-2}~\mathrm{s}$ and could therefore vary rapidly. In Scenario~B, the emission originated from the external stellar-wind region with dynamical timescales of $10$-$10^{2}~\mathrm{s}$, much shorter than the typical durations of the hard or soft states and the orbital period. This could result in modest fluctuations in the gamma-ray flux. Scenario~C, in contrast, involved a large-scale external region with a dynamical timescale of $\sim 10^8-10^9~\mathrm{s}$, much longer than both the orbital and state timescales, producing nearly steady and persistent emission.

We also investigated how the jet geometry, characterized by the jet azimuthal angle $\phi_j$ and polar angle $\vartheta_j$, influenced the predicted electromagnetic emission over the orbital phase $\phi$. Because of the large uncertainties in the modulation model, our calculations represented one plausible scenario rather than a definitive prediction. We assumed a fixed relative orientation $\Delta \phi = \phi - \phi_j$ and focused on the jet blob, since the GeV and PeV photons considered here originated from this region. A more detailed analysis, including the TeV photon contribution from the inner blob, was provided in the appendix. This simplified geometric framework could provide a possible interpretation of the energy-dependent orbital modulation reported by \citet{LHAASO-2025arXiv251216638L} among the PeV, GeV, and X-ray bands. 

In this framework, the modulation arose from phase-dependent Doppler boosting and external $\gamma\gamma$ absorption. A representative geometry with $\Delta \phi \approx -90^\circ$ reproduced the overall modulation pattern above $0.1$~PeV and was broadly consistent with the LHAASO observations. It also produced a phase shift of approximately $90^\circ$ between the $100$~GeV and $2$~PeV light curves. The nearly inverse modulation between the PeV and X-ray bands could be explained by the adopted jet geometry combined with periodic obscuration of the X-ray source by the donor star.
However, the model faced two main challenges: (1) it predicted similar $1$~GeV and $2$~PeV light curves, suggesting that either stronger external annihilation or additional mechanisms may be needed, and (2) while the Doppler-induced PeV flux variation was comparable to the observed modulation, the absorption-induced variation predicted at $100$~GeV could exceed the observed value by up to an order of magnitude. More detailed phase-dependent calculations would therefore be required in future work.

Finally, we calculated the expected neutrino event counts for all scenarios using current and next-generation detectors. Our results included muon and pion cooling, an effect not considered in previous studies, and indicated that earlier predictions may have been biased toward higher neutrino fluxes. This cooling effect also affected the gamma-ray-to-neutrino connection, which is relevant for multimessenger interpretation.
For Cygnus~X-1, detection with IceCube or IceCube-Gen2 was highly unlikely for $\Delta t_{\rm int} = 20~$yr, and remained challenging even with a 30 km$^3$ detector~\citep[e.g.,][]{Huang:2023mzt}. For Cygnus~X-3, the prospects were more promising: IceCube-Gen2 could exceed the atmospheric background around 100~TeV after a total observing time of approximately 46~yr, although the expected event number remained small. A 30 km$^3$ detector could further enhance the detectable signal from Cygnus~X-3 by a factor of $\sim 2$ in the $0.1$-$1$~PeV range.
These findings indicated that discovering neutrino emission from microquasars will likely require a future generation of more sensitive detectors.

\begin{acknowledgments}
We thank the anonymous referee for valuable comments and suggestions that helped improve this work. We thank Abraham Falcone, Derek Fox, David Radice, Jieshuang Wang, Ruo-Yu Liu, and Stephanie Wissel for their valuable suggestions, and Abhishek Das and Mainak Mukhopadhyay for their helpful support. 
B.T.Z. is supported in China by National Key R\&D program of China under the grant 2024YFA1611402.
Y.-J.W. and K.M. acknowledge support from the Institute for Gravitation and the Cosmos of The Pennsylvania State University. K.M. also thanks Mamoru Yanagisawa for his generous donation and continued support.
\end{acknowledgments}

\appendix

\section{Cooling Timescales for Leptons and Pions}
\label{sect:t_cool}

\begin{figure*}[ht!]
	\centering
	\subfigure[]{
		\includegraphics[scale=0.3]{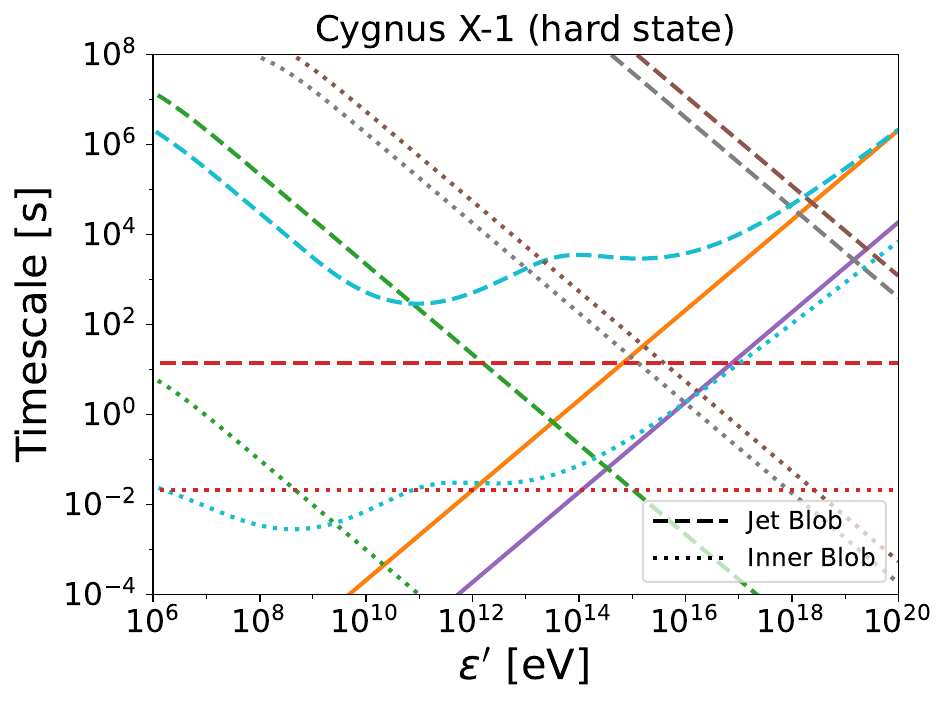}
	}%
    \subfigure[]{
		\includegraphics[scale=0.3]{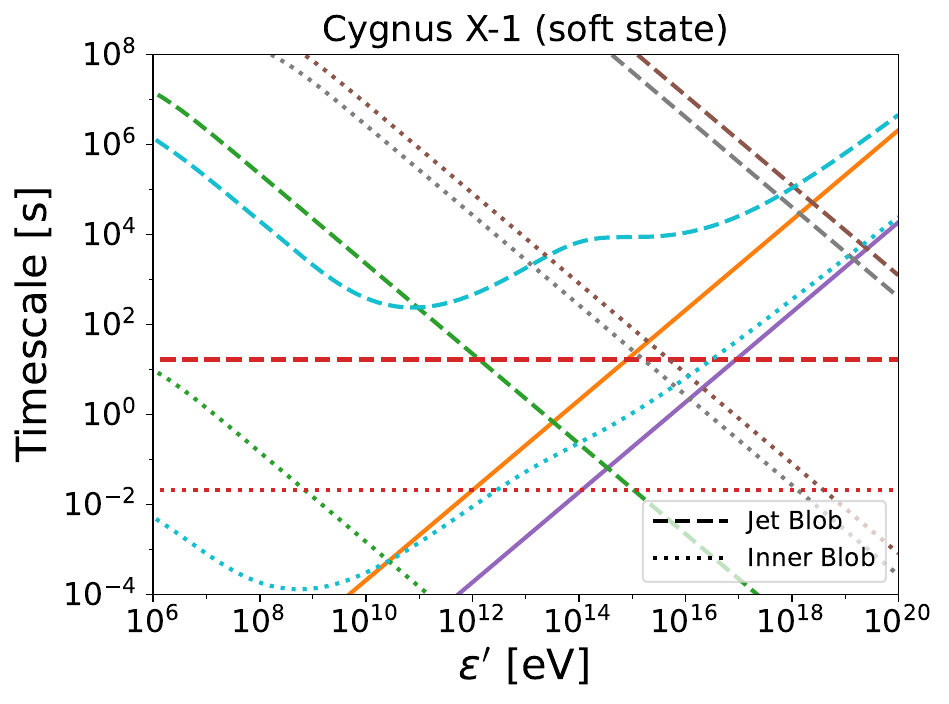}
	}%
	\subfigure[]{
		\includegraphics[scale=0.3]{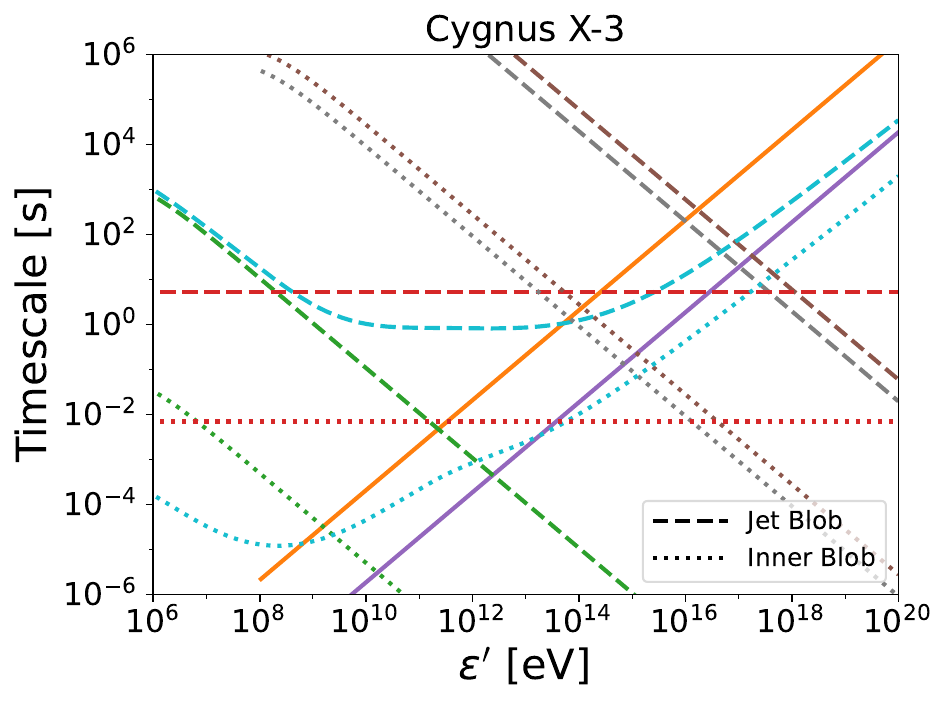}
	}%
    
	\subfigure[]{
		\includegraphics[scale=0.3]{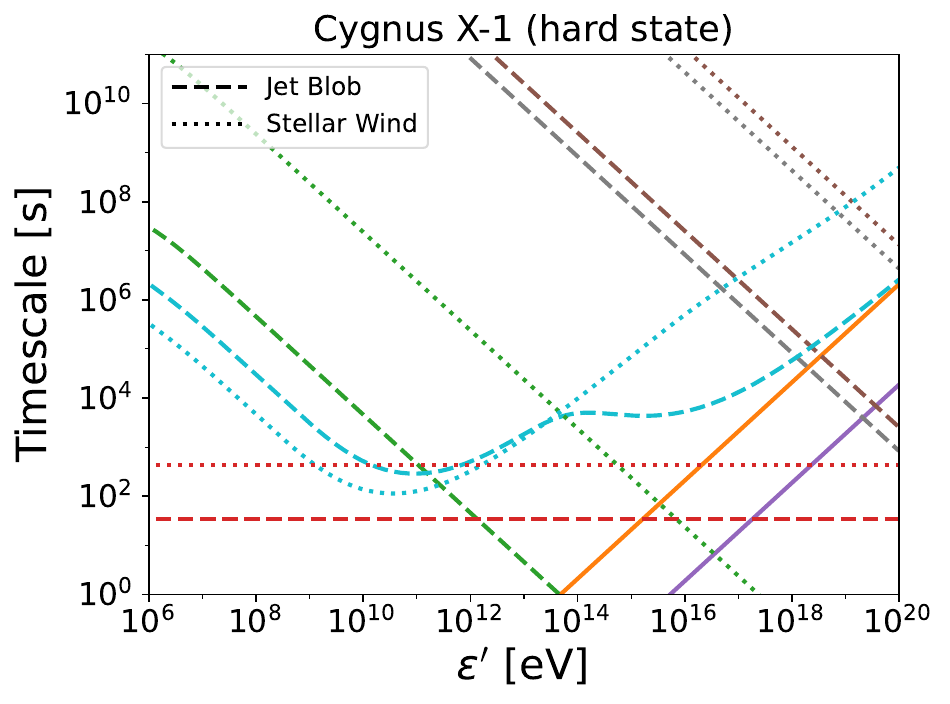}
	}%
    \subfigure[]{
		\includegraphics[scale=0.3]{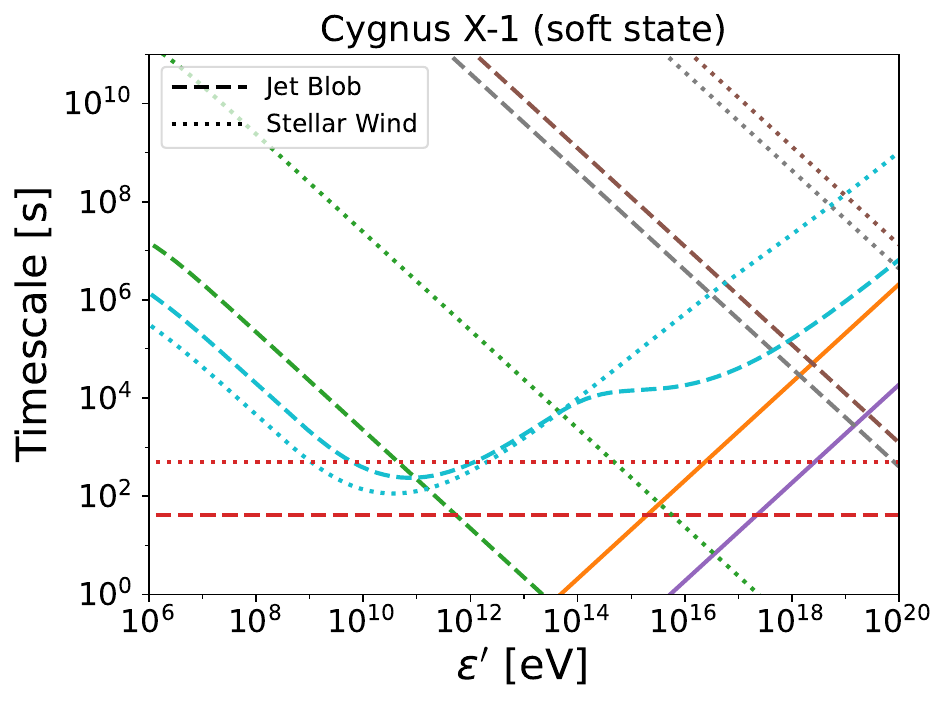}
	}%
	\subfigure[]{
		\includegraphics[scale=0.3]{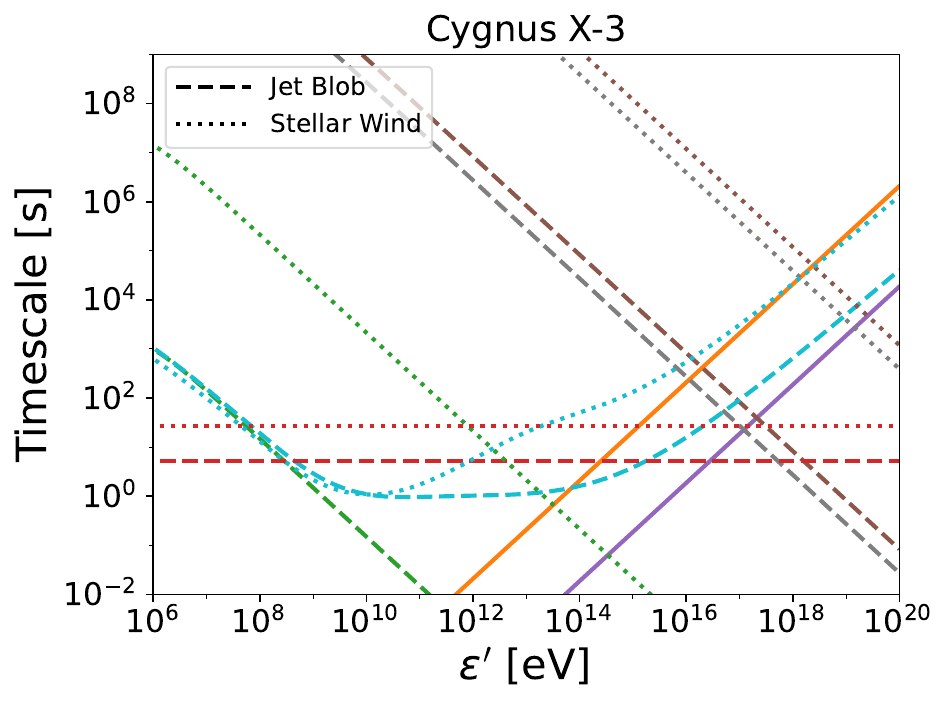}
	}%

	\subfigure[]{
		\centering
		\includegraphics[scale=0.3]{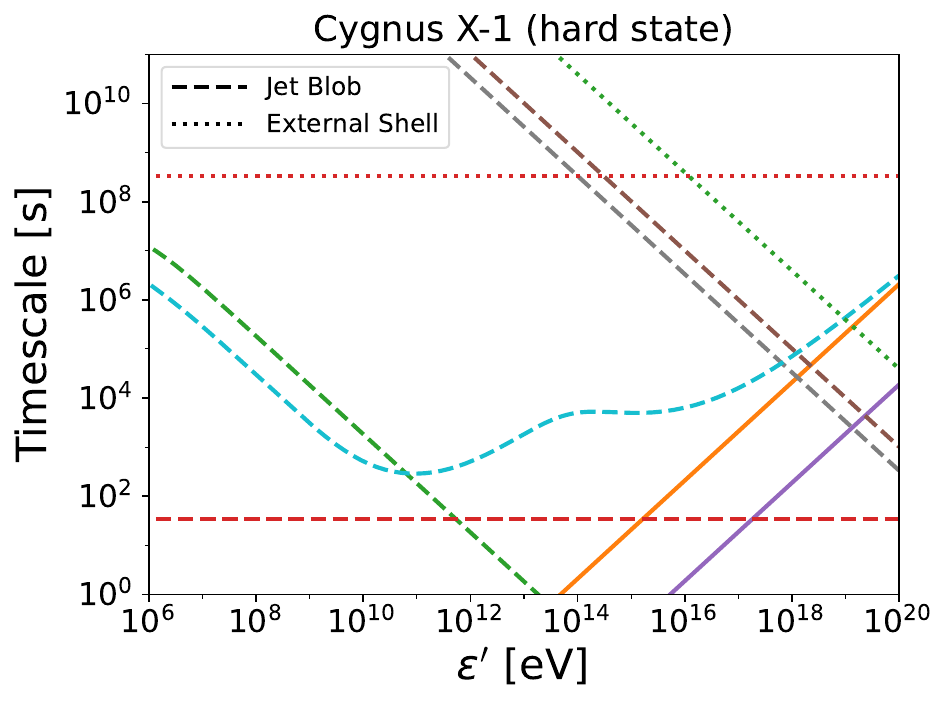}
	}%
	\centering
	\includegraphics[scale=0.3]{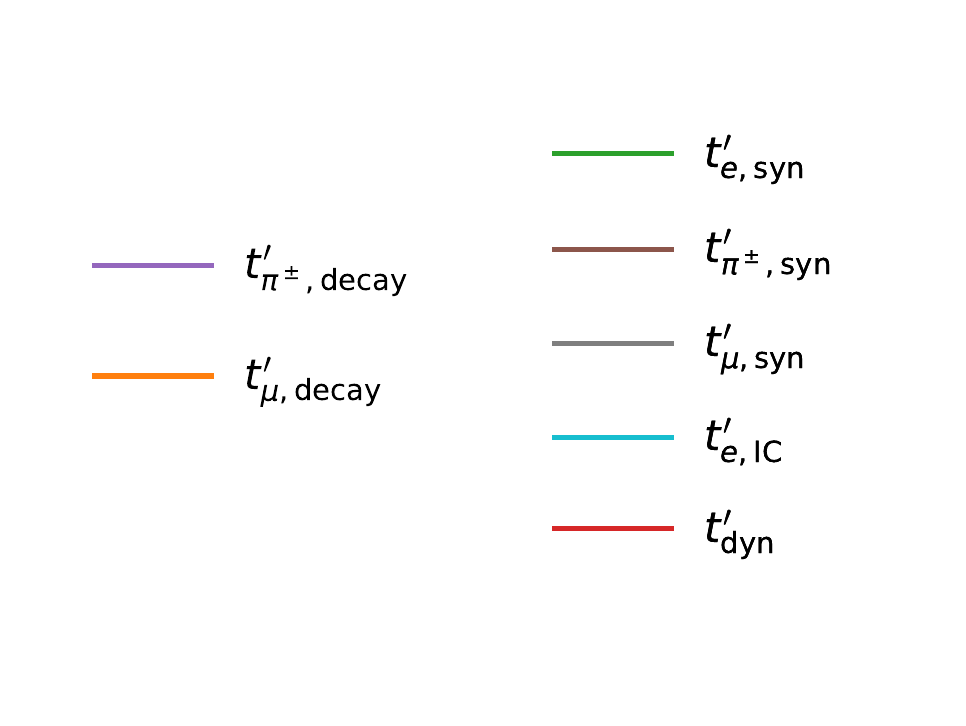}
	\subfigure[]{
		\centering
		\includegraphics[scale=0.3]{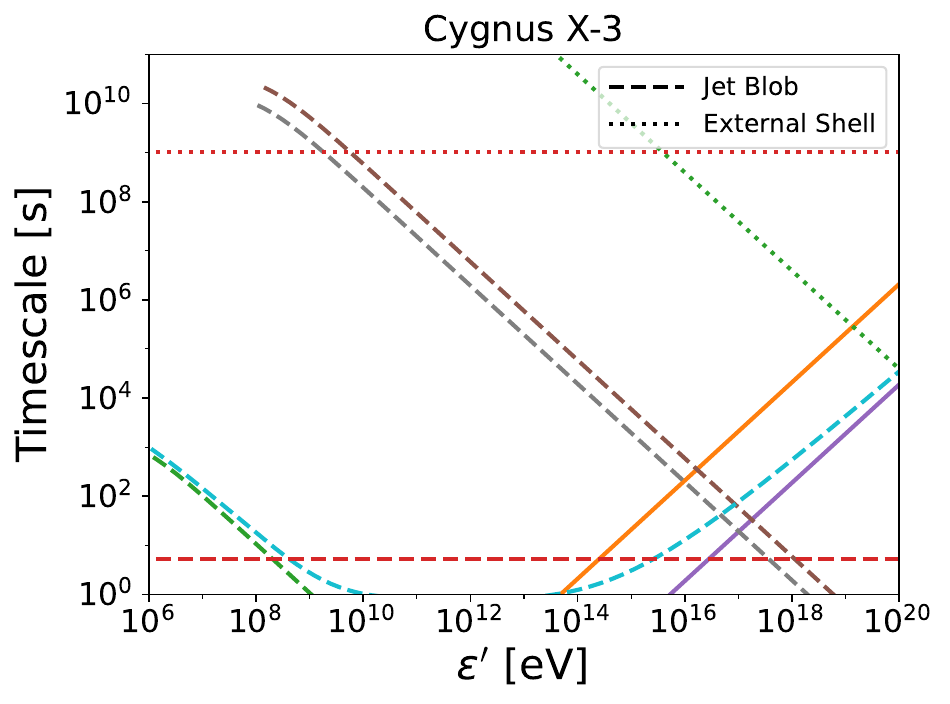}
    }%
		
	\caption{Cooling and decay timescales in the co-moving frame. Synchrotron emission of electrons, muons, and charged pions is shown in green, gray, and brown, respectively, while electron IC cooling is shown in light blue. The decay timescales of muons and charged pions are indicated by the orange and purple lines, respectively, and the dynamical timescale is shown in red. Colors represent different physical processes, while line styles denote different emission regions, with decay timescales shown as solid lines. The first, second, and third rows correspond to Scenarios~A, B, and C, respectively.}
	\label{fig:t_cool}
\end{figure*}

Fig.~\ref{fig:t_cool} shows the relevant cooling and decay timescales for leptons and pions, based on the parameters adopted in Sect.~\ref{sect:results}. Although muon IC cooling is included in our calculations, its timescale is not shown because it is subdominant over most of the energy range compared to muon synchrotron cooling.
We find that accounting for finite decay timescales, rather than assuming instantaneous decay, significantly impacts the treatment of high-energy muons and charged pions in most emission regions. The only exception is the external shell since its dynamical timescale is much longer.
Muon and pion synchrotron cooling is especially important in the inner blob region, where the magnetic field is relatively strong. Together, these effects modify the production and spectral shapes of high-energy photons and neutrinos.

\section{Influence of Geometry on Observed Electromagnetic Signals}
\label{sect:geometry}

\begin{figure*}[ht!]
	\centering
	\subfigure[]{
		\includegraphics[scale=0.3]{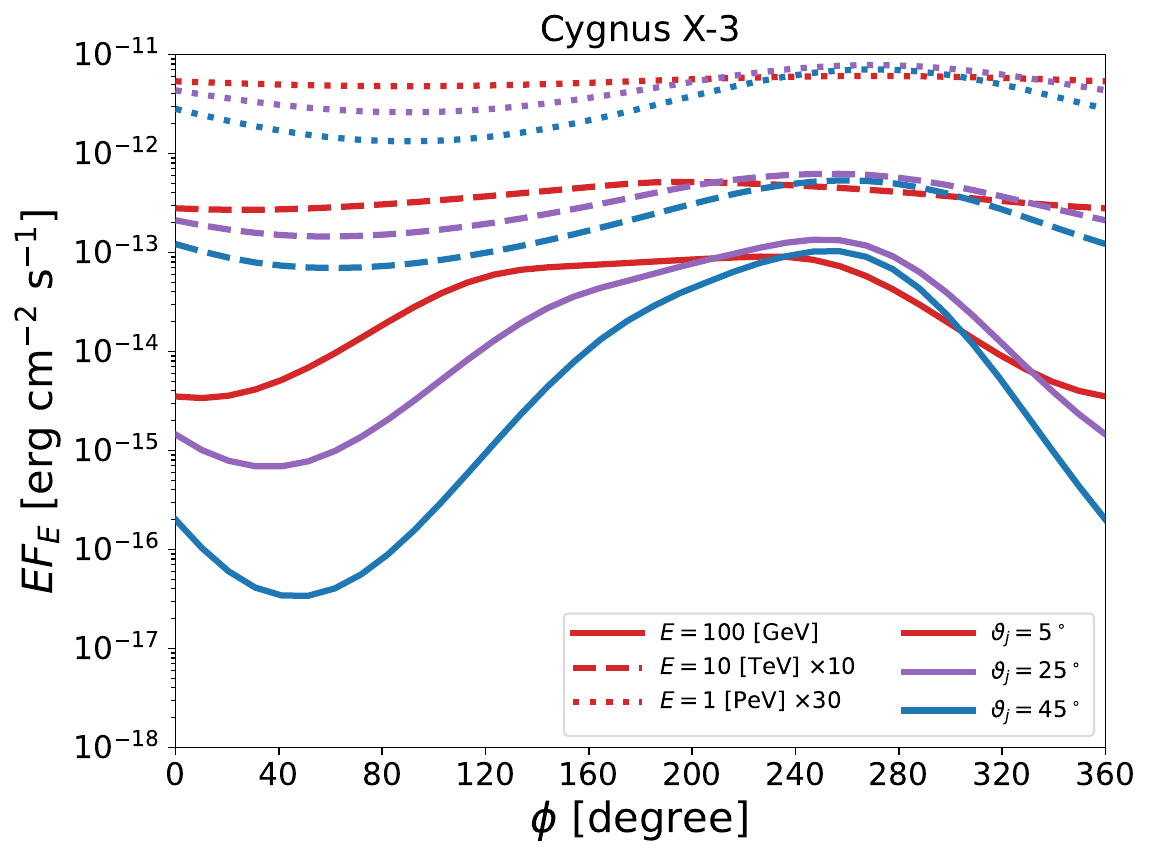}
	}%
    \subfigure[]{
		\includegraphics[scale=0.3]{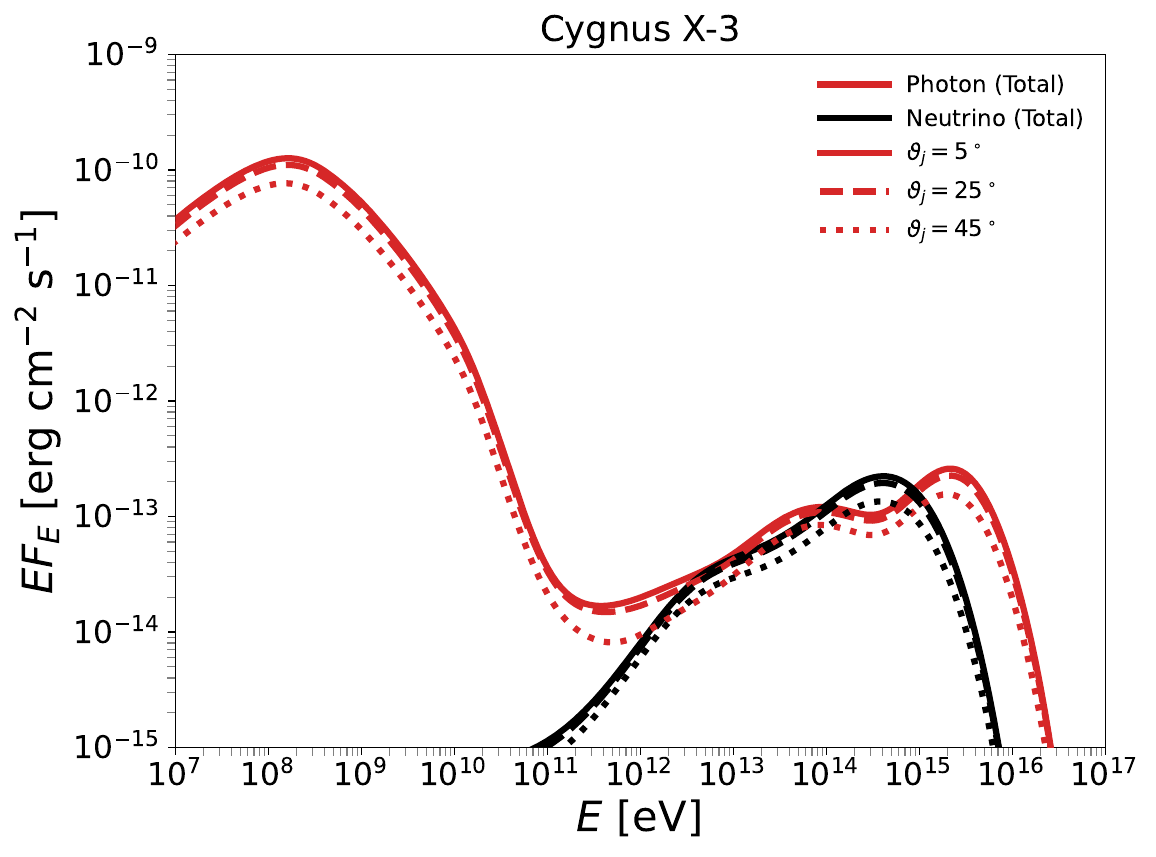}
	}%

	\subfigure[]{
		\includegraphics[scale=0.3]{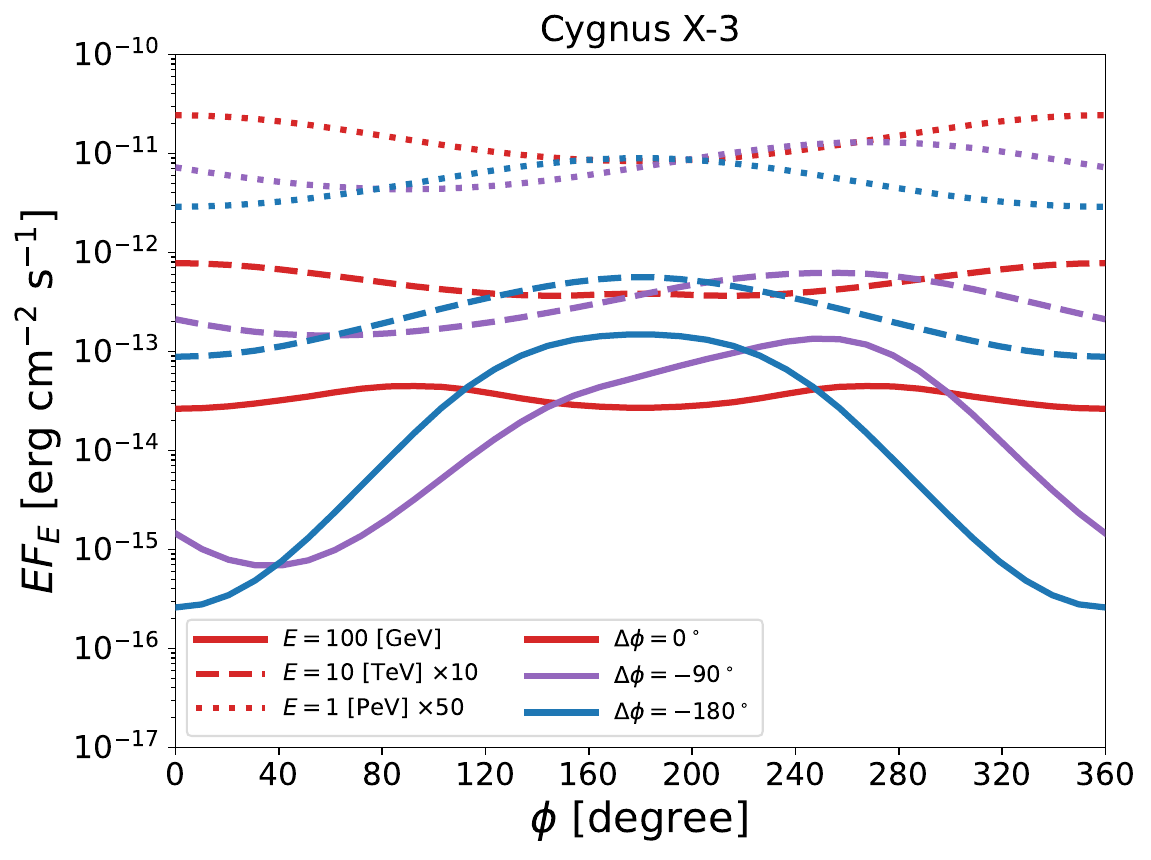}
	}%
    \subfigure[]{
		\includegraphics[scale=0.3]{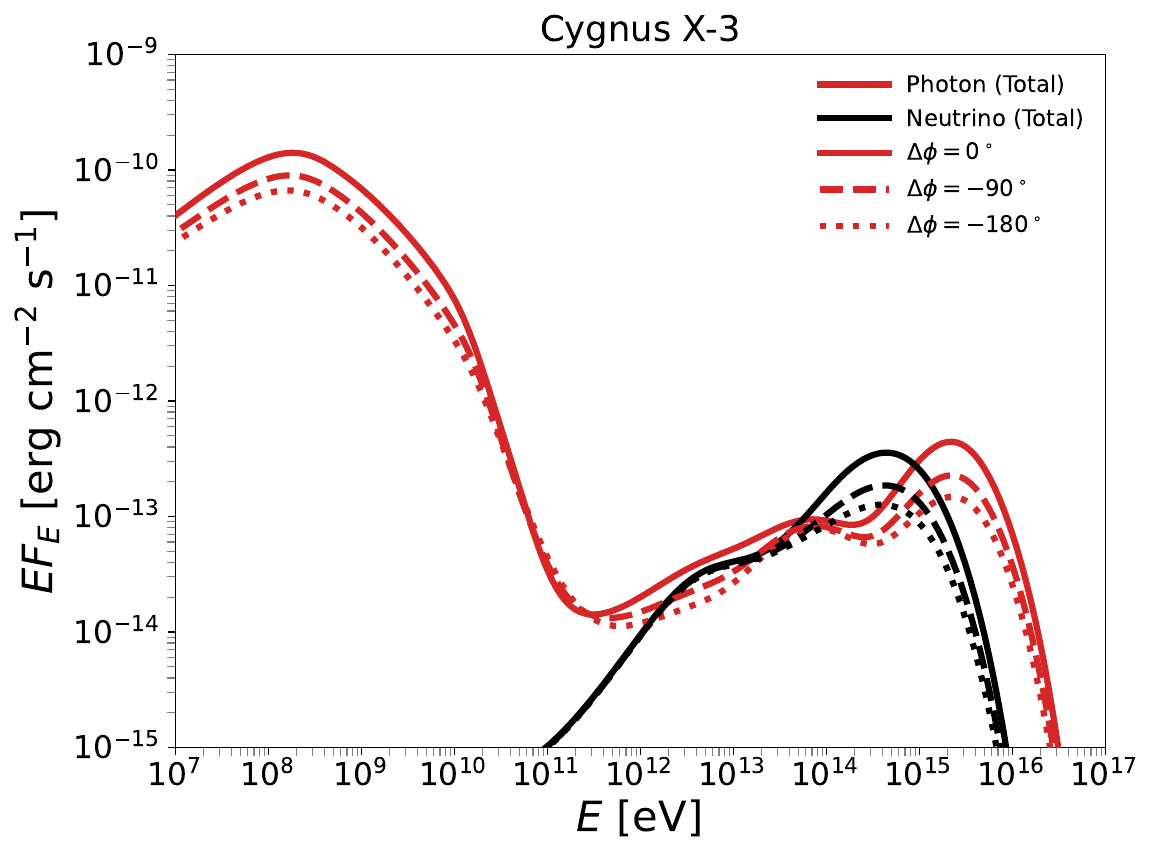}
	}%
    
	\caption{Orbital modulation of Cygnus~X-3 in Scenario~A for different jet-geometry parameters, $\vartheta_j$ and $\Delta \phi$. The first row shows results for different jet polar angles $\vartheta_j$ with fixed $\Delta \phi = -90^\circ$, while the second row shows results for different phase offsets $\Delta \phi$ with fixed $\vartheta_j = 25^\circ$. The first column shows the flux as a function of orbital phase for three representative energy bands, 100~GeV, 10~TeV, and 1~PeV, with different line styles. The second column shows the orbital-phase-averaged photon (red) and neutrino (black) spectra.
	}
\label{fig:flux_pgamma_varied}
\end{figure*}

Here, we investigate how the system geometry, characterized by $\phi$, $\phi_j$, and $\vartheta_j$, affects the observed electromagnetic emission. As in Sect.~\ref{sect:modulation}, we keep the relative phase offset $\Delta \phi = \phi - \phi_j$ fixed throughout the orbit. The flux variation is mainly driven by changes in the Doppler factor and external $\gamma\gamma$ absorption. We focus on Scenario~A, because the external emission regions in Scenarios~B and~C are static and more distant from the central engine, making their emission less sensitive to orbital geometry. The calculations presented here should be regarded as one plausible illustrative scenario rather than a definitive prediction.

Using Cygnus~X-3 as an example, we vary $\phi$, $\Delta \phi$, and $\vartheta_j$ while keeping all other physical parameters fixed to the values listed in Tables~\ref{tab:jet_blob} and~\ref{tab:pgamma}. Fig.~\ref{fig:flux_pgamma_varied} shows the resulting high-energy emission above $1$~GeV. Different choices of $\Delta \phi$ and $\vartheta_j$ can change the predicted flux by up to an order of magnitude. The modulation also differs among energy bands because of the energy dependence of external $\gamma\gamma$ absorption and the different dominant emission regions.
The first column of Fig.~\ref{fig:flux_pgamma_varied} shows the orbital-phase dependence of the flux for representative energy bands. Larger values of $\vartheta_j$ lead to stronger modulation, while different values of $\Delta \phi$ also affect the modulation pattern.
The second column shows the orbital-phase-averaged spectra over a full orbit, supporting the same conclusions.

In our models, photons from $1$ to $100$~TeV can originate from different regions, including the inner blob, the stellar-wind region, and the large-scale external region, depending on the scenario. In Scenario~A, the phase relation of the 10~TeV emission to the 100~GeV and PeV bands can be opposite, shifted, or nearly in phase, depending on the geometry, mainly because of strong external $\gamma\gamma$ absorption. 
If 10~TeV photons are produced within the jet blob under the assumption of an intrinsic X-ray luminosity ten times larger, their behavior will become similar to that at PeV energies, offering a potential observational discriminant between the models.
In contrast, Scenarios~B and~C are expected to show weaker orbital modulation because their emission arises from more extended, static regions that are less affected by orbital motion.
Future observations of high-energy gamma-ray modulation will therefore be important for constraining the jet geometry and identifying the dominant radiation mechanisms.

These results show that phase-dependent Doppler boosting and external $\gamma\gamma$ absorption can generate pronounced orbital modulation. However, our calculations do not fully include anisotropic radiative processes, which may introduce additional orbital modulation from radio to PeV energies. We also assume a circular orbit ($e = 0$), whereas an eccentric orbit could produce stronger modulation. A more comprehensive phase-dependent treatment will be required in future work.


\bibliography{ref}{}
\bibliographystyle{aasjournalv7}



\end{CJK*}

\end{document}